\documentclass[aps,prx,superscriptaddress,
notitlepage,reprint,floatfix,amsmath,amssymb]{revtex4-2}

\usepackage{SIunits}
\usepackage{xcolor}
\usepackage{graphicx}
\graphicspath{{figures/}}         

\usepackage[caption=false]{subfig} 
 \usepackage{stmaryrd} 
\usepackage{hyperref}
\usepackage{bbm}
\usepackage{mathrsfs}

\newcommand{\vb}[1]{\mathbf{#1}}

\newcommand{\Nie}[3]{\mathop{{#1}_{#2}^{#3}}}

\newcommand{\Vie}[3]{\mathop{\mathbf{#1}_{#2}^{#3}}}

\newcommand{\Exp}[1]{\mathop{\mathbb{E} \left[ #1 \right]}}

\newcommand{\Var}[1]{\mathop{\mathbb{V}\mathrm{ar} \left[ #1 \right]}}

\newcommand{\Ket}[1]{\mathop{ | #1 \rangle}}
\newcommand{\Bra}[1]{\mathop{ \langle #1 |}}

\newcommand{\BraKet}[2]{\mathop{ \langle #1 | #2 \rangle}}

\newcommand{\hv}{\hbar \omega}

\newcommand{\psic}{\Nie{\psi}{\mu  }{(c)}}
\newcommand{\psiv}{\Nie{\psi}{\nu }{(v)}}

\newcommand{\chic}{\Nie{\chi}{\mu }{(c)}}
\newcommand{\chiv}{\Nie{\chi}{\nu }{(v)}}

\begin{document}


\title{Wigner-Weyl description of light absorption in disordered semiconductor alloys using the localization landscape theory}

\author{Jean-Philippe Banon}
\affiliation{Laboratoire de Physique de la Mati\`{e}re Condens\'{e}e, CNRS, Ecole Polytechnique, Institut Polytechnique de Paris, 91120 Palaiseau, France}

\author{Pierre Pelletier}
\affiliation{Laboratoire de Physique de la Mati\`{e}re Condens\'{e}e, CNRS, Ecole Polytechnique, Institut Polytechnique de Paris, 91120 Palaiseau, France}

\author{Claude Weisbuch}
\affiliation{Laboratoire de Physique de la Mati\`{e}re Condens\'{e}e, CNRS, Ecole Polytechnique, Institut Polytechnique de Paris, 91120 Palaiseau, France}
\affiliation{Materials Department, University of California, Santa Barbara, California 93106-5050, USA}

\author{Svitlana Mayboroda}
\affiliation{School of Mathematics, University of Minnesota, Minneapolis, Minnesota 55455, USA}

\author{Marcel Filoche}
\affiliation{Laboratoire de Physique de la Mati\`{e}re Condens\'{e}e, CNRS, Ecole Polytechnique, Institut Polytechnique de Paris, 91120 Palaiseau, France}

\date{\today}

\begin{abstract}
The presence of disorder in semiconductors can dramatically change their physical properties. Yet, models faithfully accounting for it are still scarce and computationally inefficient. We present a mathematical and computational model able to simulate the optoelectronic response of semiconductor alloys of several tens of nanometer sidelength, while at the same time  accounting for the quantum localization effects induced by the compositional disorder at the nano-scale. The model is based on a Wigner-Weyl analysis of the structure of electron and hole eigenstates in phase space made possible by the localization landscape theory. After validation against eigenstate-based computations in 1D and 2D, our model is applied to the computation of light absorption in 3D~InGaN alloys of different compositions. We obtain the detailed structures of the absorption tail below the average bandgap and the Urbach energies of all simulated compositions. Moreover, the Wigner-Weyl formalism allows us to define and compute 3D~maps of the effective locally absorbed power at all frequencies. Finally the proposed approach opens the way to generalize this method to all energy-exchange processes such as radiative and non-radiative recombination in realistic devices.
\end{abstract}

\maketitle 

\section{Introduction}

Semiconductor structures used for fundamental studies or device applications most often incorporate alloy materials. The necessity of using alloys results from the incapacity of associations of pure compounds to reach the desired functions or from fabrication issues due to lattice parameters mismatch. For “common” III-V alloys, based on GaInAsP or GaInAlAs materials systems, the effects of compositional disorder inherent to random alloys on the electronic properties can be treated with a perturbative approach. This is unfortunately not the case for the more recent nitride-based GaInAlN alloys, where the changes in potential associated with the various atoms induce strong localization effects. While considerable progress has been made in past decades using such materials for high performance light generation devices, these materials and their uses in heterostructures require new tools to model their properties. Conversely, they constitute a unique laboratory to evaluate strong disorder effects due to the large difference in band gap energy and band offset between the pure compounds.

The simplest phenomena of absorption and luminescence are of primary importance for the characterization of semiconductor alloys as they reveal information about the electronic properties~\cite{Stern:book, Weisbuch:book, Singh_2003}. Absorption near the band edges is of particular interest due to its sensitivity to temperature, impurities, Coulomb interaction, and alloy disorder. It is also a much simpler phenomenon than luminescence to analyze as it directly probes the electronic band structure without the energy and momentum relaxation involved in luminescence. Phenomenological laws have been proposed to describe the behavior observed near the absorption edge for crystalline and amorphous semiconductors, such as the Tauc power laws just above the edge~\cite{Tauc:1966, Tauc:1968}, or the Urbach exponential law just below the edge~\cite{Urbach:1953}. It is now accepted that the wide variety of behaviors near the absorption edge in semiconductors may be caused by thermal effect, micro-field distribution~\cite{Dow_1972},  electron-hole Coulomb interaction~\cite{Elliott:1957}, alloy disorder~\cite{Piccardo:2017}, or the joint effect of the latter two~\cite{David:2019}. Alloy disorder can impact the absorption and emission spectra in different ways depending on the type of atomic species~\cite{Popescu:2010, Popescu:2012}, and on the type of disorder. One may encounter uncorrelated alloy disorder~\cite{Yang:2014} or correlated alloy disorder exemplified by (i) spinodal phase separation~\cite{Han:2006}, (ii) clustering~\cite{DiVito:2020}, or even (iii) the formation of pure crystalline quantum dots~\cite{Odonnell:1999, Bayliss:1999}. In practice, correlated alloys effects remain controversial in nitrides as (i) they have been shown to occur due to the degradation of the materials by the observation technique, (ii) they are debated due to the limited efficiency of atomic probe tomography (APT) in comparison with modern high resolution transmission electron microscopy,  and (iii) their observation corresponds to vastly non-optimum growth regimes, yielding microstructures never observed in industry-grade materials. 

Modeling and numerical simulation of light absorption in disordered semiconductor alloys are challenging tasks. First they require an appropriate model for the electronic structure. A hierarchy of methods exists to model the electronic structure of alloys, going from density functional theory which can be considered as a first principle method, via the tight binding method, to continuous effective models (see the tutorial~\cite{DiVito:2020} and references therein). Once a model for the electrons is chosen, there comes the second step of computing the absorption coefficient by considering the interaction with light. Several methods exist which use either directly Fermi's golden rule and require the computations of all involved electronic eigenstates of the Hamiltonian, or time dependent simulation of the polarization field~\cite{David:2019,Glutsch:1996}. \\

In the present paper, we derive a model based on the localization landscape (LL) theory~\cite{Filoche:PNAS} for disordered semiconductor alloys, which we apply to the computation of the light absorption coefficient in bulk InGaN. In Section~\ref{sec:system}, we start by introducing a continuous model of disordered band edges based on the regularization of the indium concentration from randomly placed indium atoms on a lattice. Two Schr\"odinger equations (one per band) are then written for the envelope functions of the electron and hole eigenstates. In Section~\ref{sec:theory}, exploiting this framework, we derive an exact formulation of the absorption process in phase space based on the Wigner transform of the eigenstates and the Weyl transform of the Hamiltonians. We then identify quasi-densities of states in phase space, which lead in turn to closed form approximations for the absorption coefficient and for the absorbed power density.  The results are presented in two sections: Section~\ref{sec:benchmark} is devoted to the benchmark of the landscape-based model for the absorption coefficient for 1D and 2D~systems by comparison with the eigenstate-based computation,  while Section~\ref{sec:3d} presents 3D~simulations in large samples (above \unit{100 \: 000}{nm^3}), from which we extract the characteristics of the absorption response in InGaN disordered alloys. Finally, we conclude by providing perspectives on the generalization of the presented method to a broader class of electronic processes.

\section{Disordered semiconductor alloys}\label{sec:system}

\begin{figure*}[t]
\centering
\includegraphics[width = 0.32\textwidth, trim = 1cm 0cm 1cm 0cm,clip]{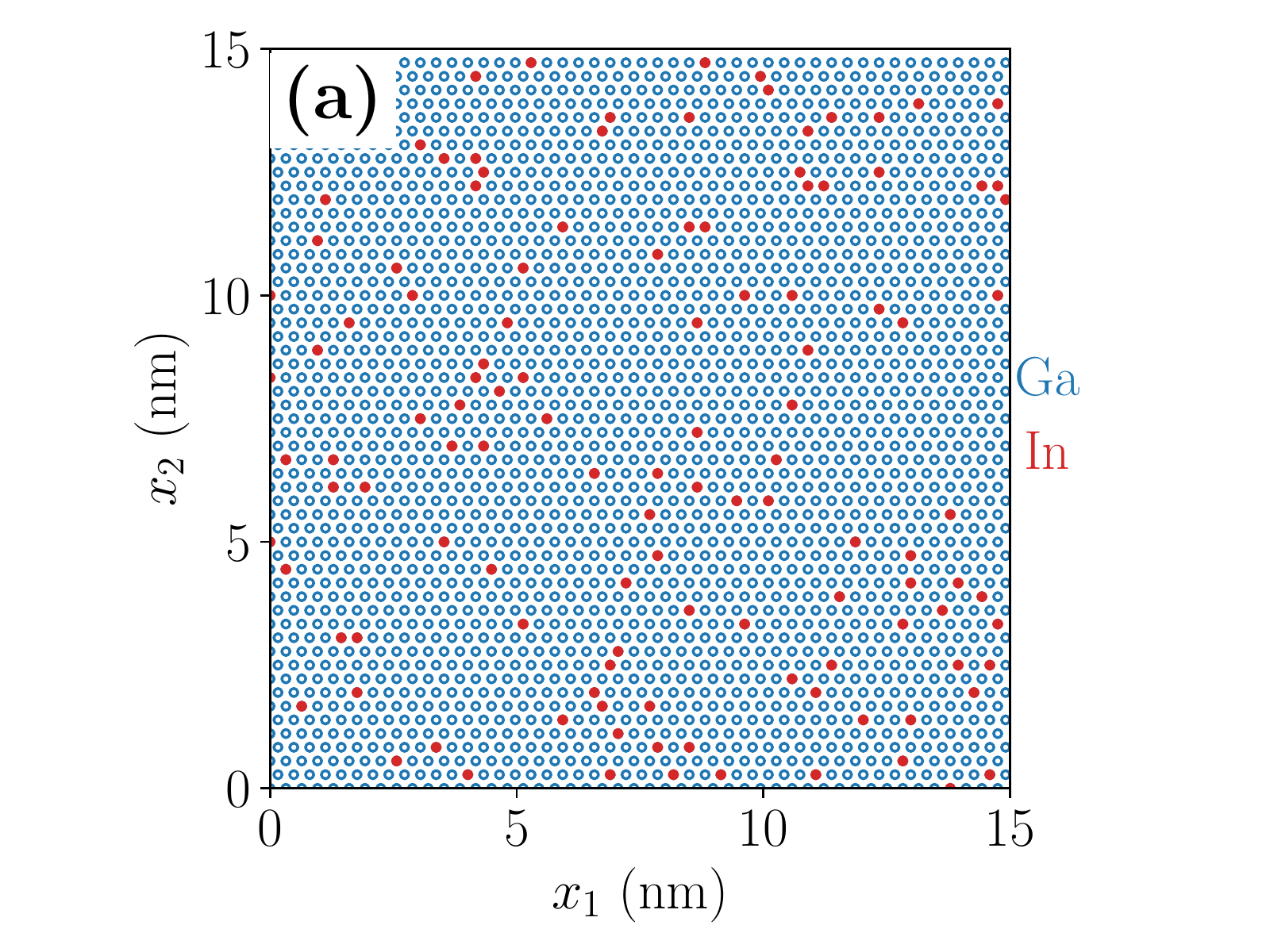}
\includegraphics[width = 0.32\textwidth, trim = 2cm 0cm 0cm 0cm,clip]{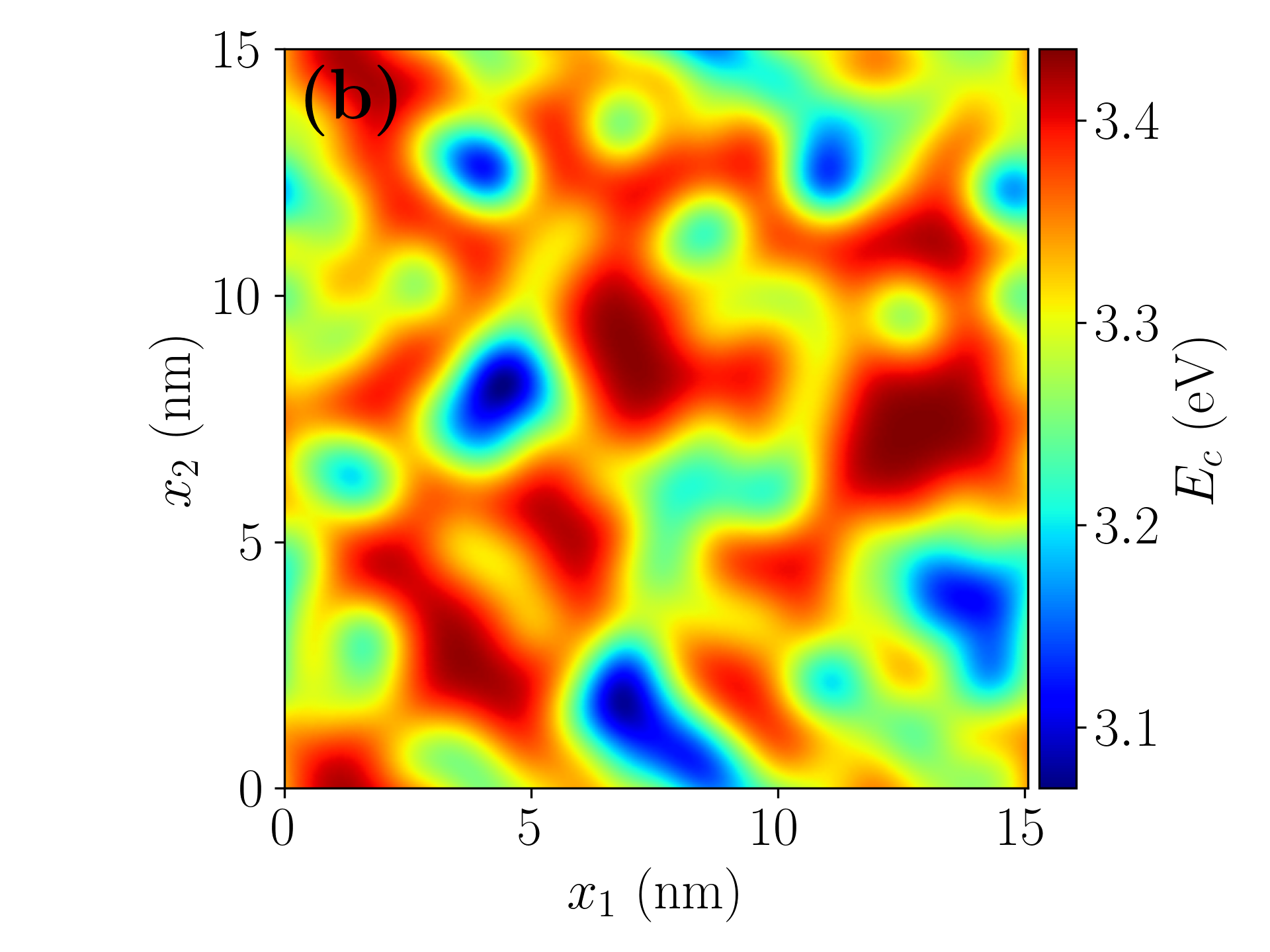}
\includegraphics[width = 0.32\textwidth, trim = 1.5cm 0cm .5cm 0cm,clip]{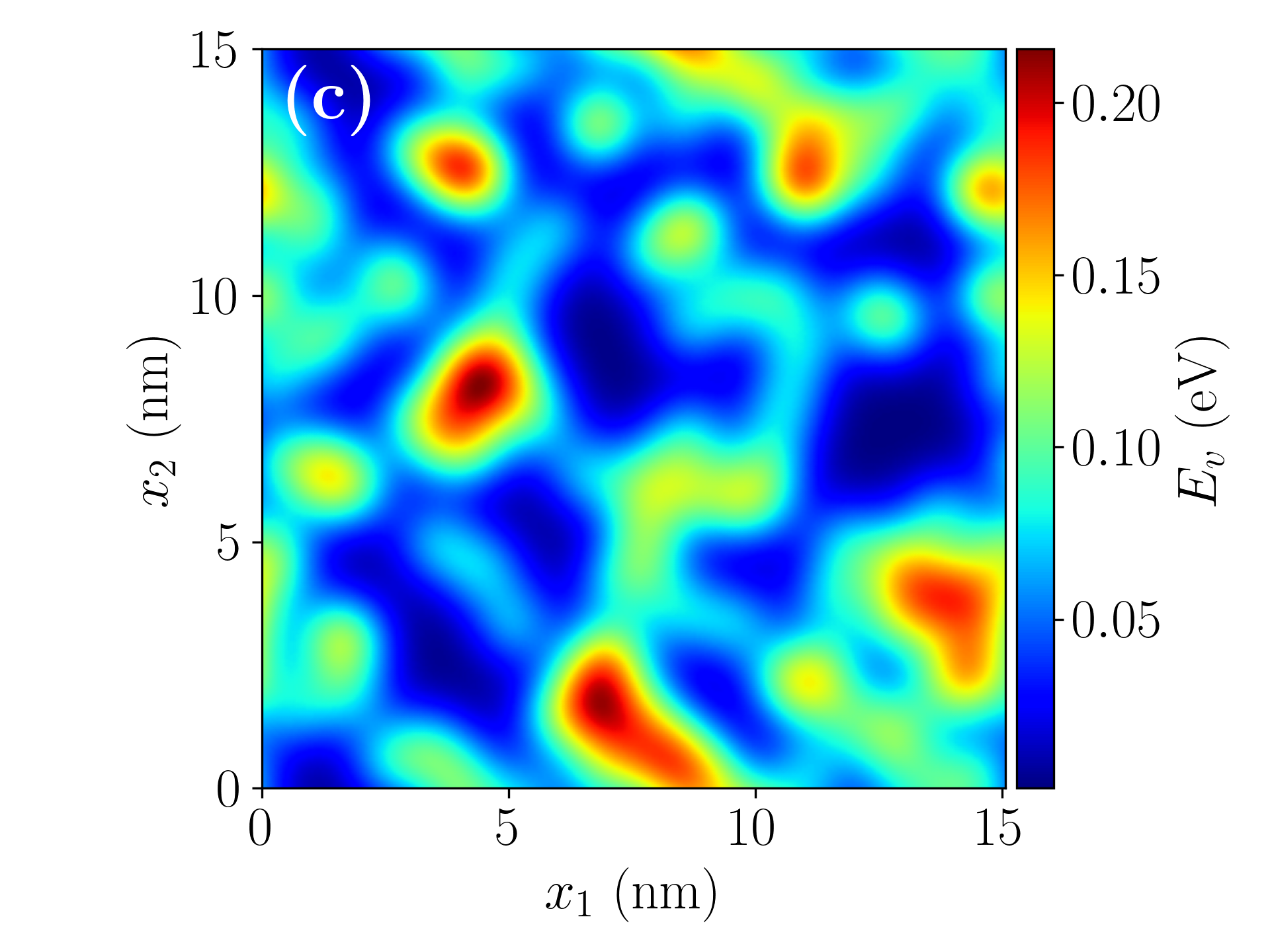}
\caption{Realization of a two-dimensional $\mathrm{In}_{0.05}\mathrm{Ga}_{0.95}\mathrm{N}$ alloy. (a) Atomic configuration. Open blue circles denote Ga atoms and red disks denote In atoms. (b) Conduction and (c) valence potentials obtained from Eqs.~(\ref{eq:Ec}) and (\ref{eq:Ev}). The smearing length was set to $\sigma = 2 a \approx 6.4$~\r{A}.}
\label{fig:alloy}
\end{figure*}

\subsection{The effective mass approximation}

We work within the framework of the effective mass approximation (EMA).  The alloy,  consisting of randomly drawn atoms on a lattice,  is described by continuous position-dependent conduction and valence potentials, and effective masses.  These profiles are obtained locally from a Gaussian averaging of the atomic composition.  In mathematical terms, we denote by $\Vie{r}{i}{}$ the position of the cation site $i \in \mathcal{I}$ on the lattice (where $\mathcal{I}$ is an arbitrary set of indices), and $X_i$ a Bernoulli random variable taking values 0 with probability $1-x$ or 1 with probability $x$ corresponding to whether a Ga atom or an In atom is found at site $i$.  We define the continuous local indium concentration $X(\vb{r})$ as
\begin{equation}
X(\vb{r}) = \frac{\displaystyle \sum_{i \in \mathcal{I}} X_i \exp \left( - \frac{|\vb{r} - \Vie{r}{i}{}|^2}{2 \sigma^2} \right) }{\displaystyle  \sum_{i \in \mathcal{I}} \exp \left(- \frac{|\vb{r} - \Vie{r}{i}{}|^2}{2 \sigma^2}  \right) } \: ,
\label{eq:gaussian:averaging}
\end{equation}
which is a Gaussian averaging of the discrete atomic composition with smearing length $\sigma$.  The $X_i$ being random variables,  this operation constitutes a continuous bounded stochastic process $X(\vb{r})$.  Note that if the random variables $X_i$ are independent and identically distributed, then the mathematical expectation of $X(\vb{r})$ is $\Exp{X(\vb{r})} = \Exp{X_i} = x$ and its variance is
\begin{equation}
\Var{X(\Vie{r}{}{})} = x(1-x) f_\sigma(\vb{r})  \: ,
\end{equation}
with
\begin{equation}
f_\sigma(\vb{r}) = \frac{\displaystyle \sum_{i \in \mathcal{ I} }  \exp \left( - \frac{|\vb{r} - \Vie{r}{i}{}|^2}{\sigma^2}  \right) }{ \displaystyle \left[ \sum_{i \in \mathcal{ I} } \exp \left( - \frac{|\vb{r} - \Vie{r}{i}{}|^2}{2 \sigma^2}  \right) \right]^2 } \: .
\end{equation}
In particular the two last equations show, through the function $f_\sigma$, that the variance is position-dependent, $X(\vb{r})$ is therefore not a stationary process. In fact, the variance of $X$ has the periodicity of the lattice. Moreover the variance decreases with increasing $\sigma$, and also decreases with the dimension of the system since the denominator grows faster than the numerator with increasing number of neighboring sites around point $\vb{r}$. Indeed, it can be easily shown that $0 \leq f_\sigma \leq 1$, and that for a cubic lattice of lattice parameter $a$ and asymptotically for $\sigma / a \gg 1$, $f_\sigma$ becomes roughly constant and we have
\begin{subequations}
\begin{equation}
f_\sigma \sim \left[ \frac{1}{2 \sqrt{\pi}} \frac{a}{\sigma} \right]^d \; ,
\label{eq:varscaling:cubic}
\end{equation}
where $d$ denotes the space dimension.  For a 3D wurtzite lattice, the above expression becomes
\begin{equation}
f_\sigma \sim \frac{\sqrt{3} a^2 c }{4 \left( 2 \sqrt{\pi} \sigma \right)^3} \; .
\label{eq:varscaling:wurtzite}
\end{equation}
\label{eq:varscaling}
\end{subequations}
Keeping in mind the decaying variance of $X(\vb{r})$ with increasing smearing length and space dimension will be useful in comparing results in Secs.~\ref{sec:benchmark} and \ref{sec:3d}. \\

From the local concentration $X(\vb{r})$,  we deduce the position-dependent band gap energy, $E_g$, the conduction and valence potentials, $E_c$ and $E_v$, and the effective masses, $m_c$ and $m_v$, as follows:
\begin{align}
E_g(\vb{r}) = &(1 - X(\vb{r})) E_{g}^{(\mathrm{GaN})} + X(\vb{r}) E_{g}^{(\mathrm{InN})} \nonumber\\
&- E_{\mathrm{bow}} X(\vb{r}) (1-X(\vb{r})) \: , \label{eq:Eg}\\
E_c (\vb{r}) = &  E_{g}^{(\mathrm{GaN})} - \gamma \big( E_{g}^{( \mathrm{GaN})}  - E_g (\vb{r}) \big) \: , \label{eq:Ec} \\
E_v (\vb{r}) = & (1 - \gamma) \big( E_{g}^{(\mathrm{GaN})}  - E_g(\vb{r}) \big) \: , \label{eq:Ev}\\
m_c (\vb{r}) = & \left[ \frac{X(\vb{r})}{m_{e}^{( \mathrm{InN})}} + \frac{1 - X(\vb{r})}{m_{e}^{ (\mathrm{GaN})}}  \right]^{-1} \: , \\
m_v (\vb{r}) = & -\left[ \frac{X(\vb{r})}{m_{h}^{ (\mathrm{InN})}} + \frac{1 - X(\vb{r})}{m_{h}^{ (\mathrm{GaN})}}  \right]^{-1} \: . \label{eq:valence:mass}
\end{align}
Note that we have chosen to give a negative sign to the valence band effective mass in Eq.~\eqref{eq:valence:mass}. This will enables us to express energies both for the valence and conduction band states on the same energy axis.
Values of the band gap energies $E_{g}^{( \mathrm{GaN})}$, $E_{g}^{(\mathrm{InN})}$, the bowing energy $E_{\mathrm{bow}}$, the effective masses $m_{e}^{(\mathrm{InN})}$, $m_{h}^{(\mathrm{InN})}$, $m_{e}^{(\mathrm{GaN})}$, $m_{h}^{ (\mathrm{GaN})}$, and the band offset factor $\gamma$ (i.e.,  the fraction of the band offset $E_g^{(\mathrm{GaN})} - E_g^{(\mathrm{InN})}$ which is attributed to the conduction band) are extracted from the literature and are given in Table~\ref{tab1} and its caption.  As an illustration, a realization of the atomic configuration and of the conduction and valence potentials for a two-dimensional InGaN alloy are shown in Fig.~\ref{fig:alloy}. In this paper,  we only take one valence band into consideration, the one for heavy holes,  and neglect the contribution from light holes,  for simplicity.  Moreover,  since we use the InGaN alloy as a proxy for a disordered semiconductor, we neglect the piezoelectric fields of the true InGaN materials.\\

\begin{table}[t]
\begin{center}
\caption{Crystal parameters ($a$ and $c$), band structure parameters (band gap $E_g$, effective masses $m_e$, $m_{hh}$ and $m_{lh}$) and energy $E_p$ associated to the momentum matrix element for wurtzite GaN and InN. Bowing parameter for InGaN: 1.4 eV. Band offset factor $\gamma = 0.63$. Parameters extracted from Refs.~\cite{Piprek:ch2,Im:1997}.}
\begin{tabular}{l l l l l l l l l}
\hline
\hline
Alloy &  $a$ & $c$ & $E_g$ & $E_p$ & $m_e$ &  $m_{h h}$  &  $m_{lh}$ \\[.1cm]
        &  (\r{A}) & (\r{A}) &  (eV) & (eV)& ($m_0$)&   ($m_0$)&  ($m_0$) \\[.1cm]
\hline
GaN & 3.189 & 5.185 &3.437 &  9.9 & 0.21  & 1.87 & 0.14 \\
InN  & 3.545 & 5.703 &0.608   & 5.7 & 0.07 & 1.61  & 0.11 \\
\hline
\hline
\end{tabular}
\label{tab1}
\end{center}
\end{table}

\subsection{Conduction and valence states}

Let the domain of study be $\Omega = [0,L[^d$ with Born-von Karman periodic boundary conditions along the three axes $x_1$, $x_2$,  and $x_3$ [to which we assign the orthonormal basis $(\Vie{e}{1}{},\Vie{e}{2}{},\Vie{e}{3}{})$]. The crystal is assumed to be oriented such that the so-called $c$-axis is aligned with the $x_3$ direction.
The band edges being disordered, the eigenstates of the Hamiltonian in the semiconductor cannot be described by Bloch waves of the form $\psi^{(v)} = u_{v,\vb{k}}(\vb{r}) \exp(i \vb{k} \cdot \vb{r}) / |\Omega|^{1/2}$ and $\psi^{(c)} = u_{c,\vb{k}}(\vb{r}) \exp(i \vb{k} \cdot \vb{r}) /|\Omega|^{1/2}$, where $u_{v,\vb{k}}$ and $u_{c, \vb{k}}$ are lattice-periodic functions, as for a homogeneous crystalline semiconductor. Instead, we can assume the states to have the form
\begin{subequations}
\begin{align}
\psic (\vb{r}) &= u_c (\vb{r}) \, \chic (\vb{r}) \: , \\
\psiv (\vb{r}) &= u_v (\vb{r}) \,  \chiv  (\vb{r}) \: ,
\end{align}
\label{eq:quasi-bloch}
\end{subequations}
where $\chic$ and $\chiv$ are envelope functions satisfying
\begin{align}
\hat{H}_c \chic &= - \frac{\hbar^2}{2} \nabla \cdot \left[ \frac{\nabla \chic}{m_c} \right] + E_c \chic = E^{(c)}_\mu \chic \: ,\\
\hat{H}_v \chiv &= -\frac{\hbar^2}{2} \nabla \cdot \left[ \frac{\nabla \chiv}{m_v} \right] + E_v \chiv = E^{(v)}_\nu \chiv \! .
\label{eq:Schrodinger:z}
\end{align}
Here $\mu$, $\nu$ are arbitrary indices associated to the eigenenergies $E^{(c)}_\mu$ and $E^{(v)}_\nu$.  
  Note that in Bloch's theorem, the lattice-periodic functions depend,  in principle, on  wave vector $\vb{k}$. However, since we are primarily interested in the band edge part of the spectrum, we have assumed here that the $u_{v,\vb{k}}$ and $u_{c, \vb{k}}$ cell functions from Bloch's theorem depend weakly on $\vb{k}$ so they can be approximated by $u_{v,\Vie{0}{}{}}$ and $u_{c, \Vie{0}{}{}}$,  and which we have simply denoted $u_{v}$ and $u_{c}$ in Eq.~(\ref{eq:quasi-bloch}). 
  
  For completeness,  we report in Appendix~\ref{app:validity} a critical discussion on the validity of the effective mass approximation and its relevance for modeling InGaN.

\section{Light absorption}\label{sec:theory}

\subsection{Absorption in the EMA}

The transition rate for the excitation of an initial state in the valence band, $\Ket{\psiv}$, to a final state in the conduction band, $\Ket{\psic}$, by absorption of a photon of energy $\hbar \omega$, is given by Fermi's golden rule \cite{Stern:book,Weisbuch:book,Singh_2003},
\begin{align}
W_{\mu \nu} = \, &\frac{2\pi}{\hbar} \left( \frac{e}{2 m_0} \right)^2 \, \Big| \Bra{\psic} \Vie{A}{0}{} \cdot
 \Vie{\hat{p}}{}{} \Ket{\psiv} \Big|^2 \nonumber\\
 &\times \delta \Big( E_\mu^{(c)} - E_\nu^{(v)} - \hbar \omega \Big) \: ,
 \label{eq:Wji_rate}
\end{align}
where $E_\nu^{(v)}$ and $E_\mu^{(c)}$ are the energies of the initial and final states, respectively. The vector $\Vie{A}{0}{}$ is the amplitude of the electromagnetic vector potential, which we take to be a plane wave, i.e.,  $\Vie{A}{}{} = \Vie{A}{0}{} \cos(\Vie{k}{0}{} \cdot \vb{r} - \omega t)$, with $\Vie{k}{0}{} =  k_0 \Vie{e}{3}{}$ being the wave vector of the plane wave in the material.  The operator $\Vie{\hat{p}}{}{} = - i \hbar \nabla$ is the momentum operator.  We assume the optical  wavelength to be significantly larger than the typical scales of variations of the potentials to work within the dipole approximation, i.e., we regard the electromagnetic field as only a varying function of time.  Assuming the light intensity to be weak and the absence of doping, we can neglect stimulated emission and the total absorption rate is obtained by summing the elementary rates, in Eq.~(\ref{eq:Wji_rate}), over all states in the valence and conduction bands,
\begin{equation}
W_\mathrm{tot}( \omega) = 2 \sum_{\mu \nu} W_{\mu \nu} \: ,
\end{equation}
where the factor of 2 accounts for the spin degeneracy. 
We have seen that within the effective mass approximation, the wave functions $\psiv$ and $\psic$ have the form given in  Eq.~(\ref{eq:quasi-bloch}). 
Provided that the envelope functions $\chic$ and $\chiv$ vary slowly over the crystal unit cell, the matrix element $M_{\mu \nu} = \Bra{\psic} \Vie{A}{0}{} \cdot
 \Vie{\hat{p}}{}{} \Ket{\psiv} $ can be factorized as~\cite{Weisbuch:book} (see Appendix~\ref{app:momentum})
 \begin{equation}
M_{\mu \nu} =  \Bra{u_c} \Vie{A}{0}{} \cdot
 \Vie{\hat{p}}{}{} \Ket{u_v} \BraKet{\chic}{\chiv} \: .
 \label{eq:Mfactorized}
 \end{equation}
Note that this assumption may be questionable if $\sigma < a$ since then the potentials $E_c$ and $E_v$ vary on a scale of the order of a few lattice constants and so may the envelope functions.  For such low values of $\sigma$,  the use of the EMA should also be questioned anyway.  The total absorption rate can then be recast as
 \begin{equation}
 W_{\mathrm{tot}} ( \omega) = \frac{\pi e^2 A_0^2 E_p }{\hbar m_0} \,  \mathcal{C} ( \hv ) \: .
 \label{eq:Wtot:final}
 \end{equation}
 Here we denote $A_0 = |\Vie{A}{0}{}|$, $\Vie{a}{}{} = \Vie{A}{0}{} / A_0$, and let $E_p = | \Bra{u_c} \Vie{a}{}{} \cdot
 \Vie{\hat{p}}{}{} \Ket{u_v} |^2 / m_0$ be the energy associated to the momentum matrix element. For practical computation, we will take for $E_p$ a linear interpolation of the values for GaN and InN weighted by the average In concentration $x$~\cite{Hermann:1977}.   Moreover,  we define the \emph{spectral coupling density} as
 \begin{equation}
\mathcal{C} (\hv) = \sum_{\mu, \nu} \big| \BraKet{\chic}{\chiv} \big|^2 \, \delta \Big( E_\mu^{(c)} - E_\nu^{(v)} - \hbar \omega \Big) \: .
\label{eq:coupling_density}
 \end{equation}
A pair of modes contributes to the spectral coupling density at energy $\hbar \omega$ if their difference of energy is equal to $\hbar \omega$ (conservation of energy) and if there is a significant coupling factor $| \BraKet{\chic}{\chiv} |^2$.  Note that in the absence of disorder, the envelope functions are plane waves and the coupling factor yields the conservation of momentum as expected for homogeneous materials (see Appendix~\ref{app:flat:band} for the derivation in the homogeneous limit).
The photon flux through the surface area $S = L^2$ is given by the flux of the average Poynting vector along the $x_3$ direction, $\Pi$, divided by $\hv$,
\begin{equation}
\Phi = \frac{\Pi S}{\hv} = \frac{\varepsilon_0 \omega n(\omega) c_0 A_0^2 S}{2 \hbar} \: ,
\label{eq:photon:flux}
\end{equation}
where $\varepsilon_0$ is the vacuum permittivity, $c_0$ is the speed of light in vacuum,  and $n(\omega)$ is the real part of the refractive index of the material. The ratio $W_\mathrm{tot} / \Phi$ is thus the fraction of absorbed photons in the volume $\Omega$ along the propagation of a distance $L$ in the $x_3$ direction, which by definition of the absorption coefficient $\alpha$ is
\begin{equation}
\frac{W_\mathrm{tot} (\omega)}{\Phi} = \alpha(\omega) L \: ,
\label{eq:alphadef}
\end{equation}
provided $\alpha L \ll 1$. From Eqs.~(\ref{eq:Wtot:final}), (\ref{eq:photon:flux}), and (\ref{eq:alphadef}) we deduce the following expression of the absorption coefficient:
\begin{equation}
\alpha(\omega) = \frac{2 \pi e^2 E_p }{ m_0 \varepsilon_0 \omega c_0 n(\omega)} \,\frac{ \mathcal{C}(\hv) }{|\Omega|} \: .
\label{eq:alphaC}
\end{equation}

\subsection{Wigner transform and Weyl law}

According to Eq.~(\ref{eq:coupling_density}), the spectral coupling density for a given realization of the alloy requires computing the eigenstates $\chic$ and $\chiv$, which can be numerically costly, especially for 3D alloys.  Instead, we look for an alternative way to evaluate $\mathcal{C}(\hv)$ without resorting to solving the Schr\"{o}dinger equations. This will be achieved in two steps. First, we will rewrite Eq.~(\ref{eq:coupling_density}) using the Wigner transform of $\chic$ and $\chiv$, and reinterpret the spectral coupling density in terms of quasi-densities of states in phase space. Second, we will approximate the quasi-densities of states in phase space by exploiting the properties of the localization landscape.\\

To begin, we recall that the Wigner transform $W_{\psi}$ of a function $\psi$ is a distribution in phase space and is defined by~\cite{Mallat:ch4,Wigner}
\begin{equation}
W_{\psi} (\vb{r}, \vb{k}) =\int \psi^* \Big(\vb{r} - \frac{\Vie{x}{}{}}{2} \Big) \psi \Big(\vb{r} + \frac{\Vie{x}{}{}}{2} \Big) \, \exp \big( - i \vb{k} \cdot \Vie{x}{}{} \big) \: \mathrm{d}^d x  \: .
\end{equation}
%
There exist several conventions for the definition of the Wigner transform in the literature, differing in factors $2 \pi$ and $\hbar$ depending on whether one works with the wave vector $\vb{k}$ or the momentum $\Vie{p}{}{} = \hbar \vb{k}$. Here we have chosen the convention used in Ref.~\cite{Mallat:ch4}. 
 The square modulus of the scalar product, $|\BraKet{\chic}{\chiv}|^2$, appearing in Eq.~(\ref{eq:coupling_density}), can equivalently be written in terms of the Wigner transforms $W_{\chic}$ and $W_{\chiv}$ of $\chic$ and $\chiv$. This is done via Moyal's formula~\cite{Mallat:ch4}:
\begin{equation}
\left| \BraKet{\chic}{\chiv} \right|^2 = \iint W_{\chic}(\vb{r}, \vb{k}) W_{\chiv}(\vb{r}, \vb{k}) \: \frac{\mathrm{d}^d r \, \mathrm{d}^d k}{(2 \pi)^d}  \: .
\label{eq:moyal}
\end{equation}
Inserting Eq.~(\ref{eq:moyal}) into Eq.~(\ref{eq:coupling_density}), we obtain
\begin{widetext}
\begin{equation}
\mathcal{C} (\hv) =  \iint  \sum_{\mu, \nu}  W_{\chic} (\vb{r},\vb{k}) \, W_{\chiv} (\vb{r},\vb{k})  \, \delta(E_\mu^{(c)} - E_\nu^{(v)} - \hv) \: \frac{\mathrm{d}^d r \, \mathrm{d}^d k}{(2 \pi)^d} \: .
\label{eq:cmoyal} 
\end{equation}
\end{widetext}
Now,  to decouple the sums over $\mu$ and $\nu$, we write the Dirac mass $\delta(E_\mu^{(c)} - E_\nu^{(v)} - \hv)$ as the convolution product
\begin{equation}
\delta(E_\mu^{(c)} - E_\nu^{(v)} - \hv) = \int \delta(E_\mu^{(c)} - \hv - \varepsilon) \delta(E_\nu^{(v)} - \varepsilon) \: \mathrm{d}\varepsilon \: .
\label{eq:delta:conv}
\end{equation}
By inserting Eq.~(\ref{eq:delta:conv}) into Eq.~(\ref{eq:cmoyal}), we obtain
\begin{equation}
\mathcal{C}(\hv) = \iiint \mathcal{D}^{(c)}(\vb{r},\vb{k},\varepsilon + \hv)   \, \mathcal{D}^{(v)}(\vb{r},\vb{k},\varepsilon)  \: \mathrm{d}\varepsilon \,  \frac{\mathrm{d}^d r \, \mathrm{d}^d k}{(2 \pi)^d}  \: ,
\label{eq:coupling:Rdos}
\end{equation} 
where we have defined
\begin{subequations}
\begin{align}
\mathcal{D}^{(c)} (\vb{r},\vb{k}, E) &= \sum_\mu  W_{\chic} (\vb{r},\vb{k})  \, \delta(E_\mu^{(c)} - E) \: , \label{eq:ps:ldos:c}\\
\mathcal{D}^{(v)} (\vb{r},\vb{k}, E) &= \sum_\nu  W_{\chiv} (\vb{r},\vb{k})  \, \delta(E_\nu^{(v)} - E) \: . \label{eq:ps:ldos:v}
\end{align}
\label{eq:ps:ldos}
\end{subequations}
Since each Wigner function involved in the sum in Eq.~\eqref{eq:ps:ldos} corresponds to a quasi-probability density in phase space associated with an eigenstate,  the quantities $\mathcal{D}^{(c)} (\vb{r},\vb{k}, E)$ and $\mathcal{D}^{(v)} (\vb{r},\vb{k}, E)$ can be interpreted as \emph{quasi-densities of states in phase space} at energy $E$ for the conduction and valence band, respectively.  The quasi-density of states in phase space is in fact tightly linked with the usual densities of states, such as the local and integrated density of states and the spectral function,  since the latter are recovered as marginal integrations of the quasi-density of states in phase space as shown in Appendix~\ref{app:IDOS} and expressed below in Eq.~\eqref{eq:LDOSP:IDOS}.

\begin{figure*}[t]
\centering
\includegraphics[width = 0.4\textwidth, trim = .3cm .3cm .5cm .4cm,clip]{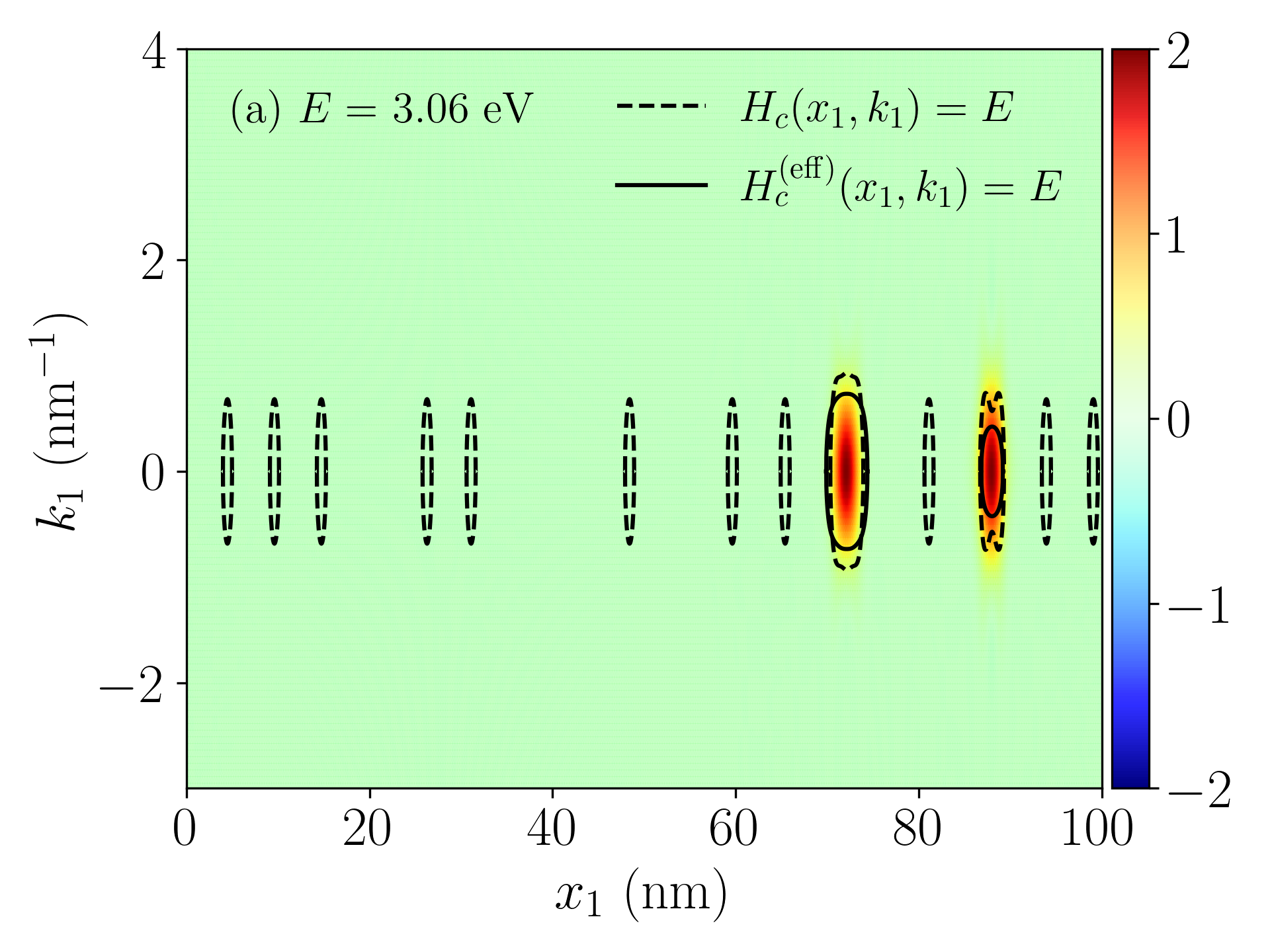}
\includegraphics[width = 0.4\textwidth, trim = .3cm .3cm .5cm .4cm,clip]{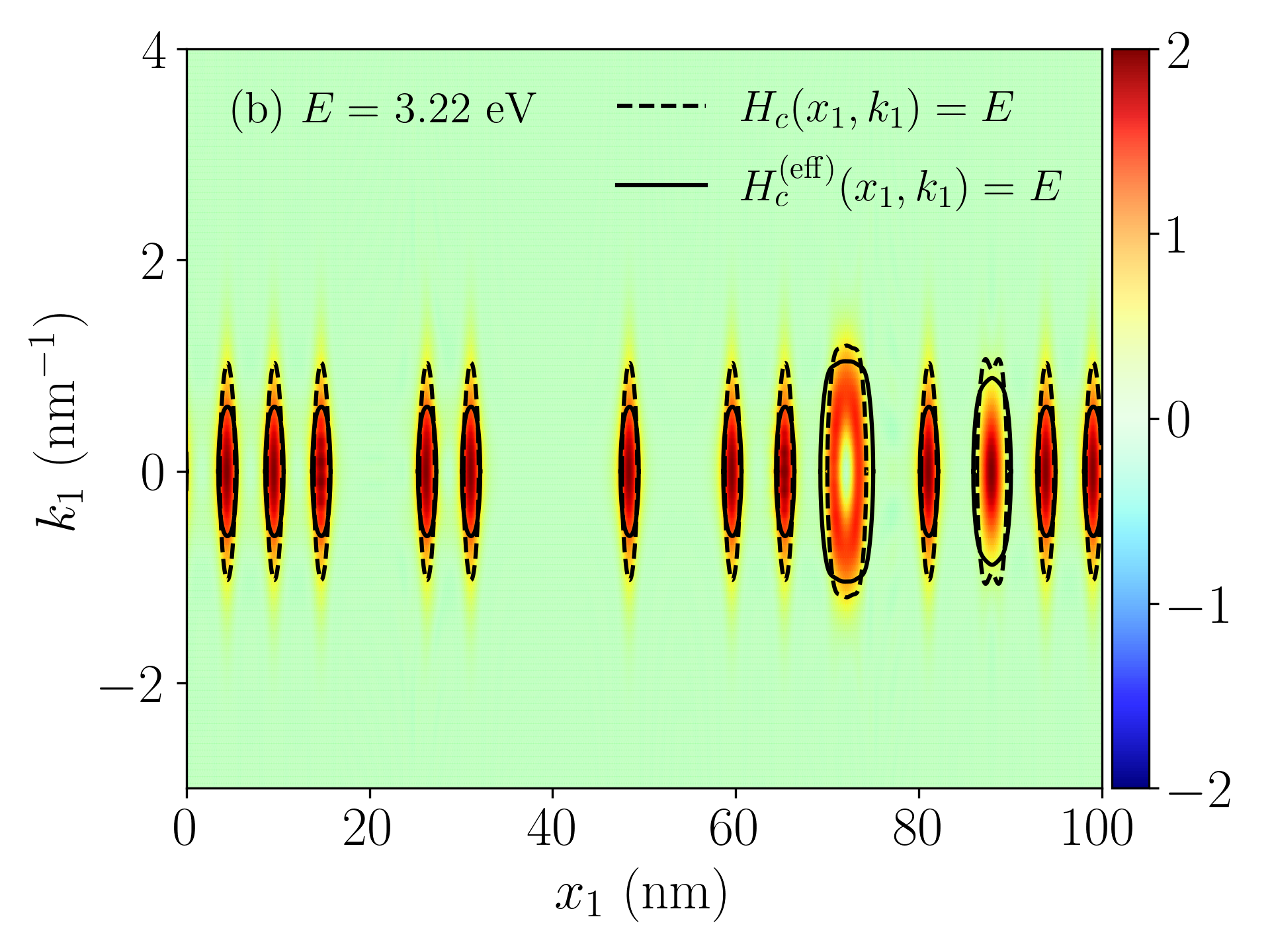}
~
\includegraphics[width = 0.4\textwidth, trim = .3cm .3cm .5cm .4cm,clip]{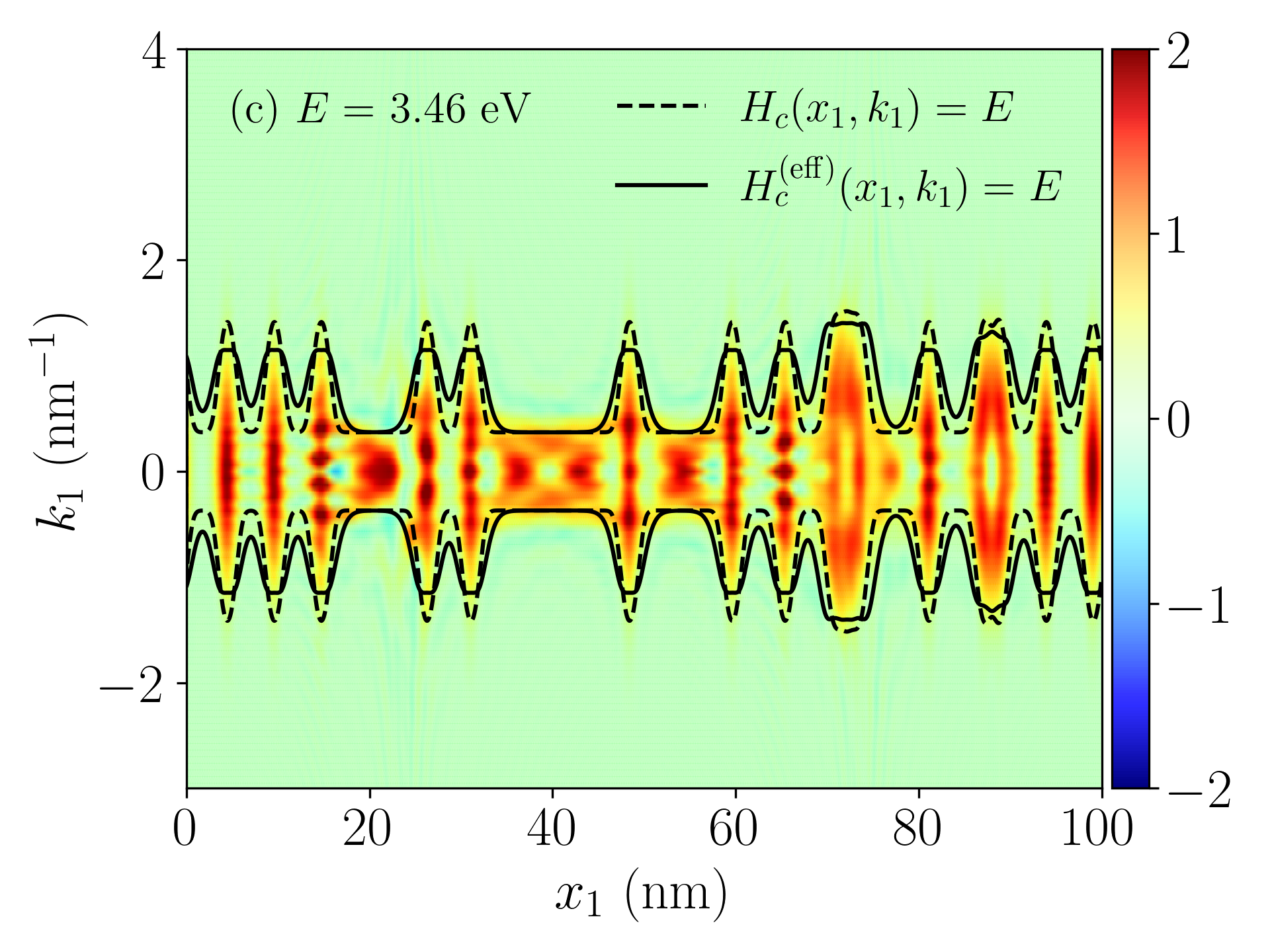}
\includegraphics[width = 0.4\textwidth, trim = .3cm .3cm .5cm .4cm,clip]{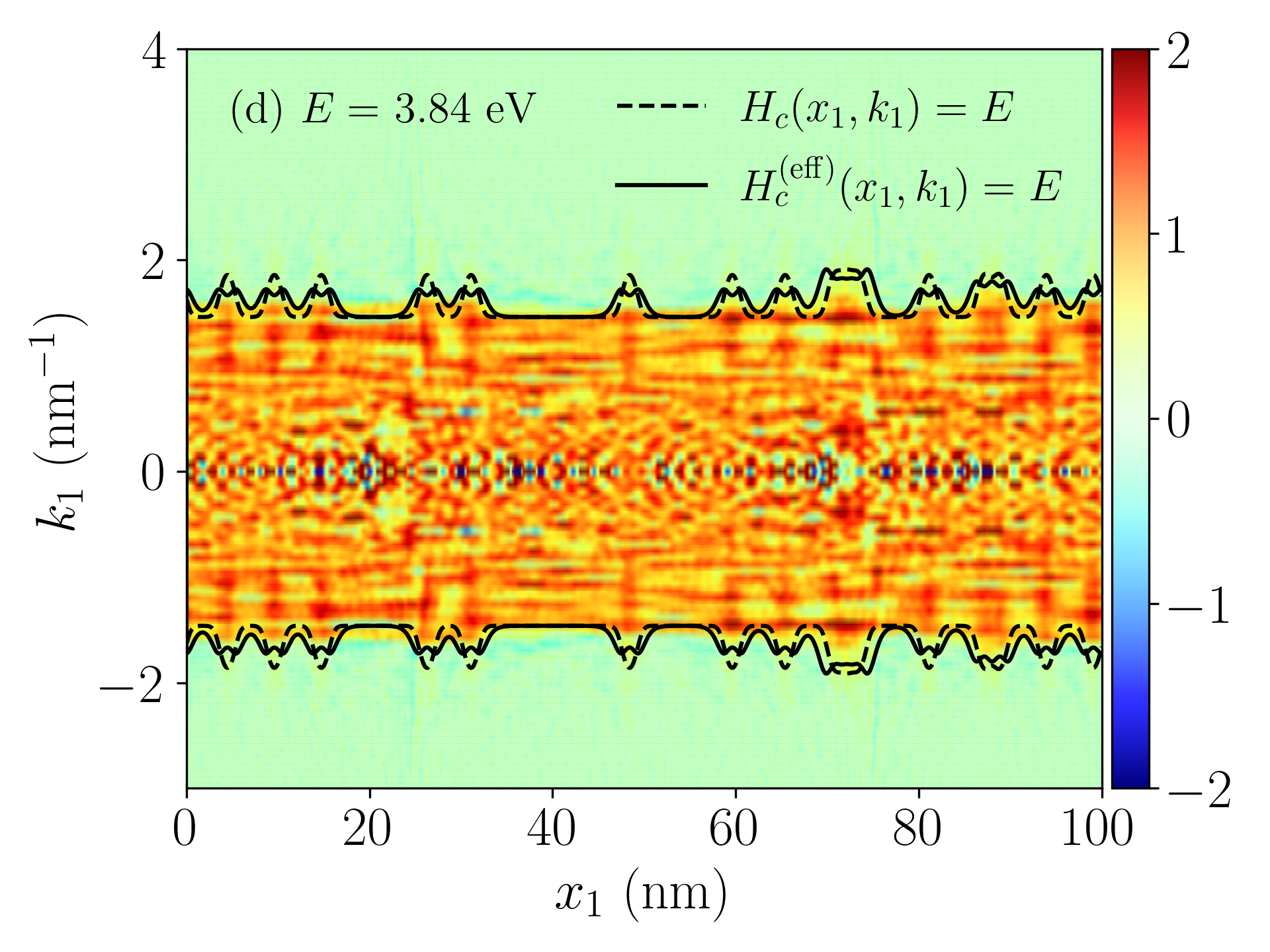}
\vspace*{-.2cm}
\caption{Integrated density of states in phase space, i.e.  the sum of the Wigner transforms $W_{\chic}$ of eigenstates whose eigenenergy lie below a given energy $E$ [see Eq.~(\ref{eq:intD})] for a one-dimensional alloy with In concentration $x=5$\% and $L=100$~nm. The smearing length was set to $\sigma = 2 a$. (a) $E = 3.06$~eV (first two states), (b) $E=3.22$~eV (first 14 states),   (c) $E =3.46$~eV (first 20 states), and (d) $E =3.84$~eV (first 50 states). The dashed lines are the contour $H_c(x_1,k_1) = E$ and the solid lines are the contour $H_c^{(\mathrm{eff})}(x_1,k_1) = E$.  The color scale is held fixed for the sake of visibility.}
\label{fig:phase_space}
\end{figure*}

Equation~(\ref{eq:coupling:Rdos}) therefore provides an alternative but equivalent picture of the spectral coupling density to that given by Eq.~(\ref{eq:coupling_density}). Equation~(\ref{eq:coupling_density}) states that to contribute to the spectral coupling density at energy $\hv$, a pair of states $\chic$ and $\chiv$ must be such that their difference of energies is equal to $\hv$ and that they have significant overlap integral. Equation~(\ref{eq:coupling:Rdos}) states, instead,  that the coupling spectral density evaluated at $\hv$ can be viewed as summing over the whole phase space the convolution product of the conduction and valence quasi-density of states in phase space at energy $\hv$. The conservation of energy is encoded in the convolution product, i.e., we scan in energy $\mathcal{D}^{(c)}$ and $\mathcal{D}^{(v)}$ simultaneously but with a fixed energy difference equal to $\hv$.  The coupling weight encoded in the square modulus of the scalar product in direct space, is now encoded in the sum over phase space of the product of $\mathcal{D}^{(c)}$ and $\mathcal{D}^{(v)}$.

The steps we have taken so far are exact. We now need to assess the quasi-densities of states in phase space, $\mathcal{D}^{(c)}$ and $\mathcal{D}^{(v)}$.
To that end, we integrate \eqref{eq:ps:ldos:c} over energy,
\begin{equation}
\int_{-\infty}^{E} \mathcal{D}^{(c)} (\vb{r},\vb{k},\varepsilon) \: \mathrm{d}\varepsilon = \sum_\mu W_{\chic} (\vb{r},\vb{k}) \, \Theta (E - E_\mu^{(c)})  \: ,
\label{eq:intD}
\end{equation}
where $\Theta$ is the Heaviside step function.  We refer to the above quantity as the \emph{integrated density of states in phase space}.  
The idea now is to observe that, on the one hand, the sum over phase space of the above integral is \emph{exactly} equal to the integrated density of states (IDOS) for \emph{all} $E$:
\begin{equation}
\iint \int_{-\infty}^E \mathcal{D}^{(c)}(\vb{r},\vb{k},\varepsilon) \: \mathrm{d}\varepsilon \, \frac{\mathrm{d}^d r \, \mathrm{d}^d k}{(2\pi)^d} = \mathrm{IDOS}^{(c)}(E) \: .
\label{eq:LDOSP:IDOS}
\end{equation}
Equation~(\ref{eq:LDOSP:IDOS}) is proven in Appendix~\ref{app:IDOS}.
On the other hand,  Weyl's law for the IDOS associated with the Hamiltonian $\hat{H}_c$ states that, asymptotically for $E \to \infty$, the IDOS is proportional to the volume of the region of phase space $H_c (\vb{r},\vb{k}) = \frac{\hbar^2 k^2}{2 m_c (\vb{r})} + E_c(\vb{r}) < E$, i.e.,
\begin{equation}
\mathrm{IDOS}^{(c)} (E) \sim \int_{H_c(\vb{r},\vb{k}) < E} \frac{\mathrm{d}^d r \, \mathrm{d}^d k}{(2 \pi)^d} \: .
\label{eq:Weyl:law}
\end{equation}
Equations~(\ref{eq:LDOSP:IDOS}) and (\ref{eq:Weyl:law}) suggest that, asymptotically for $E \to \infty$, the function $\int_{-\infty}^{E} \mathcal{D}^{(c)} (\vb{r},\vb{k},\varepsilon) \: \mathrm{d}\varepsilon$,  can be approximated by a plateau function equal to 1 within the domain $H_c(\vb{r},\vb{k}) < E$, i.e.,
\begin{equation}
\int_{-\infty}^{E} \mathcal{D}^{(c)} (\vb{r},\vb{k},\varepsilon) \: \mathrm{d}\varepsilon \approx \Theta \Big(E - H_c(\vb{r},\vb{k}) \Big) \: .
\label{eq:plateau:c}
\end{equation}
We have verified numerically that the approximation Eq.~(\ref{eq:plateau:c}) is indeed satisfied for $E$ sufficiently large.  Figure~\ref{fig:phase_space} illustrates the plateau function approximation for a one-dimensional alloy,  by comparing the integrated density of states in phase space,  Eq.~(\ref{eq:intD}), with the level line $H_c (x_1, k_1) = E$ for different values of $E$ (dashed lines).   For $E$ large enough the dashed contour line captures well the volume occupied by the integrated density of states in phase space [see Figs.~\ref{fig:phase_space}(c) and \ref{fig:phase_space}(d); lower values of $E$ will be discussed at the end of the section].  We note, however, that the suggested approximation may not be mathematically valid point-wise but rather in a weaker sense, as can be seen by the high frequency oscillations on the line $k_1=0$, which develop for sufficiently large values of $E$ [see Fig.~\ref{fig:phase_space}(d)]. These oscillations result from the interference in the Wigner transform of high energy states which are quasi-plane waves $\approx \exp(\pm i k x_1)$.  
Clearly,  a similar approximation holds for the valence band:
\begin{equation}
\int_{E}^{\infty} \mathcal{D}^{(v)} (\vb{r},\vb{k},\varepsilon) \: \mathrm{d}\varepsilon \approx \Theta \Big(H_v(\vb{r},\vb{k}) - E \Big) \: ,
\label{eq:plateau:v}
\end{equation}
with $H_v(\vb{r},\vb{k}) = \frac{\hbar^2 k^2}{2m_v (\vb{r})} + E_v(\vb{r})$. The change of order in the bounds of the integral is due to the negative effective mass $m_v$, i.e., higher order excited states have decreasing energy. Differentiating with respect to the energy $E$ in Eqs.~(\ref{eq:plateau:c}) and (\ref{eq:plateau:v}) readily yields 
\begin{subequations}
\begin{align}
\mathcal{D}^{(c)} (\vb{r},\vb{k},E) &\approx \delta \Big( E - H_c(\vb{r},\vb{k}) \Big) \: , \\
\mathcal{D}^{(v)} (\vb{r},\vb{k},E) &\approx \delta \Big( H_v(\vb{r},\vb{k}) - E \Big) \: .
\end{align}
\end{subequations}
It follows that the integral over $\varepsilon$ in Eq.~(\ref{eq:coupling:Rdos}) can be approximated by
\begin{align}
 &\int \mathcal{D}^{(c)}(\vb{r},\vb{k},\varepsilon + \hv)   \mathcal{D}^{(v)}(\vb{r},\vb{k},\varepsilon) \: \mathrm{d}\varepsilon \approx \nonumber \\
& \delta\Big( \hv - H_c(\vb{r},\vb{k}) + H_v(\vb{r},\vb{k}) \Big) \: . 
\end{align}
Inserting the above equation into Eq.~(\ref{eq:coupling:Rdos}) we finally obtain
\begin{align}
\mathcal{C} (\hv) &\approx \iint \delta\Big( \hv - H_c(\vb{r},\vb{k}) + H_v(\vb{r},\vb{k}) \Big)  \: \frac{\mathrm{d}^d r \, \mathrm{d}^d k}{(2\pi)^d}  \nonumber\\
&\approx   \frac{\mathrm{d}}{\mathrm{d} E} \iint_{ H^{(c)}(\vb{r},\vb{k}) - H^{(v)}(\vb{r},\vb{k}) < E }  \: \frac{\mathrm{d}^d r \, \mathrm{d}^d k}{(2\pi)^d} \, \Bigg|_{E=\hv} \: ,
\label{eq:C:weyl}
\end{align}
the two right-hand sides being equivalent ways of writing the same quantity. Equation~(\ref{eq:C:weyl}) can be referred to as a Weyl law for the spectral coupling density.
 Furthermore, the integration over $\vb{k}$ can be performed analytically. Indeed, we have
\begin{equation}
H_c (\vb{r},\vb{k}) - H_v (\vb{r},\vb{k}) = \frac{\hbar^2 |\vb{k}|^2}{2 m_r(\vb{r})} + E_g(\vb{r}) \: ,
\end{equation}
\\where the reduced mass $m_r$ is given by
\begin{equation}
\frac{1}{m_r} = \frac{1}{m_c} - \frac{1}{m_v} \: .
\end{equation}
Thus, we have
\begin{widetext}
\begin{equation}
\iint_{ H_c (\vb{r},\vb{k}) - H_v (\vb{r},\vb{k}) < E} \, \frac{\mathrm{d}^d r \, \mathrm{d}^d k}{(2\pi)^d} = \iint_{ |\vb{k}|^2  < 2 m_r (E-  E_g) / \hbar^2 }  \, \frac{\mathrm{d}^d r \, \mathrm{d}^d k}{(2\pi)^d} = \frac{v_d}{(2 \pi)^d} \int \left[ \frac{2 m_r(\vb{r}) (E - E_g(\vb{r}) )}{\hbar^2} \right]_+^{d/2} \: \mathrm{d}^d r \: ,
\end{equation}
where $v_d = \pi^{d/2} / \Gamma(d/2+1)$ is the volume of the $d$-dimensional unit ball, and the $+$ subscript denotes the positive part function, i.e., $x_+ = \max(x,0)$. By differentiation with respect to $E$, we obtain the following expressions for the spectral coupling density in any dimension, and for the absorption coefficient in 3D~\footnote{We specify the absorption coefficient in 3D only, as the prefactor in Eq.~(\ref{eq:alphaC}) is only valid in 3D (see the formula for the flux of the Poynting vector).}:
\begin{equation}
\mathcal{C}_\mathrm{WW} (\hv) = \frac{d v_d }{2 (2 \pi)^d} \int_\Omega \left[ \frac{2 m_r(\vb{r})}{\hbar^2} \right]^{d/2} \Big( \hbar \omega - E_g(\vb{r}) \Big)_+^{d/2-1} \: \mathrm{d}^d r \: ,
\label{eq:C:weyl:final}
\end{equation}
\begin{equation}
\alpha_\mathrm{WW} (\omega) = \frac{e^2 E_p v_3}{m_0 \varepsilon_0 c_0 \omega n(\omega) (2 \pi)^{2} |\Omega|} \frac{3}{2} \int_\Omega  \left[ \frac{2 m_r(\vb{r})}{\hbar^2} \right]^{3/2} \Big( \hbar \omega - E_g(\vb{r}) \Big)_+^{1/2} \: \mathrm{d}^3 r \: .
\label{eq:absorption:Weyl}
\end{equation}
\end{widetext}
Note that in the homogeneous limit (where $m_r$ and $E_g$ are constant), we recover the well-known expression for the absorption coefficient (see Appendix~\ref{app:flat:band}). We give in Appendix~\ref{app:deriv2} an alternative derivation of the above result based on the Weyl transform of a two-particle Hamiltonian. Although the result is identical, this second approach gives a complementary picture in terms of electron-hole pairs, and may be a good starting point for further developments, in particular, for taking the electron-hole interaction into account.\\

We have now found a closed form approximation for the absorption coefficient, which in view of our analysis appears to be accurate either in the limit of vanishing disorder or for large enough values of the photon energy.  But what about the bottom of the spectrum?  For lower values of $E$,  we observe that the level line $H_c (x_1, k_1) = E$ overestimates the volume occupied by the integrated density of states in phase space [see Fig.~\ref{fig:phase_space}(a)]. The level lines $H_c (x_1, k_1) = E$ enclose a significant volume in phase space which does not hold any eigenstate at the considered energy.  
Eigenstates appear in these pockets of phase space only at slightly larger energy values [see Fig.~\ref{fig:phase_space}(b)], this phenomenon being a manifestation of the tails in the density of states characteristic of a disordered system at low energy. We thus foresee here two limitations of the plateau function approximation: (i) the set $H_c (x_1, k_1) < E$ overestimates the volume in phase space occupied by eigenstates as the phase space pockets appear too early energy wise and (ii) the volume enclosed by the level lines grows continuously with energy while eigenstates appear at discrete energies. We show in the next paragraph how the so-called localization landscape overcomes the first limitation.

\subsection{Localization landscape and effective potential} 

In Refs.~\cite{Arnold:2016, Arnold:2019}, the authors introduced an object called the \emph{effective potential}, defined as the reciprocal of the localization landscape (LL) $\mathcal{L}$, which is the solution to the equation $\hat{H} \mathcal{L} = 1$ ($\hat{H}$ being the Hamiltonian and the right-hand side being the constant function equal to one)~\cite{Filoche:PNAS}. In particular, they showed numerically that for a wide class of potentials, one could obtain a very accurate approximation of the integrated density of states over the entire spectrum by replacing in Weyl's asymptotic law the original potential by this effective potential. Following this work, we introduce the localization landscapes $\mathcal{L}_c$ and $\mathcal{L}_v$ associated with the conduction band and the valence band potentials, respectively, which we define by
\begin{subequations}
\begin{align}
- \frac{\hbar^2}{2} \nabla \cdot \left[ \frac{1}{m_c} \nabla \mathcal{L}_c \right] + (E_c - \min E_c) \mathcal{L}_c &= 1 \: , \label{eq:landc}\\
\frac{\hbar^2}{2} \nabla \cdot \left[ \frac{1}{m_v} \nabla \mathcal{L}_v \right] - (E_v - \max E_v) \mathcal{L}_v &= 1  \: . \label{eq:landv}
\end{align}
\end{subequations}
Note the change of sign in the Schr\"{o}dinger operator in Eq.~(\ref{eq:landv}) to comply with the hypothesis of the positiveness of the operator from the localization landscape theory (i.e., a change of orientation of the energy axis). Also note that the reference of energy is set in such a way that the potentials $E_c - \min E_c$ and $-(E_v -\max E_v)$ are non-negative. The effective potentials (expressed in the original energy frame) are then deduced from the localization landscapes as
\begin{subequations}
\begin{align}
E_c^{(\mathrm{eff})} (\vb{r})  &= \min E_c + \frac{1}{\mathcal{L}_c (\vb{r}) } \: , \\
E_v^{(\mathrm{eff})} (\vb{r})  &= \max E_v - \frac{1}{\mathcal{L}_v (\vb{r}) } \: ,
\end{align}
\end{subequations}
and their difference defines the effective band gap profile
\begin{equation}
E_g^{(\mathrm{eff})} (\vb{r})  = E_c^{(\mathrm{eff})} (\vb{r}) - E_v^{(\mathrm{eff})} (\vb{r}) \: .
\end{equation}
We thus obtain an approximation for the absorption coefficient $\alpha$ (or, equivalently, for $\mathcal{C}$) by replacing $E_g$ by $E_g^{(\mathrm{eff})}$ in Eq.~(\ref{eq:absorption:Weyl}) [or Eq.~(\ref{eq:C:weyl:final})], 
\begin{widetext}
\begin{equation}
\mathcal{C}_\mathrm{WWL} (\hv) = \frac{d v_d }{2 (2 \pi)^d} \int_\Omega \left[ \frac{2 m_r(\vb{r})}{\hbar^2} \right]^{d/2} \Big( \hbar \omega - E_g^{(\mathrm{eff})}(\vb{r}) \Big)_+^{d/2-1} \: \mathrm{d}^d r \: ,
\label{eq:C:weylLL:final}
\end{equation}
\begin{equation}
\alpha_\mathrm{WWL} (\omega) = \frac{e^2 E_p v_3}{m_0 \varepsilon_0 c_0 \omega n(\omega) (2 \pi)^{2} |\Omega|} \frac{3}{2} \int_\Omega  \left[ \frac{2 m_r(\vb{r})}{\hbar^2} \right]^{3/2} \Big( \hbar \omega - E_g^{(\mathrm{eff})} (\vb{r}) \Big)_+^{1/2} \: \mathrm{d}^3 r \: .
\label{eq:absorption:WeylLL}
\end{equation}
\end{widetext}
%

Equation~(\ref{eq:C:weylLL:final}) is asymptotically equivalent to Eq.~(\ref{eq:C:weyl:final}) as $\hv \to \infty$. The advantage of the approximation based on the localization landscape can be appreciated at the bottom of the spectrum.  Indeed, Eq.~(\ref{eq:C:weylLL:final}) corresponds to replacing the conduction potential $E_c$ by its effective counterpart $E^{(\mathrm{eff})}_c$ in the plateau function approximation Eq.~(\ref{eq:plateau:c}) and similarly for the valence potential in Eq.~(\ref{eq:plateau:v}):
\begin{subequations}
\begin{align}
\int_{-\infty}^{E} \mathcal{D}^{(c)} (\vb{r},\vb{k},\varepsilon) \: \mathrm{d}\varepsilon \approx \Theta \left(E - H_c^{(\mathrm{eff})}(\vb{r},\vb{k}) \right) \: ,\\
\int_{E}^{\infty} \mathcal{D}^{(v)} (\vb{r},\vb{k},\varepsilon) \: \mathrm{d}\varepsilon \approx \Theta \left(H_v^{(\mathrm{eff})} (\vb{r},\vb{k}) - E \right) \: ,
\end{align}
\end{subequations}
with $H_c^{(\mathrm{eff})}(\vb{r},\vb{k}) = \frac{\hbar^2 \vb{k}^2}{2 m_c(\vb{r})} + E_c^{(\mathrm{eff})} (\vb{r})$ and $H_v^{(\mathrm{eff})}(\vb{r},\vb{k}) = \frac{\hbar^2 \vb{k}^2}{2 m_h(\vb{r})} + E_v^{(\mathrm{eff})} (\vb{r})$. The level lines $H_c^{(\mathrm{eff})} = E$ (resp. $H_c = E$) are shown as solid (respectively, dashed) black lines in Fig.~\ref{fig:phase_space}. While both lines capture correctly the volume in phase hosting eigenstates at large energies [see Figs.~\ref{fig:phase_space}(c) and \ref{fig:phase_space}(d)], the quality of the approximation provided by the effective potential appears clearly at lower energy [Figs.~\ref{fig:phase_space}(a) and \ref{fig:phase_space}(b)]. Figure~\ref{fig:phase_space}(a) is particularly illustrative of the fact that phase space pockets enclosed by the lines $H_c^{(\mathrm{eff})} = E$ appear at higher energy than those obtained with $H_c = E$ due to the quantum confinement energy that is accounted for by the localization landscape. As a general rule, for a given energy value~$E$, the phase space pockets are slightly broader along $x_1$ and less broad along $k_1$ because the effective potential implicitly incorporates the uncertainty principle~\citep{Arnold:2019}. The plateau function based on the effective potential is thus expected to be a more faithful continuous approximation of the integrated density of states in phase space which evolves in jumps as the energy is increased. This property is reminiscent of that found for the integrated density of states in Refs.~\cite{Arnold:2016, Arnold:2019}.\\

\subsection{Spatial distribution of the absorbed power}

The absorption coefficient appearing in Eq.~\eqref{eq:absorption:WeylLL} is expressed as an integral over the volume $\Omega$ of a function proportional to $( \hbar \omega - E_g^{(\mathrm{eff})} (\vb{r}))_+^{1/2} $. We note that due to the positive part function, only the volume $E_g^{(\mathrm{eff})} < \hv$ contributes to the integral. This suggests an energetic picture, namely, that the power brought by a given photon of energy~$\hv$ is absorbed inside the volume $E_g^{(\mathrm{eff})} < \hv$. Let us attempt to make this idea more precise by defining an absorbed power density at frequency $\omega$, $\mathcal{P}(\vb{r}, \omega)$. Coming back to Fermi's golden rule expressed in Eq.~\eqref{eq:Wji_rate}, the transition rate~$W_{\mu \nu}$ gives the number of transitions from state~$\psiv$ to state~$\psic$ per unit time, i.e., the number of photons of energy $\hv$ absorbed by this transition per unit time. Hence $\hv W_{\mu \nu}$ is the (time-) average absorbed power by the transition~$\nu \to \mu$. If we ask now where this energy is absorbed during one such transition, we could answer that the energy $\hv$ of the absorbed photon is transferred to the electron during the time of the interaction as the wave function of the electron, $\psi (\vb{r},t)$, evolves from the initial state $\psiv$ to the final state $\psic$. In virtue of the Poynting theorem from classical electrodynamics~\cite{Jackson}, the instantaneous absorbed power density is equal to the work transferred to the electron moving in the electric field, which is of the form
\begin{equation}
p_{\mu \nu} (\vb{r}, t) = \Vie{J}{\mu \nu}{}(\vb{r},t) \cdot \vb{E} (\vb{r},t) \: ,
\end{equation}
where $\Vie{J}{}{} = \frac{e }{m} \mathrm{Re} (\psi^* \Vie{\hat{p}}{}{} \psi)$ is the electric current density and $\vb{E}$ is the electric field.  In fact, the absorbed power density can easily be induced by rewriting Fermi's golden rule Eq.~(\ref{eq:Wji_rate}) involved in the time-averaged absorbed power:
\begin{widetext}
\begin{align}
\hv W_{\mu \nu} &= \int_\Omega \left( \frac{e}{2 m_0} \right)^2 \, {\Bra{\psic} \Vie{A}{0}{} \cdot \Vie{\hat{p}}{}{} \Ket{\psiv} }^* {\psic}^*(\vb{r}) \Vie{\hat{p}}{}{} \psiv (\vb{r}) \cdot \omega \Vie{A}{0}{} 2 \pi \delta \Big( E_\mu^{(c)} - E_\nu^{(v)} - \hv\Big) \, \mathrm{d}^d r \label{eq:Intpmunu} \\
&= \int_{-\infty}^\infty \int_\Omega \left( \frac{e}{2 m_0} \right)^2 \, \frac{1}{\hbar} {\Bra{\psic} \Vie{A}{0}{} \cdot \Vie{\hat{p}}{}{} \Ket{\psiv} }^* {\psic}^*(\vb{r}) \Vie{\hat{p}}{}{} \psiv (\vb{r}) \exp(i \omega_{\mu \nu} t) \cdot \omega \Vie{A}{0}{} \exp(- i \omega t) \, \mathrm{d}^d r \, \mathrm{d} t \: , \nonumber\\
&= \int_{-\infty}^\infty \int_\Omega \Vie{J}{\mu \nu}{}(\vb{r}, t) \cdot \Vie{E}{}{} (t) \, \mathrm{d}^d r \, \mathrm{d} t \: .
\label{eq:hvW}
\end{align}
\sloppy Here we have introduced the short-hand notation $\omega_{\mu \nu} = (E_\mu^{(c)} - E_\nu^{(v)})/\hbar$ and used the relation $2 \pi \delta(\omega) = \int \, \exp(i \omega t) \mathrm{d}t $ in the second step. In Eq.~(\ref{eq:hvW}), we have identified the electric field $\Vie{E}{}{} = \omega \Vie{A}{0}{} \exp(-i \omega t)$ and the current density associated to the transition $\nu \to \mu$:
\begin{equation}
\Vie{J}{\mu \nu}{}(\vb{r}, t) = \left( \frac{e}{2 m_0} \right)^2 \, \frac{1}{\hbar} {\Bra{\psic} \Vie{A}{0}{} \cdot \Vie{\hat{p}}{}{} \Ket{\psiv} }^* {\psic}^*(\vb{r}) \Vie{\hat{p}}{}{} \psiv (\vb{r}) \exp(i \omega_{\mu \nu} t) \: .
\label{eq:current}
\end{equation}
Equations~(\ref{eq:hvW}) and (\ref{eq:current}) are interesting as they link the concept of transition rate between stationary states given by Fermi's golden rule from quantum mechanics, and the electromagnetic power from classical electrodynamics. Here we are rather interested in the time-averaged power density associated to the transition for a photon of energy $\hv$, $p_{\mu \nu}(\vb{r}, \omega)$, which in view of Eq.~(\ref{eq:Intpmunu}) can be defined as
\begin{equation}
p_{\mu \nu}(\vb{r}, \omega) = \left( \frac{e}{2 m_0} \right)^2 \, {\Bra{\psic} \Vie{A}{0}{} \cdot \Vie{\hat{p}}{}{} \Ket{\psiv} }^* {\psic}^*(\vb{r}) \Vie{\hat{p}}{}{} \psiv (\vb{r}) \cdot \omega \Vie{A}{0}{} 2 \pi \delta \Big( E_\mu^{(c)} - E_\nu^{(v)} - \hv\Big) \: .
\label{eq:pmunu}
\end{equation}
If we wish to consider the absorbed power density at the scale of the envelope functions, i.e., without resolving the contributions of the periodic functions $u_c$ and $u_v$ in Eq.~(\ref{eq:pmunu}), we can integrate $p_{\mu \nu}$ over a lattice unit cell and follow the same steps as the ones presented in Appendix~\ref{app:momentum} for the factorization of the matrix element $M_{\mu \nu}$. This gives the following cell averaged absorbed power density:
\begin{equation}
\bar{p}_{\mu \nu}(\vb{r}, \omega) = \int_{\Omega_{\mathrm{cell}}} p_{\mu \nu}( \vb{r} - \Vie{r}{}{\prime} , \omega) \, \frac{\mathrm{d}^d r^\prime}{\Omega_{\mathrm{cell}}} = \frac{\pi \omega e^2 A_0^2 E_p}{2 m_0} \, {\BraKet{\chic}{\chiv} }^* {\chic}^*(\vb{r}) \chiv (\vb{r}) \, \delta \Big( E_\mu^{(c)} - E_\nu^{(v)} - \hv\Big) \: .
\label{eq:pbmunu}
\end{equation}
The above result gives a clear intuitive picture of the localization of the absorbed power. The absorbed power is distributed proportionally to the product of the envelope functions of the initial and final states. 
 The total power density absorbed at frequency $\omega$, normalized by the incident photon power $\Pi S$, is obtained by summing Eq.~(\ref{eq:pbmunu}) over all transitions:
\begin{equation}
\mathcal{P}(\vb{r},\omega) = \frac{2}{\Pi S} \sum_{\mu \nu} \bar{p}_{\mu \nu} (\vb{r}, \omega) 
= \frac{2 \pi e^2 E_p}{m_0 \varepsilon_0 \omega n(\omega) c_0 S} \sum_{\mu \nu} { \BraKet{\chic}{\chiv} }^* {\chic}^*(\vb{r}) \chiv (\vb{r}) 
 \delta \Big( E_\mu^{(c)} - E_\nu^{(v)} - \hbar \omega \Big) \: .
\label{eq:Power}
\end{equation}
We note the close resemblance with the expression for the absorption coefficient given in Eq.~\eqref{eq:alphaC}. As for the absorption coefficient, evaluating the above expression for $\mathcal{P}$ is numerically costly since it requires the knowledge of the eigenstates. We can nevertheless make a simple guess for an approximation of $\mathcal{P}$ based on the Wigner-Weyl approach. Consider the integral of $\mathcal{P}$ over the volume $\Omega$. It is clear by integration of Eq.~\eqref{eq:Power} that
\begin{equation}
\int_\Omega \mathcal{P}(\vb{r},\omega) \, \mathrm{d}^d r = \alpha(\omega) L \: .
\end{equation}
This result was expected, of course, since by definition of $\mathcal{P}$ its integral should agree with $\hv W_\mathrm{tot} / \hv \Phi$ (by construction). The interesting point is that since we have an approximation for $\alpha$, e.g., Eq.~\eqref{eq:absorption:WeylLL}, we directly obtain an approximation for $\int \mathcal{P}$, namely,
\begin{equation}
\int_\Omega \mathcal{P}(\vb{r},\omega) \, \mathrm{d}^d r \approx \frac{e^2 E_p v_d}{m_0 \varepsilon_0 c_0 \omega n(\omega) (2 \pi)^{d-1} S} \frac{d}{2} \int_\Omega \left[ \frac{2 m_r(\vb{r})}{\hbar^2} \right]^{d/2} \Big( \hbar \omega - E_g^{(\mathrm{eff})} (\vb{r}) \Big)_+^{d/2-1} \: \mathrm{d}^d r  \: .
\end{equation}
The above equation states that two integrals over $\Omega$ are equal for all frequencies $\omega$.  We therefore propose to induce that the corresponding integrands are equal (a derivation which is not mathematically correct in general).  This yields the following approximation for $\mathcal{P}$:
\begin{equation}
\mathcal{P}_\mathrm{WWL} (\vb{r},\omega) = \frac{e^2 E_p v_d}{m_0 \varepsilon_0 c_0 \omega n(\omega) (2 \pi)^{d-1} S} \frac{d}{2} \left[ \frac{2 m_r(\vb{r})}{\hbar^2} \right]^{d/2} \Big( \hbar \omega - E_g^{(\mathrm{eff})} (\vb{r}) \Big)_+^{d/2-1}   \: .
\label{eq:PowerWLL}
\end{equation}
\end{widetext}
Equation~\eqref{eq:PowerWLL} translates mathematically our initial intuition at the beginning of the paragraph: the absorbed power at frequency $\omega$ is deposited in the volume $E_g^{(\mathrm{eff})} < \hv$, and the associated power density is proportional to $[ 2 m_r(\vb{r}) / \hbar^2 ]^{d/2} ( \hbar \omega - E_g^{(\mathrm{eff})} (\vb{r}) )_+^{d/2-1}$. For photon energies $\hv < \min E_g^{\mathrm{(eff)}}$,  there is no energy transfer since no photon is absorbed. For $\min E_g^{(\mathrm{eff})} < \hv < \max E_g^{(\mathrm{eff})}$, the energy is absorbed in the part of the volume hosting somewhat localized states (either in the valence or in the conduction band) whose energies lie between the minimum and maximum of the effective potentials and which contribute to photon absorption. For $\hv > \max E_g^{(\mathrm{eff})}$,  the whole volume contributes. Note that the absorbed power density around a given point $\vb{r}$ changes with $\hv$ in a way which is reminiscent of the density of states. This encodes the fact that several states may contribute to the power density at a given point and given frequency $\omega$.

\section{Numerical benchmark}\label{sec:benchmark}

\subsection{Numerics}\label{sec:numerics}

\emph{Indium concentration map} --- The local indium concentration $X(\vb{r})$ as given by Eq.~(\ref{eq:gaussian:averaging}) can be expressed in terms of convolution products
\begin{equation}
X = \frac{g_\sigma * \displaystyle \sum_{i \in \mathcal{I}} X_i \delta_{\Vie{r}{i}{}}}{g_\sigma * \displaystyle \sum_{i \in \mathcal{I}} \delta_{\Vie{r}{i}{}}} \: ,
\label{eq:Xconvolution}
\end{equation}
where $g_\sigma(\vb{r}) = \exp(- |\vb{r}|^2/ 2\sigma)$. The convolution products in Eq.~(\ref{eq:Xconvolution}) are conveniently computed numerically by the use of the fast Fourier transform (FFT)~\cite{fftw}. Given an almost cubic box of size $L_1 \times L_2 \times L_3$, where $L_1, L_2, L_3$ are the nearest integer multiples of lattice constant lengths along $x_1$, $x_2$, and $x_3$ to a desired length $L$, we construct a rectangular grid with discretization steps $\Delta x_1$, $\Delta x_2$, and $\Delta x_3$ significantly smaller than the lattice constants, and commensurate with the cation lattice sites (i.e., that lattice sites exactly fall on grid points).
Based on this spatial discretization grid, we can construct three arrays. An array $G$ for $g_\sigma$ evaluated at the grid points, an array $\Lambda$ for the indicator of the lattice of the Ga and In sites (i.e., equal to one for lattice points $\Vie{r}{i}{}$ and zero otherwise) and an array $I$ for the indicator of the In sites [which depends on the realization of $(X_i)_{i \in \mathcal{I}}$]. The discrete Fourier transforms $\hat{G} = \mathrm{FFT}[G]$, $\hat{\Lambda}= \mathrm{FFT}[\Lambda]$, and $\hat{I} = \mathrm{FFT}[I]$ are computed with the FFT, and the indium concentration array evaluated on the grid, $X_{ijk} = X (\Vie{r}{ijk}{})$, is given by
\begin{equation}
X = \frac{\mathrm{FFT}^{-1}[\hat{G} \hat{I}] }{ \mathrm{FFT}^{-1}[\hat{G} \hat{\Lambda}] } \: ,
\end{equation}
where the product of arrays is performed point wise. This method is significantly faster than the naive method consisting in summing Eq.~(\ref{eq:gaussian:averaging}) on the sampled grid points as it benefits from the low complexity of the FFT. Furthermore, the resulting map automatically satisfies periodic boundary conditions. For alloys of average indium concentration $x$, the wurtzite lattice parameters are chosen following Vegard's law, i.e., to be a linear interpolation of the InN and GaN parameters $a = x \, a_\mathrm{InN} + (1-x) a_\mathrm{GaN}$ and $c = x \, c_\mathrm{InN} + (1-x) c_\mathrm{GaN}$ (see Table~\ref{tab1} for values of the lattice parameters for InN and GaN).\\

\emph{Finite element computation of eigenstates and localization landscapes} --- The computation of the localization landscapes and of the eigenstates~\footnote{ Eigenstates are used only for comparing the Wigner-Weyl law for $\mathcal{C}$, Eq.~(\ref{eq:C:weyl}), to the exact formula Eq.~(\ref{eq:coupling_density}) in 1D and 2D as a benchmark.} is achieved by using the finite element method. Meshes are generated with Gmsh~\cite{gmsh} and we have used the finite element solver GetDP~\cite{getdp:1998,getdp}. The band-edge data (potentials and effective masses) are interpolated on the nodal points. The discretized linear system is solved either by using a direct method or the iterative method of generalized minimal residual (GMRES).\\

\emph{Computation of the absorption coefficient} --- The absorption coefficient, or equivalently $\mathcal{C}$, is computed either according to Eqs.~(\ref{eq:coupling_density}), (\ref{eq:C:weyl:final}) or (\ref{eq:C:weylLL:final}). 
 For summing Eq.~(\ref{eq:coupling_density}), the Dirac masses are regularized as
\begin{equation}
\delta_\varepsilon(E_\mu^{(c)} - E_\nu^{(v)} - \hv) = \frac{ \exp \Big[ - \frac{(E_\mu^{(c)} - E_\nu^{(v)} - \hv)^2 }{ 2 \varepsilon^2} \Big] }{\sqrt{2 \pi } \varepsilon} \: ,
\end{equation}
with an energy width $\varepsilon = 5$~meV (unless specified otherwise), which we have experienced to be small enough to resolve some sharp physically meaningful peaks (see Sec.~\ref{sec:benchmark}). The absorption coefficient is averaged over $N$ realizations of the random alloy.\\

\emph{Numerical parameters} --- Material parameters used for the computation are summarized in Table~\ref{tab1}. Numerical parameters such as the size $L$ of the box, and the discretization steps are summarized in Table~\ref{tab2} for the different simulations. The real part of the refractive index is taken to be the experimentally measured refractive index of GaN for simplicity~\cite{Yu:1997}.

\begin{table}[t]
\begin{center}
\caption{Numerical parameters used in the simulations: Simulation box size $L$, finite element mesh step $\Delta x$, number of degrees of freedom DoF, number of eigenstates per band $M$ (only used for 1D and 2D benchmark), number of alloy realizations $N$, and CPU speed-up between eigenstates and landscape computation. The two different values of the speed-up correspond to the use of the direct linear solver or the iterative method GMRES.}
\begin{tabular}{c c c c c c c c}
\hline
\hline
Simulations &  $L$~(nm)  & $\Delta x$~(\r{A}) & DoF  & $M$ &$N$ & Speed-up \\[.1cm]
\hline
1D Eig./WWL & 200.0  & 0.5 & $4.0 \times 10^3$ & 1000 & 100 & $178|178$ \\
2D Eig./WWL & 40.0   & 3.0 & $2.1 \times 10^4$ & 750 & 100 & $235|321$\\
3D WWL 	     & 20.0   & 3.0 & $3.5 \times 10^5$ & - & 50 & -\\
\hline
\hline
\end{tabular}
\label{tab2}
\end{center}
\end{table}


\begin{figure*}[t]
\centering
\includegraphics[width = 0.45\textwidth, trim = 0cm 0cm 0cm 0cm,clip]{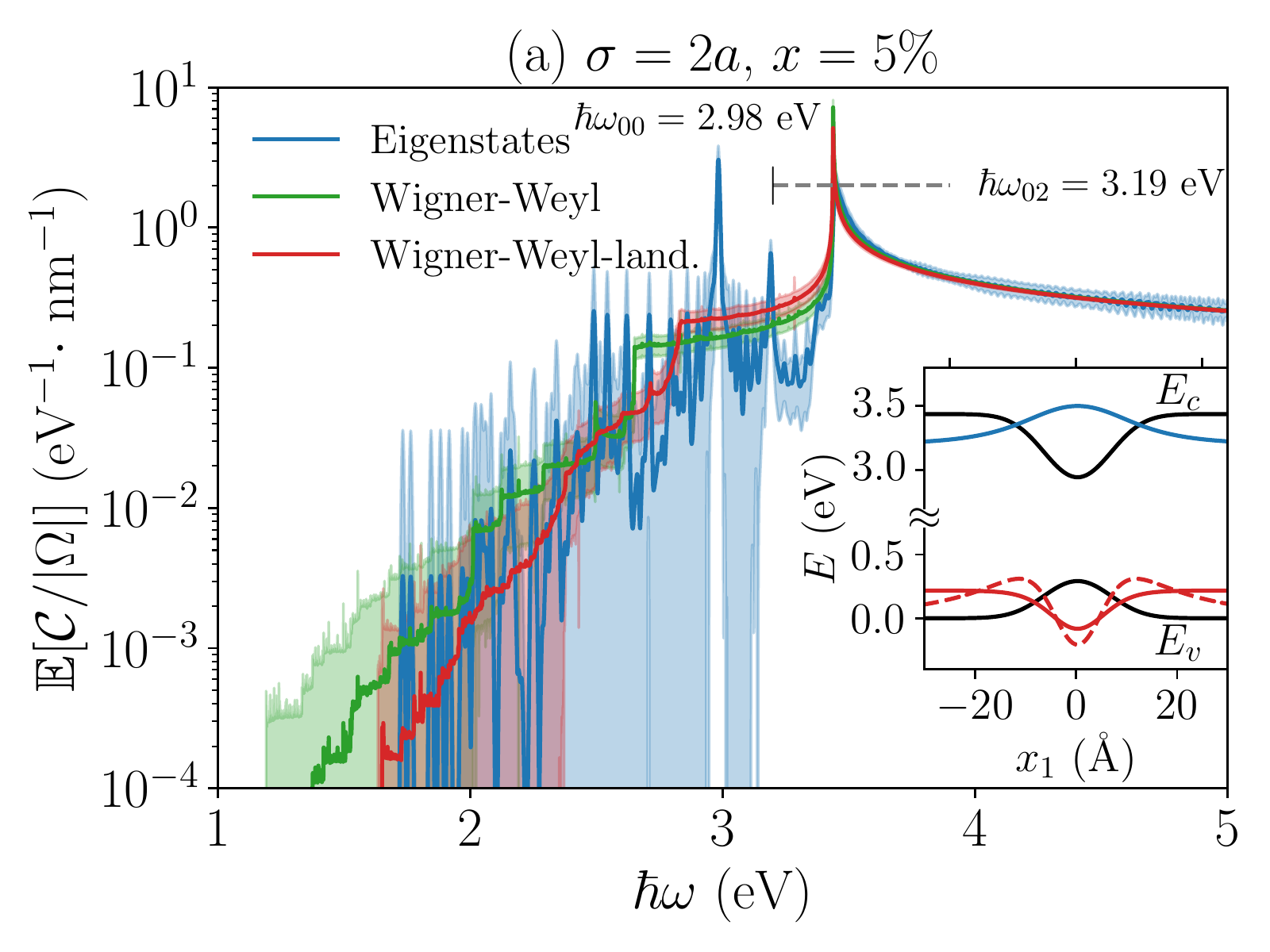}
\includegraphics[width = 0.45\textwidth, trim = 0cm 0cm 0cm 0cm,clip]{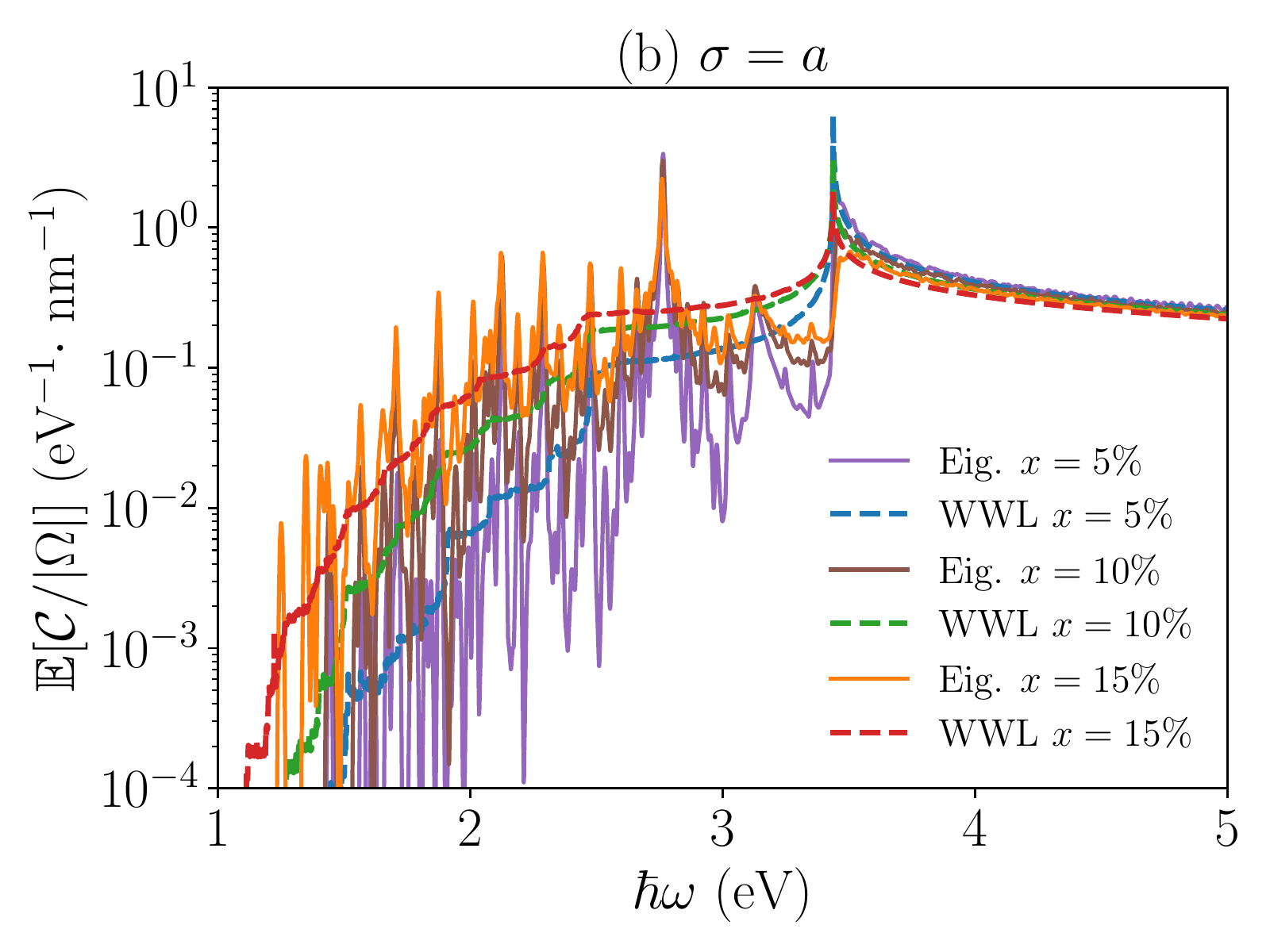}

\includegraphics[width = 0.45\textwidth, trim = 0cm 0cm 0cm 0cm,clip]{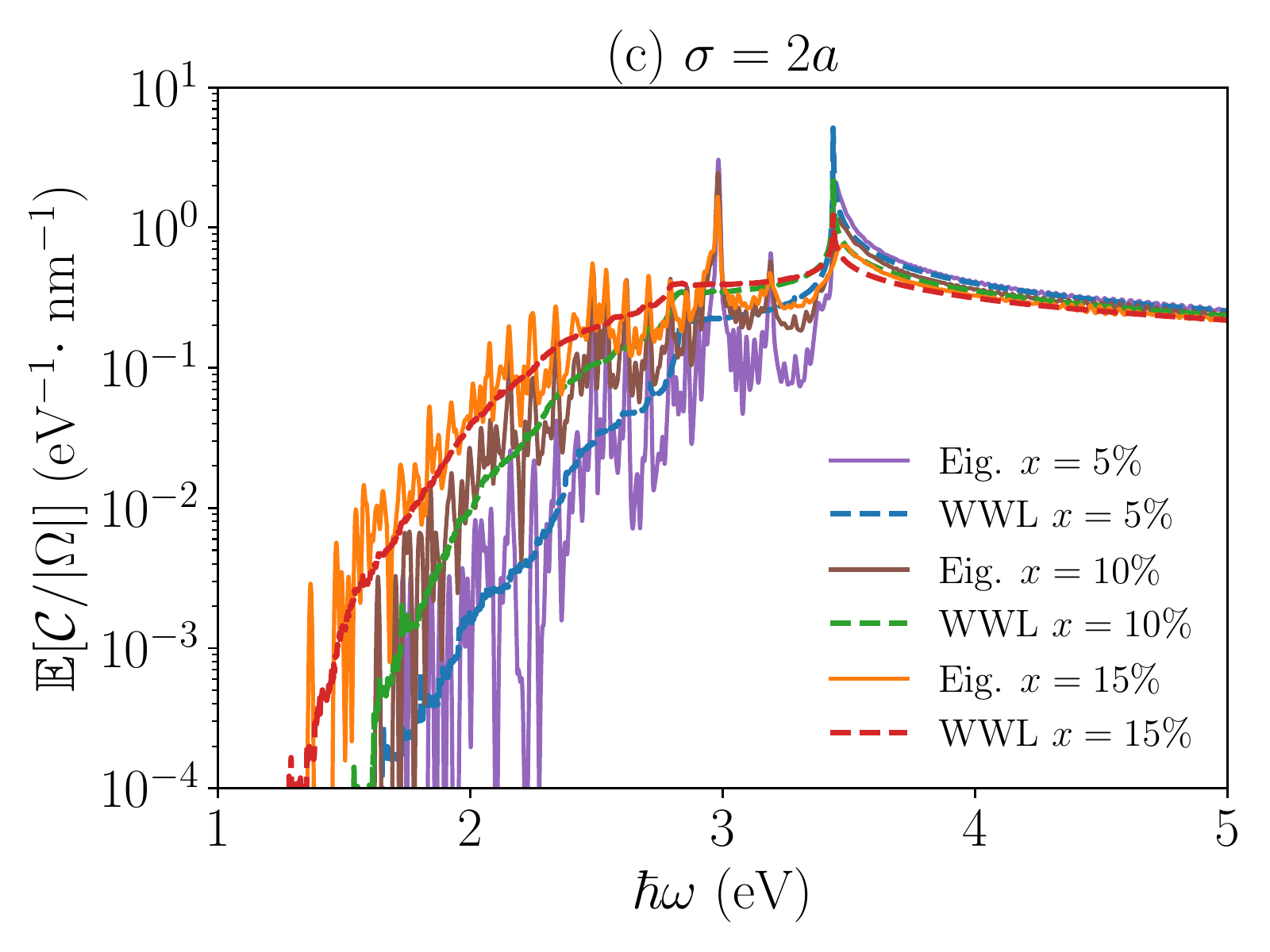}
\includegraphics[width = 0.45\textwidth, trim = 0cm 0cm 0cm 0cm,clip]{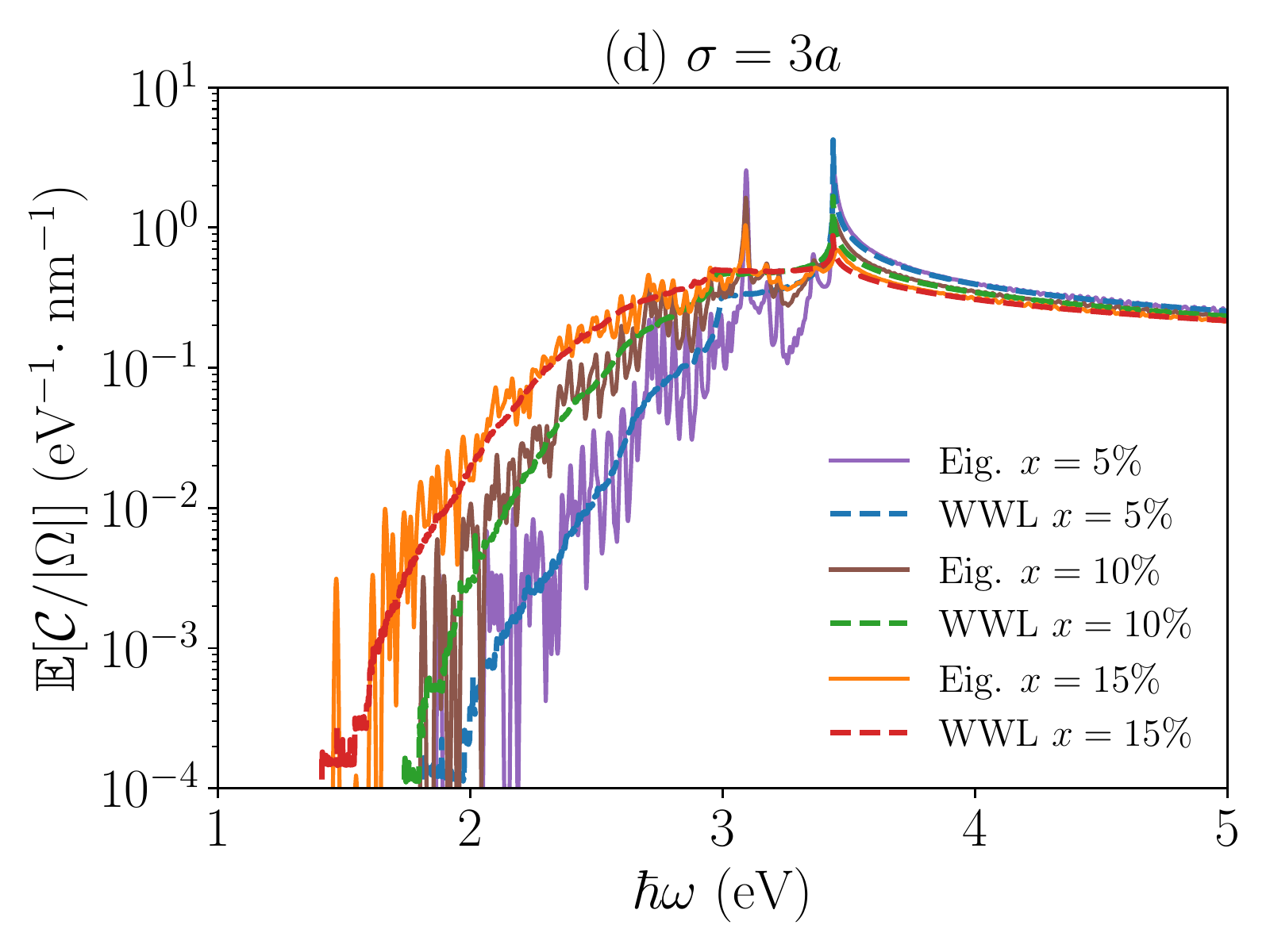}
\caption{Average spectral coupling density per unit length, $\Exp{\mathcal{C}/ |\Omega|}$, for one-dimensional In$_x$Ga$_{1-x}$N alloys. (a) In-concentration fixed $x=5\%$, and smearing length fixed $\sigma = 2a$. (b)-(d)  Varying In-concentration $x \in \{5\%, 10\%, 15\% \}$ and fixed smearing length (b) $\sigma=a$, (c) $\sigma = 2a$, and (d) $\sigma = 3a$. The results were obtained by using the eigenstate-based expression (eig.), Eq.~(\ref{eq:coupling_density}) , the Wigner-Weyl expression (WW),  Eq.~(\ref{eq:C:weyl:final}), and the Wigner-Weyl localization landscape expression (WWL), Eq.~(\ref{eq:C:weylLL:final}),  averaged over $N = 100$ realizations of the alloy chain of length $L = 200$~nm. The shaded areas correspond to one standard deviation around the average.}
\label{fig:1D}
\end{figure*}

\begin{figure*}[t]
\centering
\includegraphics[width = 0.45\textwidth, trim = 0cm 0cm 0cm 0cm,clip]{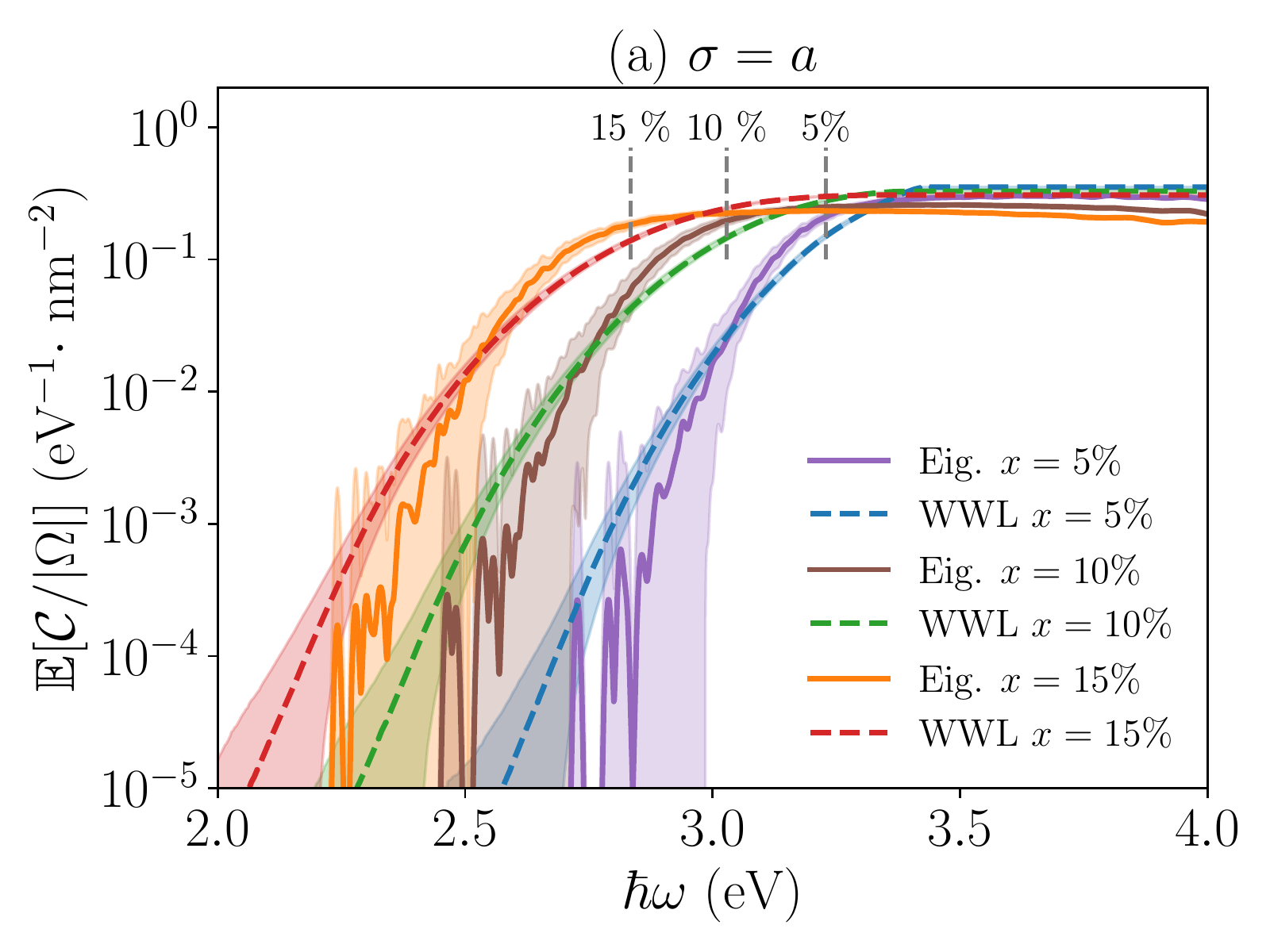}
\includegraphics[width = 0.45\textwidth, trim = 0cm 0cm 0cm 0cm,clip]{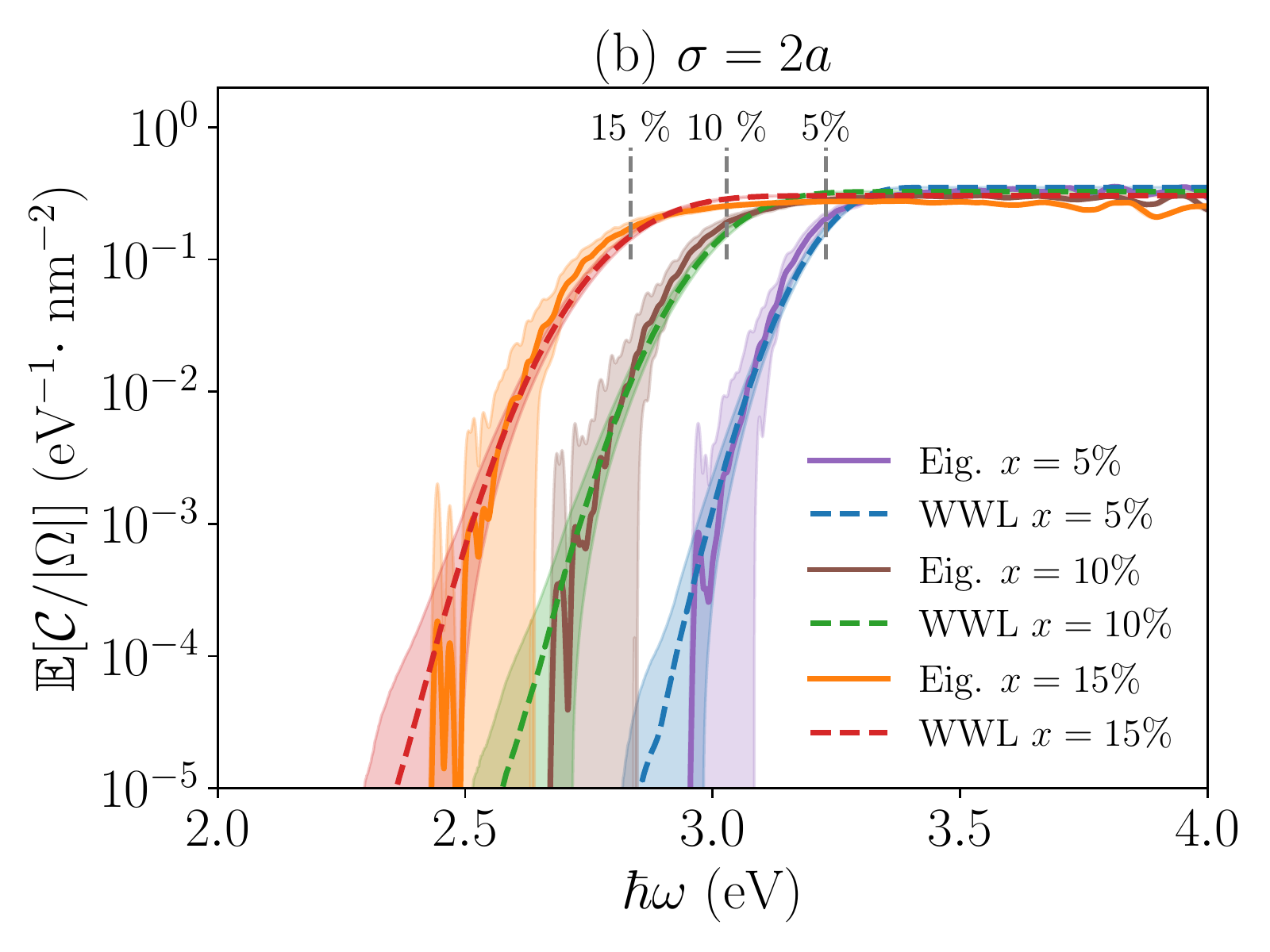}
\includegraphics[width = 0.45\textwidth, trim = 0cm 0cm 0cm 0cm,clip]{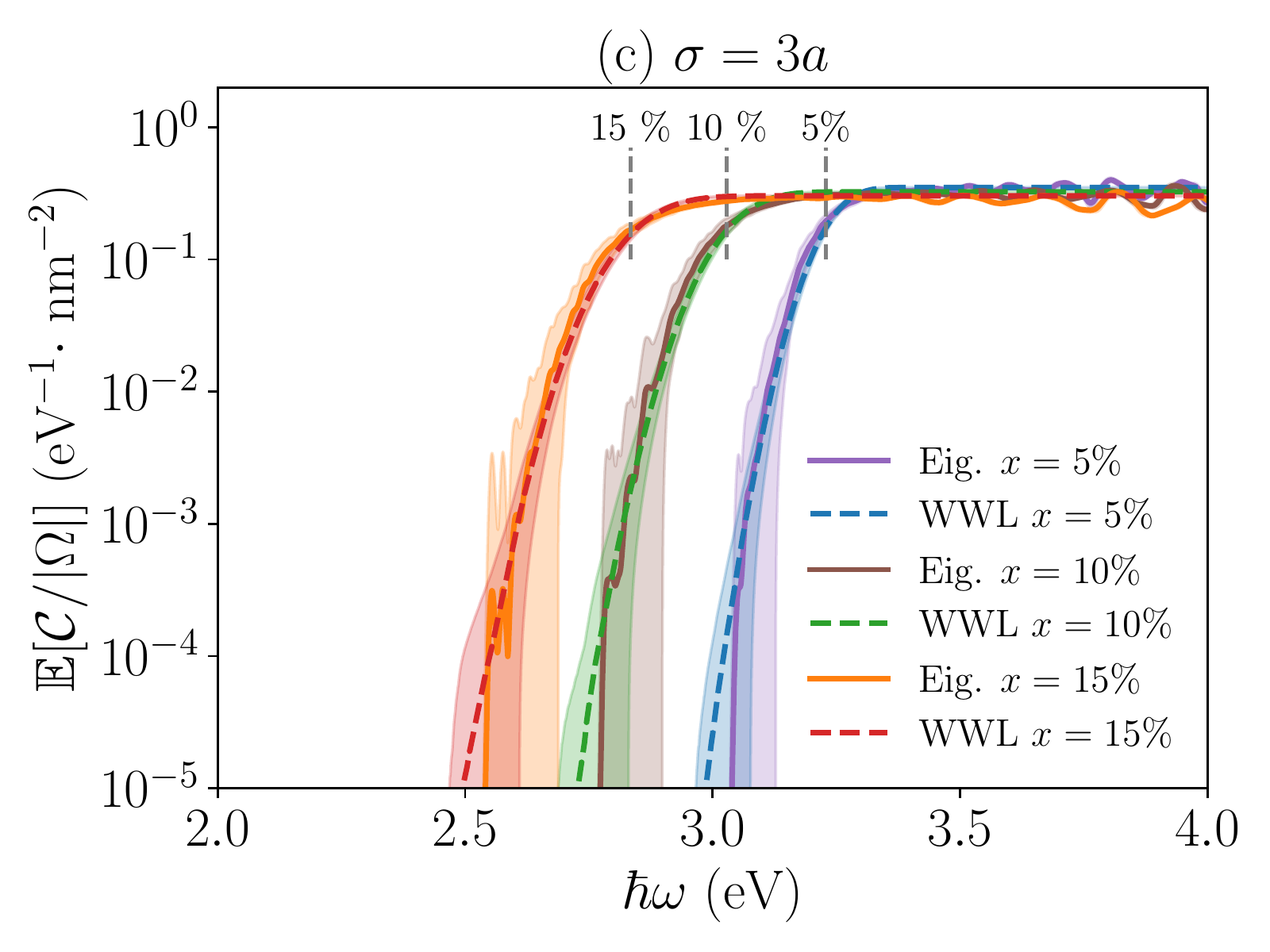}
\includegraphics[width = 0.45\textwidth, trim = 0cm 0cm 0cm 0cm,clip]{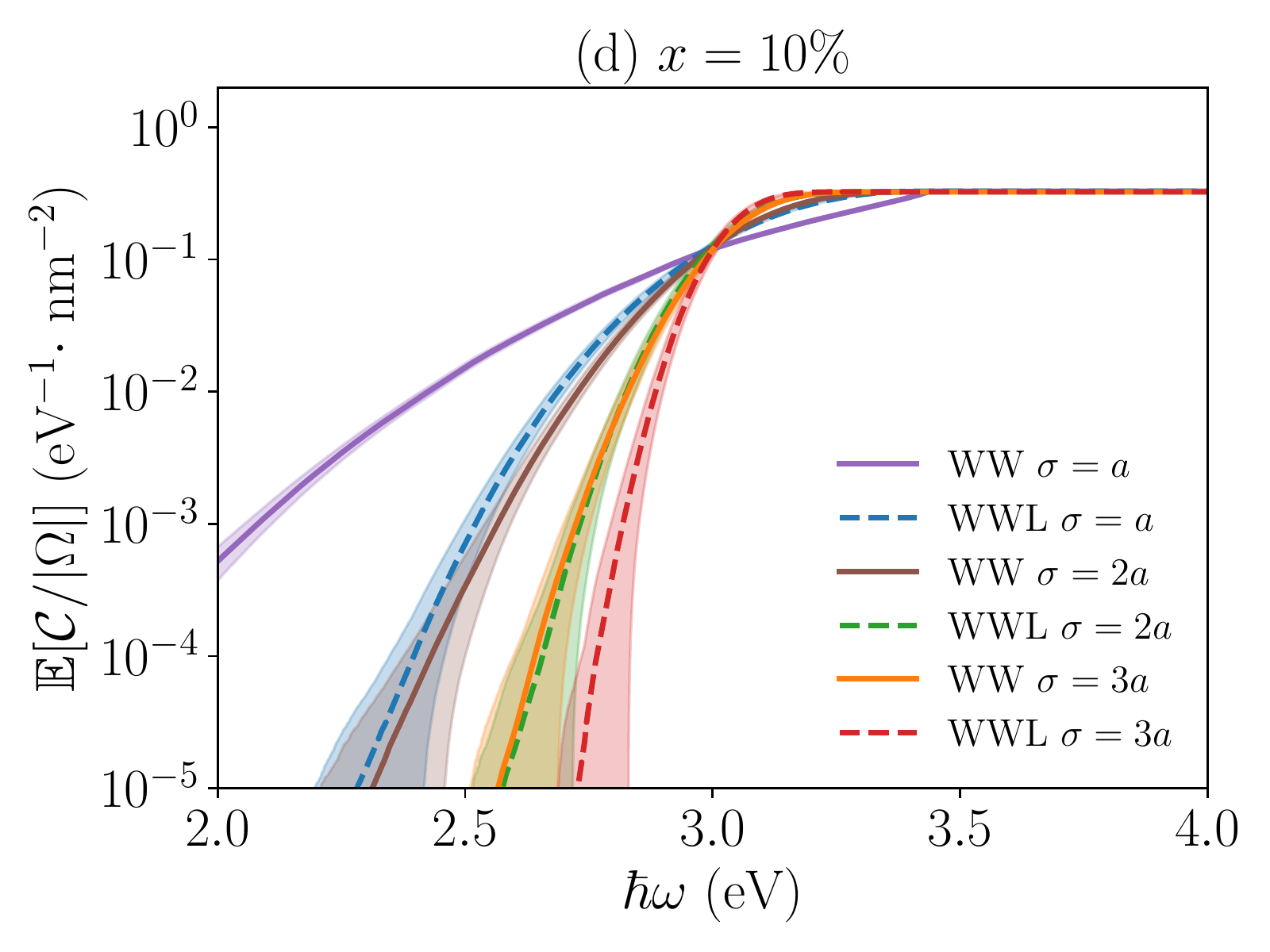}
\caption{Average spectral coupling density per unit area, $\Exp{\mathcal{C}/ |\Omega|}$, for two-dimensional In$_x$Ga$_{1-x}$N alloys.  (a)-(c)  Comparison between the eigenstates based formula [eig.,  Eq.~(\ref{eq:coupling_density})] and the Wigner-Weyl law based on the localization landscape [WWL, Eq.~(\ref{eq:C:weylLL:final})] for varying In concentration $x \in \{5\%, 10\%, 15\%\}$ and fixed smearing length (a) $\sigma=a$, (b) $\sigma = 2a$, and (c) $\sigma = 3a$. (d) Comparison between the usual Wigner-Weyl law [WW, Eq.~(\ref{eq:C:weyl:final})] and the Wigner-Weyl law based on the localization landscape [WWL, Eq.~(\ref{eq:C:weylLL:final})] for a fixed In concentration $x = 10\%$, and varying smearing length $\sigma \in \{a, 2a, 3a\}$. The results were obtained by averaging over $N = 100$ realizations of the alloy of area $L \times L = 40$~nm~$\times~40$~nm. The shaded areas correspond to one standard deviation around the average. The vertical dashed lines indexed with a percentage corresponds to the band gap energy obtained with the bowing formula, Eq.~(\ref{eq:Eg}), for $X$ set to the average concentration $x$. }
\label{fig:2D}
\end{figure*}

\subsection{Absorption spectra in 1D and 2D}

We first consider one- and two-dimensional systems, i.e., either a chain or a monolayer of InGaN with randomly drawn Ga and In atoms. Since the parameters given in Table~\ref{tab1} are relevant for three-dimensional materials, we should not attempt to interpret our results in terms of realistic one- or two-dimensional materials. Provided such materials could be made, the band gap would be \emph{a priori} different, etc.  Furthermore, as noted in Eq.~(\ref{eq:absorption:Weyl}), the prefactor in the absorption coefficient is only valid in 3D.  Nevertheless, we use the parameters from Table~\ref{tab1}, and our only concern in the present section is to assess the quality of our approximations, Eqs.~(\ref{eq:C:weyl:final}), and (\ref{eq:C:weylLL:final}), against the exact formula based on the computation of the eigenstates, Eq.~(\ref{eq:coupling_density}), for the \emph{spectral coupling density} per unit ($d$-dimensional) volume, $\mathcal{C} / |\Omega|$.\\

Figure~\ref{fig:1D}(a) displays the spectral coupling density per unit length averaged over $N=100$~realizations of the alloy chain. The indium concentration and the smearing length are held fixed to $x = 5 \%$ and $\sigma = 2a$, respectively. The exact computation of the spectral coupling density (denoted eigenstates) can be decomposed into three regimes:

(i) Above the band gap energy of GaN, $E_{g}^{ (\mathrm{GaN})} = 3.44$~eV, $\Exp{\mathcal{C}/ |\Omega|}$ exhibits an inverse square-root behavior, $\Exp{\mathcal{C}/ |\Omega|} \propto (\hv - E_{g}^{(\mathrm{GaN})})^{-1/2}$ characteristic of the one-dimensional density of states for a homogeneous material. This is to be expected since for sufficiently large values of $\hv$ Weyl's law applies. This can be interpreted from the fact that the eigenstates at large enough energies are weakly affected by the potential and are perturbed plane waves.

(ii) Within an intermediate range of photon energy $2.5~\mathrm{eV} < \hv < 3.44$~eV, the spectrum exhibits a plateau with two peaks located at $\hv_{00} = 2.98 $~eV and $\hv_{02} = 3.19$~eV (indices 0 and 2 refer to the local ground and second excited states in a well as will become clear below). The plateau can be interpreted as the contribution of transitions between states in the valence band and in the conduction band whose energies are roughly between the minimum and the maximum of each band potential. In other words, this can be seen as the average broadening width of the band edges due to disorder. The two peaks correspond to transitions from states in the valence band to states in the conduction band which are localized on isolated In atoms, and form sets of quasi-degenerate eigenstates, as will be seen below. Note the small standard deviation at the two peaks as indicated by the shaded area, which is a signature of the robustness of these quasi-degenerate eigenenergies from one realization to the other, and comforts the idea that the transitions are indeed between states localized on isolated In atom wells. The peak of lowest energy, $\hv_{00} =2.98 $~eV, corresponds to a transition from the local ground state of an isolated In well in the valence band to the local ground state of the \emph{same} isolated In well in the conduction band [see the red and blue solid lines in the inset of Fig.~\ref{fig:1D}(a)]. The second peak, at photon energy $\hv_{02} = 3.19$~eV, corresponds to a transition between the local second excited state of an isolated In well in the valence band to the local ground state of the same isolated In well in the conduction band [see the dashed red line in the inset of Fig.~\ref{fig:1D}(a)]. The first excited state of the isolated well in the valence band does not couple significantly to the local ground state in the conduction band due to the different parity of the wave functions, and what would be the first excited state in a local well in the conduction band is slightly delocalized compared to that of the valence band due to the difference in effective masses. There is no significant coupling between those as compared to coupling between local ground states. The inset in Fig.~\ref{fig:1D}(a) pictures the aforementioned states and we verify that the differences between their respective eigenenergies indeed match the two peaks energy in the spectrum.

(iii) Finally, for photon energies $\hv < 2.5$~eV, we observe a rapid decay of $\Exp{\mathcal{C}/ |\Omega|}$ with decreasing photon energy, also called the Urbach tail. Transitions contributing to the Urbach tail correspond to low-energy states, respectively, close to the minimum of the disordered conduction potential and the maximum of the disordered valence potential. These are mainly occurring where In atoms occupy several neighboring sites, thus generating deep and broad wells. The probability of occurrence of successive sites occupied by In atoms is exponentially small with increasing number of consecutive sites, and explains the somewhat exponential trend of the Urbach tail.

For one-dimensional systems, we observe that the approximations to $\mathcal{C}$ based on the Weyl law with the original or effective potentials (denoted WW and WWL) both agree with the exact result above the band gap of GaN, as expected asymptotically. The plateau regime and the Urbach tail are also captured, although the Wigner-Weyl-landscape model is in closer agreement with the exact computation in the Urbach tail in terms of trend. However, both approximations fail to capture the peaks which are characteristic of quasi-degenerate states. The reason for this behavior can be understood in the sense that the derived approximations use continuous, smoothly varying potentials or effective potentials. The phase-space Hamiltonian functions $H_c(\vb{r},\vb{k})$ and $H_v(\vb{r},\vb{k})$, or their effective counterparts, are smooth representations of the energy landscape in phase space in a semi-classical picture. To capture the individual peaks in the spectral coupling density, one would need an approach in which the quantized flavor of the states energies is, in some sense, preserved. A simple heuristic to give a correction to the Wigner-Weyl-landscape model and, for example, capture the peak associated to the transitions between local-well ground states could be the following: one could approximate the ground-state energy of a local well by using the rule of thumb~\cite{Arnold:2016} $E_\mu^{(c)} \approx (1 + d/4) \min E_c^{(\mathrm{eff})} $, where the minimum is taken locally for the considered well (and similarly for $E_\nu^{(v)}$), and then consider the probability density (histogram) normalized by the wells volume of the approximated energy differences $(1 + d/4) ( \min E_c^{(\mathrm{eff})} - \max E_v^{(\mathrm{eff})})$ over the domain $\Omega$ (not shown here).
 Figures~\ref{fig:1D}(b)-\ref{fig:1D}(d) show the average spectral coupling density computed both based on the eigenstates or by using the Wigner-Weyl-landscape model for different values of the smearing length $\sigma$, and for different average indium concentration $x$. We observe overall good agreement between the exact calculation and the Wigner-Weyl-landscape model for all considered values of $\sigma$ and $x$.\\

For two-dimensional systems, we observe that due to the weaker variability of the potentials compared to the one-dimensional case [see Eq.~(\ref{eq:varscaling})], there is no peak associated to transitions between quasi-degenerate states. Figure~\ref{fig:2D} shows that the average spectral coupling density increases monotonically with photon energy $\hv$, to reach a constant value when $\hv \to \infty$ as expected from the Weyl law in 2D. Note that the slow decay and the possible oscillations of the spectral coupling density with $\hv$ at high energy for the eigenstate-based computation [Fig.~\ref{fig:2D}(a)-\ref{fig:2D}(c)] comes from the limited number of eigenstates accounted for in the computation. We have indeed observed that, for a computation with fewer realizations,  the high energy behavior becomes constant for a large enough number of eigenstates taken into account. The comparison between the eigenstates based computation of the average spectral coupling density and the Wigner-Weyl-landscape model in Fig.~\ref{fig:2D}(a)-\ref{fig:2D}(c) shows overall good agreement over the whole spectrum and for all the considered indium concentration $x$. The agreement seems to be better for increasing values of the smearing length $\sigma$. The Wigner-Weyl-landscape model seems to slightly overestimate the Urbach tail for $\sigma = a$. Figure~\ref{fig:2D}(d) shows a comparison of the spectral coupling density computed with the Wigner-Weyl and Wigner-Weyl-landscape models for $x=10 \%$ and different values of $\sigma$. We observe that the Wigner-Weyl model clearly overestimates the Urbach tail compared to the Wigner-Weyl-landscape model. This gives a clearer illustration, here in 2D compared to 1D, that the Wigner-Weyl-landscape model indeed performs better than the model based on the usual Weyl law. \\

\subsection{Urbach energy}

\begin{figure*}[t]
\centering
\includegraphics[width = 0.32\textwidth, trim = 1.5cm 0cm .5cm 0cm,clip]{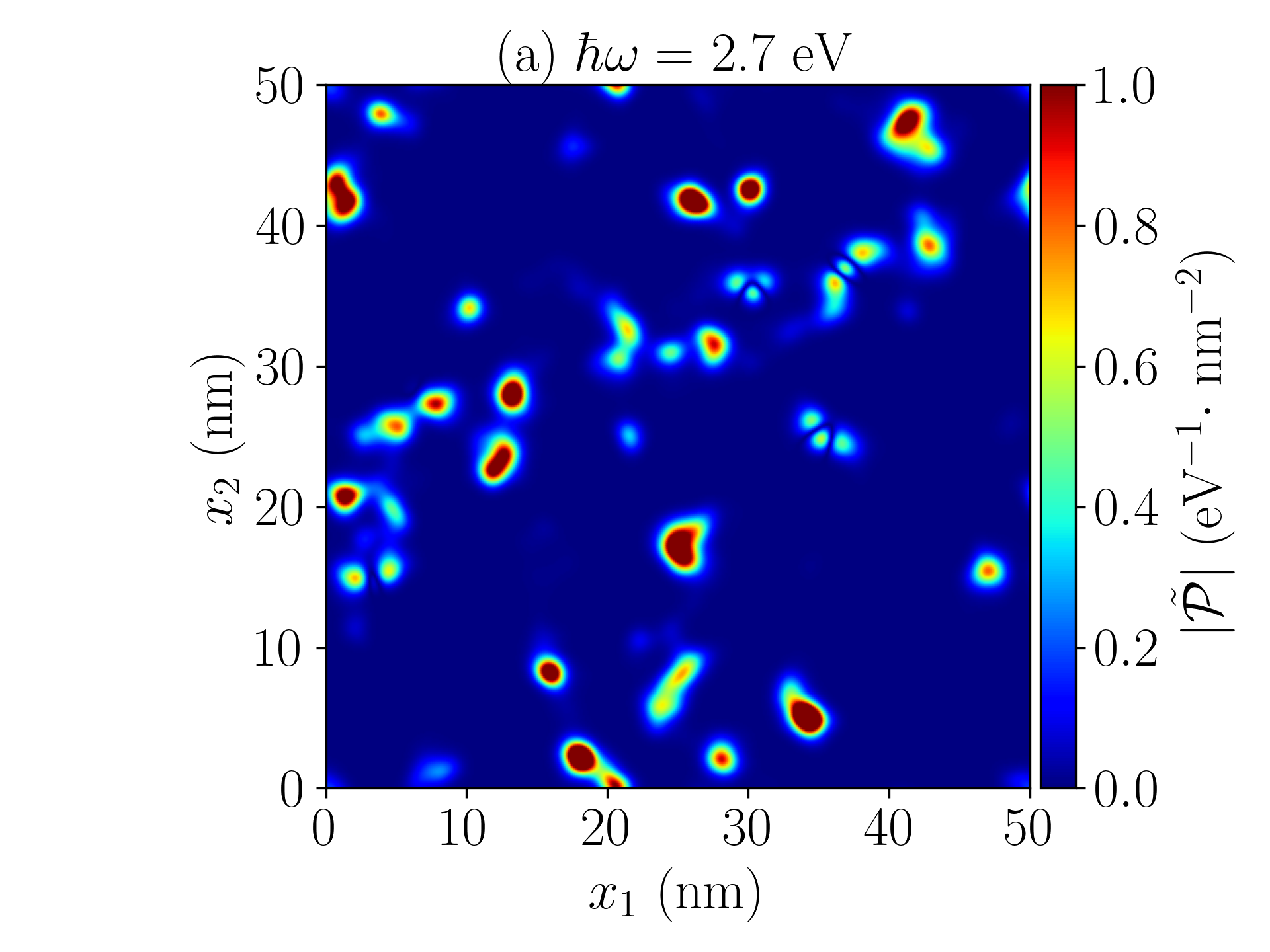}
\includegraphics[width = 0.32\textwidth, trim = 1.5cm 0cm .5cm 0cm,clip]{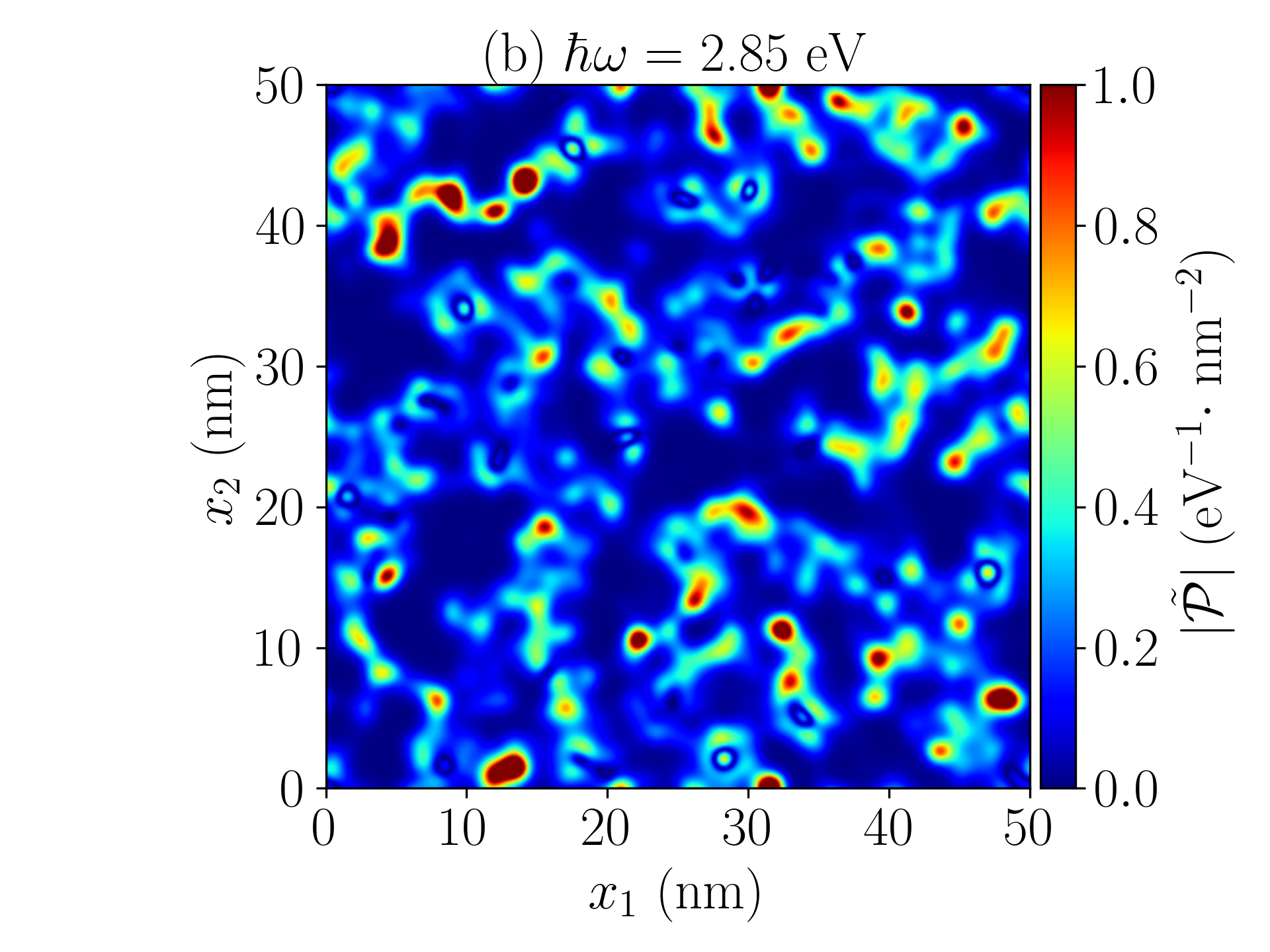}
\includegraphics[width = 0.32\textwidth, trim = 1.5cm 0cm .5cm 0cm,clip]{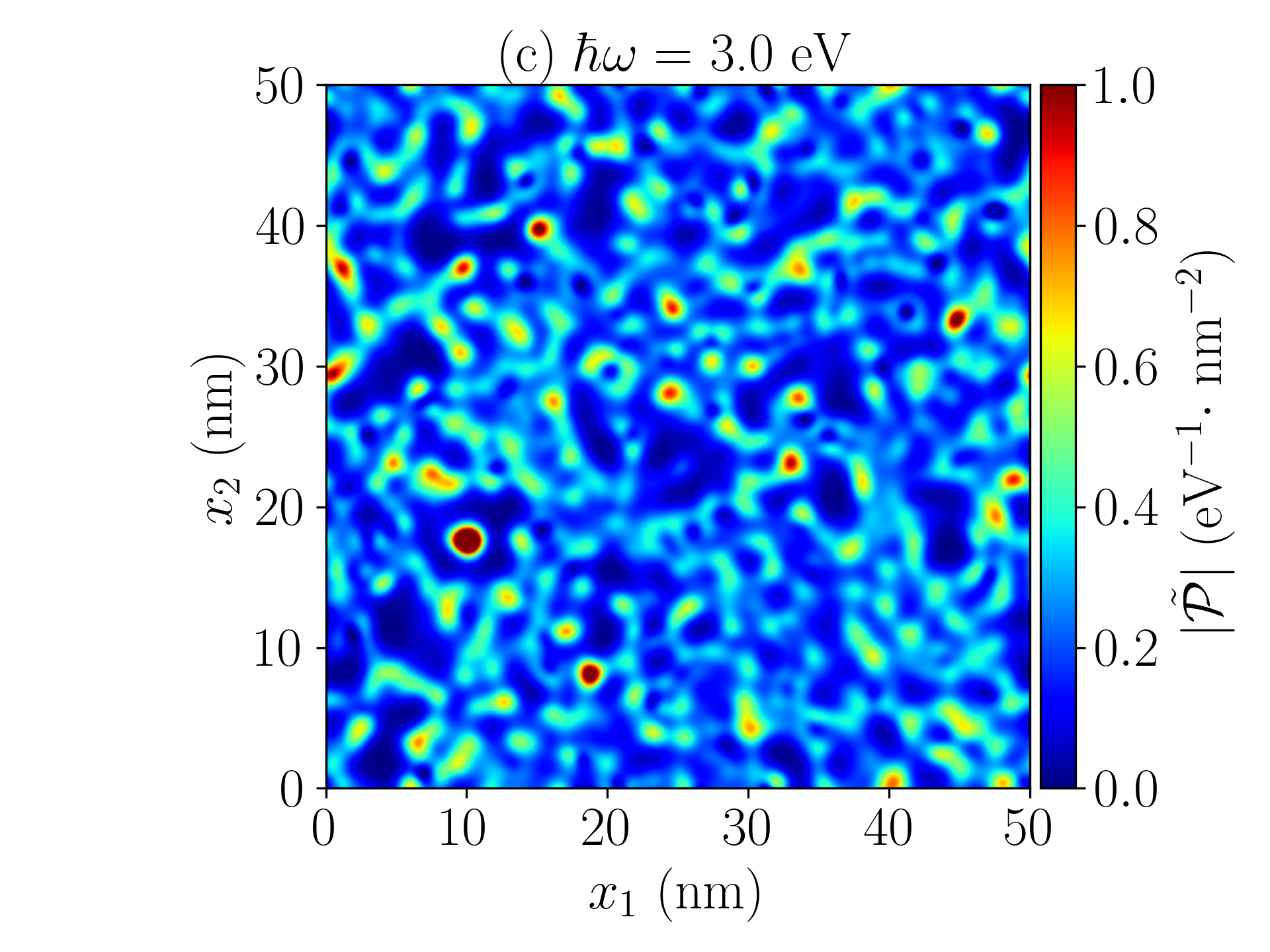}

\includegraphics[width = 0.32\textwidth, trim = 1.5cm 0cm .5cm 0cm,clip]{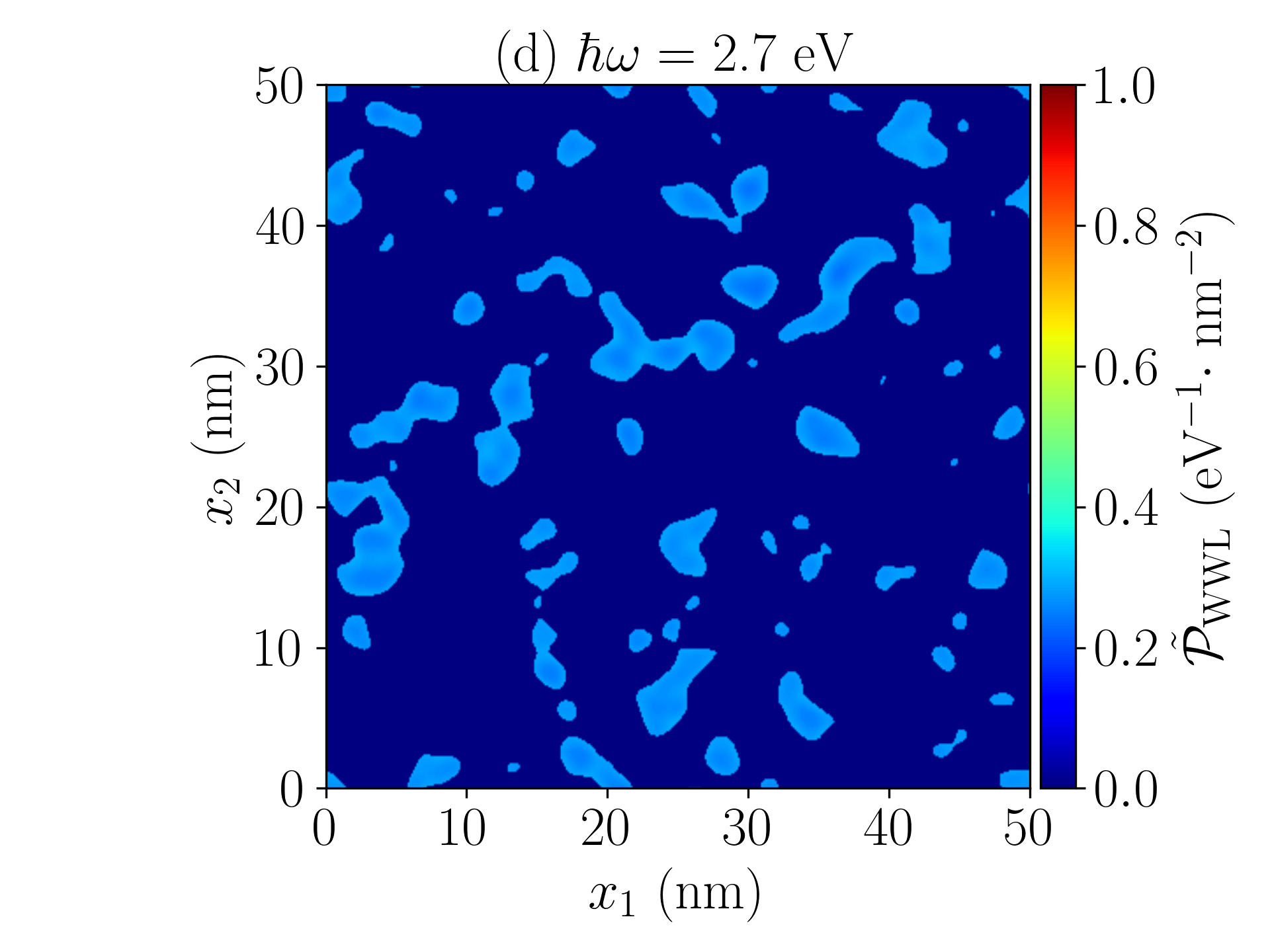}
\includegraphics[width = 0.32\textwidth, trim = 1.5cm 0cm .5cm 0cm,clip]{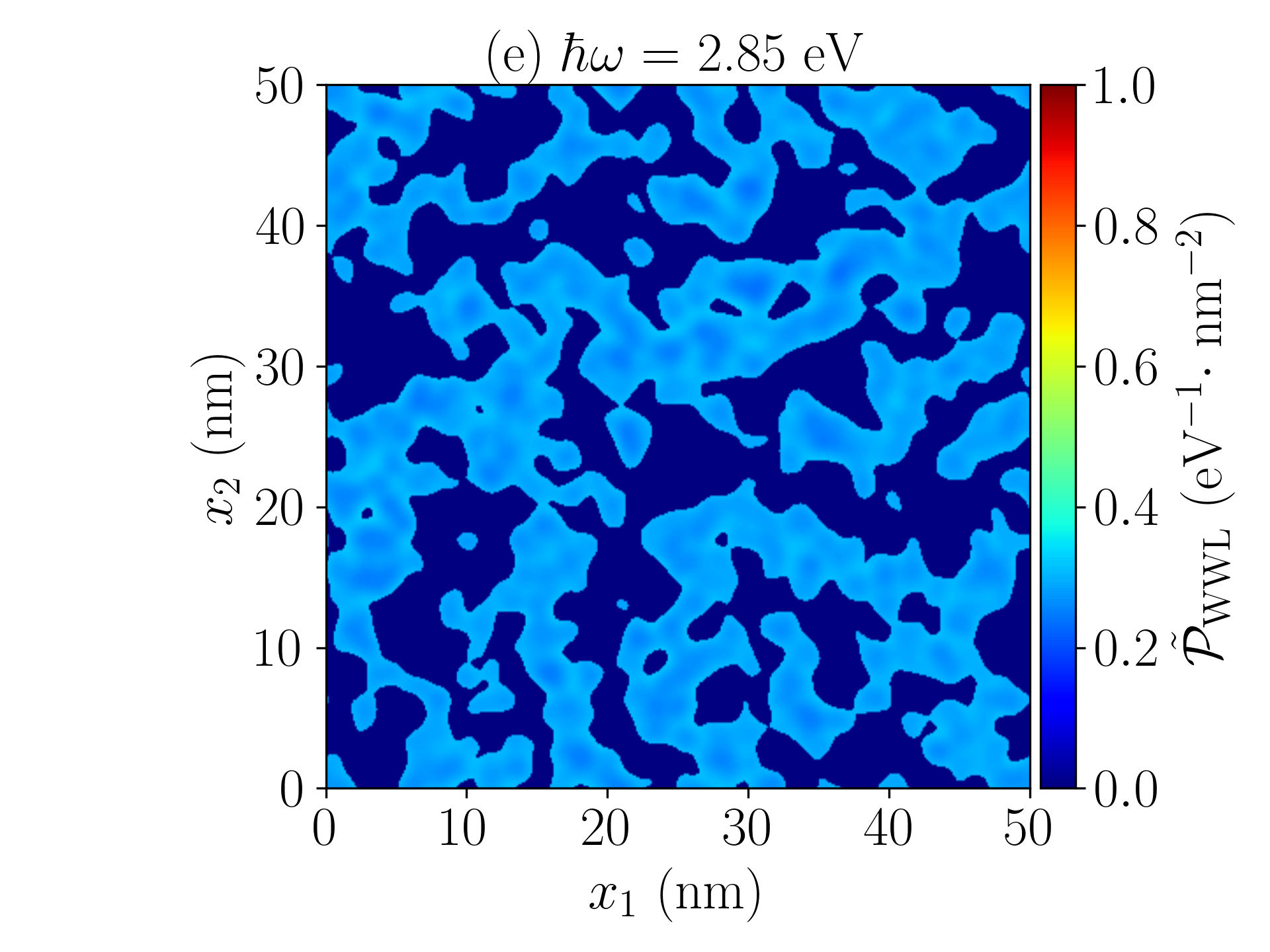}
\includegraphics[width = 0.32\textwidth, trim = 1.5cm 0cm .5cm 0cm,clip]{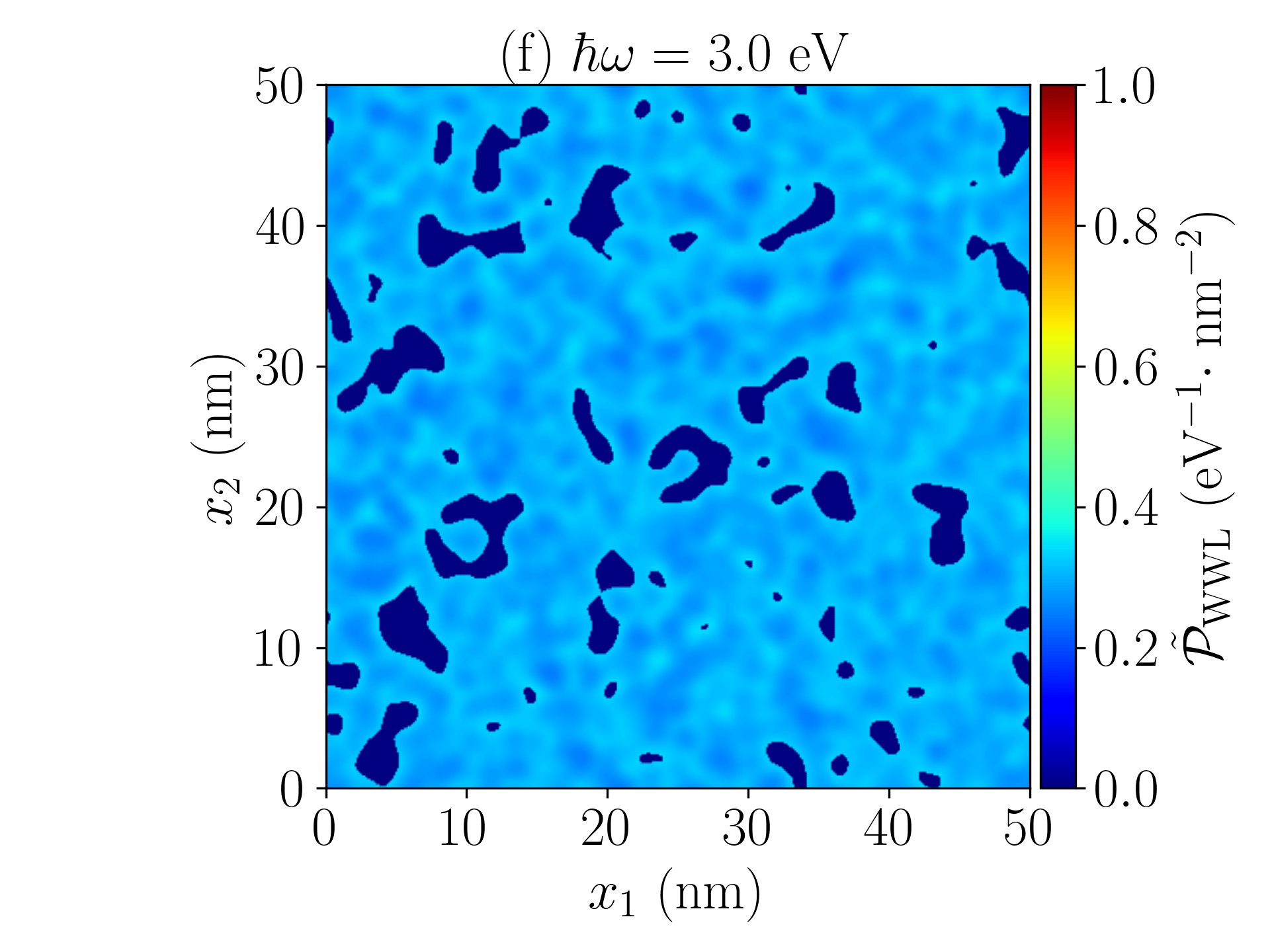}
\caption{(a)-(c) Absolute value of the reduced absorbed power density obtained with Eq.~(\ref{eq:ReducedPow}) for $\varepsilon = 20$~meV and (d)-(f) reduced absorbed approximated by the Wigner-Weyl-localization landscape model, Eq.~(\ref{eq:ReducedPWLL}), for a two-dimensional $\mathrm{In}_{0.15}\mathrm{Ga}_{0.85}\mathrm{N}$ alloy,  with $\sigma = 2a$.   The maps are shown for different values of the photon energy, (a,d) $\hv = 2.7$~eV, (b,e) $\hv = 2.85$~eV, (c, f) $\hv = 3.0$~eV. }
\label{fig:pow2D}
\end{figure*}

\begin{table}[t]
\begin{center}
\caption{Urbach energy $E_U$ for two-dimensional alloys deduced by fitting an exponential function $\alpha \propto \exp(\hv / E_U)$ to the tail of the average spectral coupling density. }
\begin{tabular}{c c c c  }
\hline
\hline
$\sigma / a$ & $x$~(\%)  & $E_U^{(\mathrm{eig})}$~(meV)& $E_U^{(\mathrm{WWL})}$~(meV)  \\[.1cm]
\hline
1.0 & 5 & 48 $\pm$ 2 & 49 $\pm$ 3 \\
1.0 & 10 & 57 $\pm$ 2 & 52 $\pm$ 3 \\
1.0 & 15 & 56 $\pm$ 2 & 52 $\pm$ 3 \\
\hline
2.0 & 5 & 27 $\pm$ 3 & 34 $\pm$ 3 \\
2.0 & 10 & 41 $\pm$ 2 & 40 $\pm$ 3  \\
2.0 & 15 & 46 $\pm$ 4 & 43 $\pm$ 3 \\
\hline
3.0 & 5 & 21 $\pm$ 2 & 23 $\pm$ 3 \\
3.0 & 10 & 20 $\pm$ 2 & 28 $\pm$ 3 \\
3.0 & 15 & 27 $\pm$ 2 & 30 $\pm$ 3 \\
\hline
\hline
\end{tabular}
\label{tab3}
\end{center}
\end{table}

We obtain the Urbach energies (reported in Table~\ref{tab3} for two-dimensional systems) by fitting an exponential function in the Urbach tail both for the eigenstate-based and the landscape-based computations. The fit is performed by minimizing a least-squares cost function
\begin{equation}
\chi^2 = \frac{1}{N_\omega - p} \sum_{n=1}^{N_\omega} \left[ \frac{\ln ( \Exp{\alpha (\hbar \omega_n) }) - \ln (\alpha_{\mathrm{exp}} (\hbar \omega_n)) }{ \Sigma(\hv_n) } \right]^2 \: .
\end{equation}
Here $N_\omega$ is the number of discrete frequency points $\omega_n$ taken into consideration, $\Exp{\alpha}$ is estimated by the empirical average of the absorption coefficient obtained from simulations, $\alpha_\mathrm{exp}(\hv) = \alpha_0 \exp(\hv/E_U)$ is the Urbach tail exponential model where $\alpha_0$ and $E_U$ are free parameters, $p=2$ is the number of free parameters, and
\begin{equation}
\Sigma (\hv_n) = \sqrt{\frac{\Var{\alpha(\hv_n)}}{N}} \frac{1}{\Exp{\alpha(\hv_n)}} 
\end{equation}
is the uncertainty on the logarithm of the empirical average ($\sqrt{ \Var{\alpha} / N}$ is the uncertainty on the average) where $\Var{\alpha}$ is estimated by the empirical variance.
The uncertainty on the Urbach energy, $E_U$, is estimated from the diagonal element of the Hessian matrix of the cost function corresponding to the parameter $E_U$~\cite{nr}. From Table~\ref{tab3}, we see that the Urbach energies $E_U^{(\mathrm{WWL})}$ obtained with the Wigner-Weyl-landscape model are in very good agreement with that obtained with the rigorous model. We note that the Urbach energy tends to increase for increasing indium concentration, which is intuitively understandable since the disorder increases. Moreover, the Urbach energy tends to decrease with increasing smearing length, which we can understand as well since an increasing $\sigma$ means a decreasing strength (variance) of the conduction and valence potentials.

\subsection{Absorbed power density}

Figure~\ref{fig:pow2D} shows the reduced two-dimensional absorbed power density
\begin{align}
\tilde{\mathcal{P}} (\vb{r},\omega)  = &\sum_{\mu \nu} { \BraKet{\chic}{\chiv} }^*  {\chic}^*(\vb{r}) \chiv (\vb{r}) \nonumber\\
&\times \delta \Big( E_\mu^{(c)} - E_\nu^{(v)} - \hbar \omega \Big) \: ,
\label{eq:ReducedPow}
\end{align}
and its approximation [see Eq.~(\ref{eq:PowerWLL})]
\begin{equation}
\tilde{\mathcal{P}}_\mathrm{WWL} (\vb{r},\omega) = \frac{d \, v_d}{2 (2 \pi)^{d}} \left[ \frac{2 m_r(\vb{r})}{\hbar^2} \right]^{d/2} \Big( \hbar \omega - E_g^{(\mathrm{eff})} (\vb{r}) \Big)_+^{d/2-1} \: ,
\label{eq:ReducedPWLL}
\end{equation}
for different values of the photon energy $\hv$.  Note that for the exact reduced power density $\tilde{\mathcal{P}}$ we have used an energy smearing width of $\varepsilon = 20$~meV.  The reason for choosing a rather large energy smearing width is that the approximation $\tilde{\mathcal{P}}_\mathrm{WWL}$ is intrinsically smooth with $\hv$ while for a finite-size system,  $\tilde{\mathcal{P}}$ exhibits contributions at discrete photon energies.  The chosen value is  arbitrary for the comparison but it reflects the intrinsic energy smearing of the effective potentials in this case.  
We observe that the exact reduced power density,  $\tilde{\mathcal{P}}$,  is localized in a small volume at low photon energy [Fig.~\ref{fig:pow2D}(a)] which corresponds to the contribution of local fundamental states. As the photon energy increases, more delocalized states contribute and the power density spreads over a larger volume to eventually become roughly uniform over the whole volume [Fig.~\ref{fig:pow2D}(b)-\ref{fig:pow2D}(c)]. 
Similarly, the approximate power density based on the localization landscape, $\tilde{\mathcal{P}}_\mathrm{WWL}$, exhibits an almost constant value in an increasingly larger domain with increasing photon energy [Fig.~\ref{fig:pow2D}(d)-\ref{fig:pow2D}(f)]. The fact that the density is almost piecewise constant is a particularity of the spatial dimension $d=2$. Indeed, for $d=2$, $\tilde{\mathcal{P}}_{\mathrm{WWL}}(\vb{r},\omega) $ vanishes for $\hv < E_g^{(\mathrm{eff})}(\vb{r})$ and is proportional to $m(\vb{r})$ for  
$\hv > E_g^{(\mathrm{eff})}(\vb{r})$ [note that $(\hv - E_g^{(\mathrm{eff})}(\vb{r}))^0 = 1$].  This local effective mass does not vary much in view of the close values of the effective masses for InN and GaN.  The interesting feature of $\tilde{\mathcal{P}}_{\mathrm{WWL}}$ is that it predicts remarkably well the \emph{volume} in which the eigenstates contribute by comparison with the exact absorbed power density.  The approximated density may be interpreted as a smoothing in energy space,  in some sense,  of the exact power density.

\section{3D absorption}\label{sec:3d}

\subsection{Absorption spectra} 
 
\begin{figure*}[t]
\centering
\includegraphics[width = 0.42\textwidth, trim = 0cm 0cm 0cm 0cm,clip]{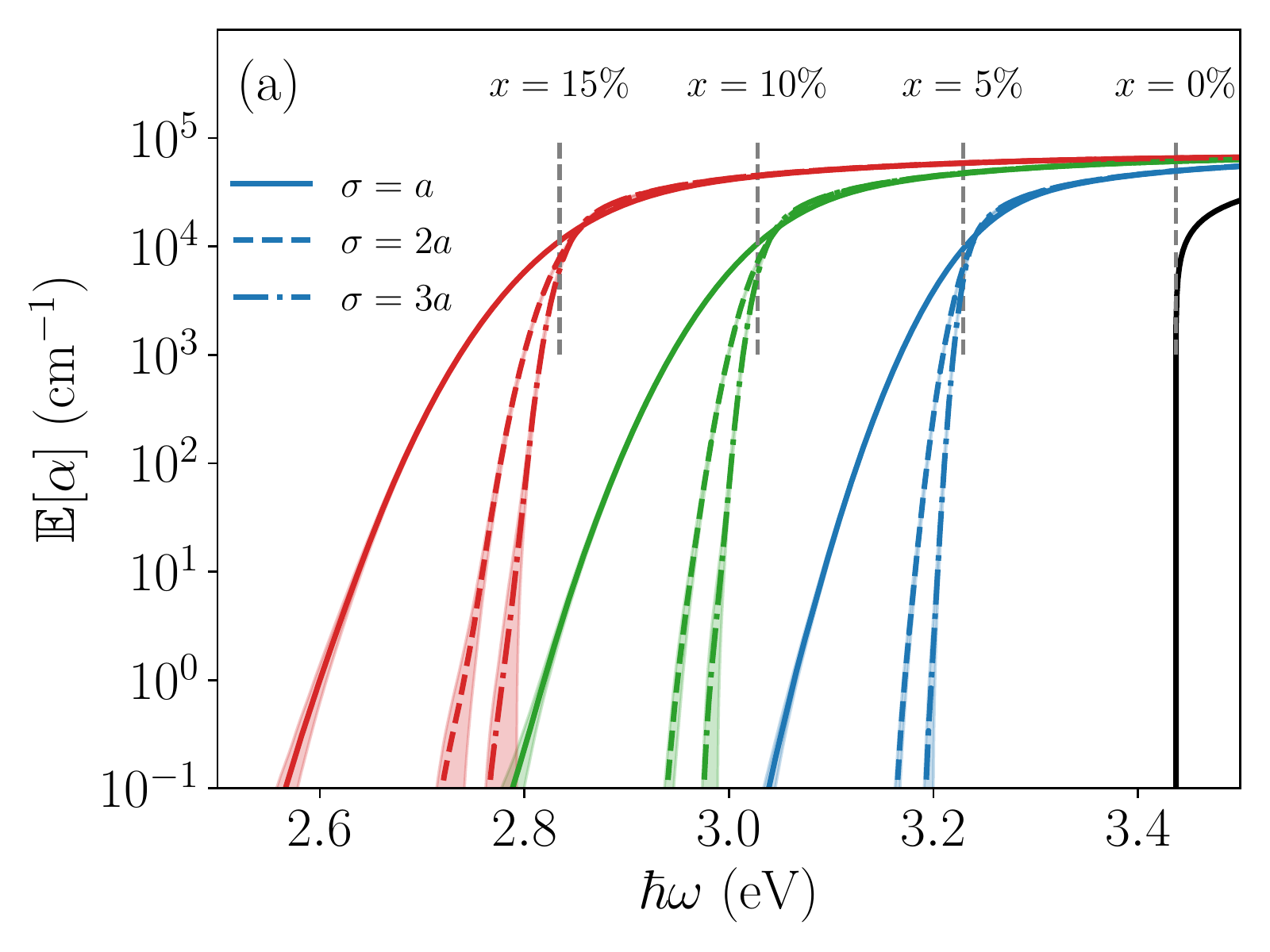}
\includegraphics[width = 0.42\textwidth, trim = 0cm 0cm 0cm 0cm,clip]{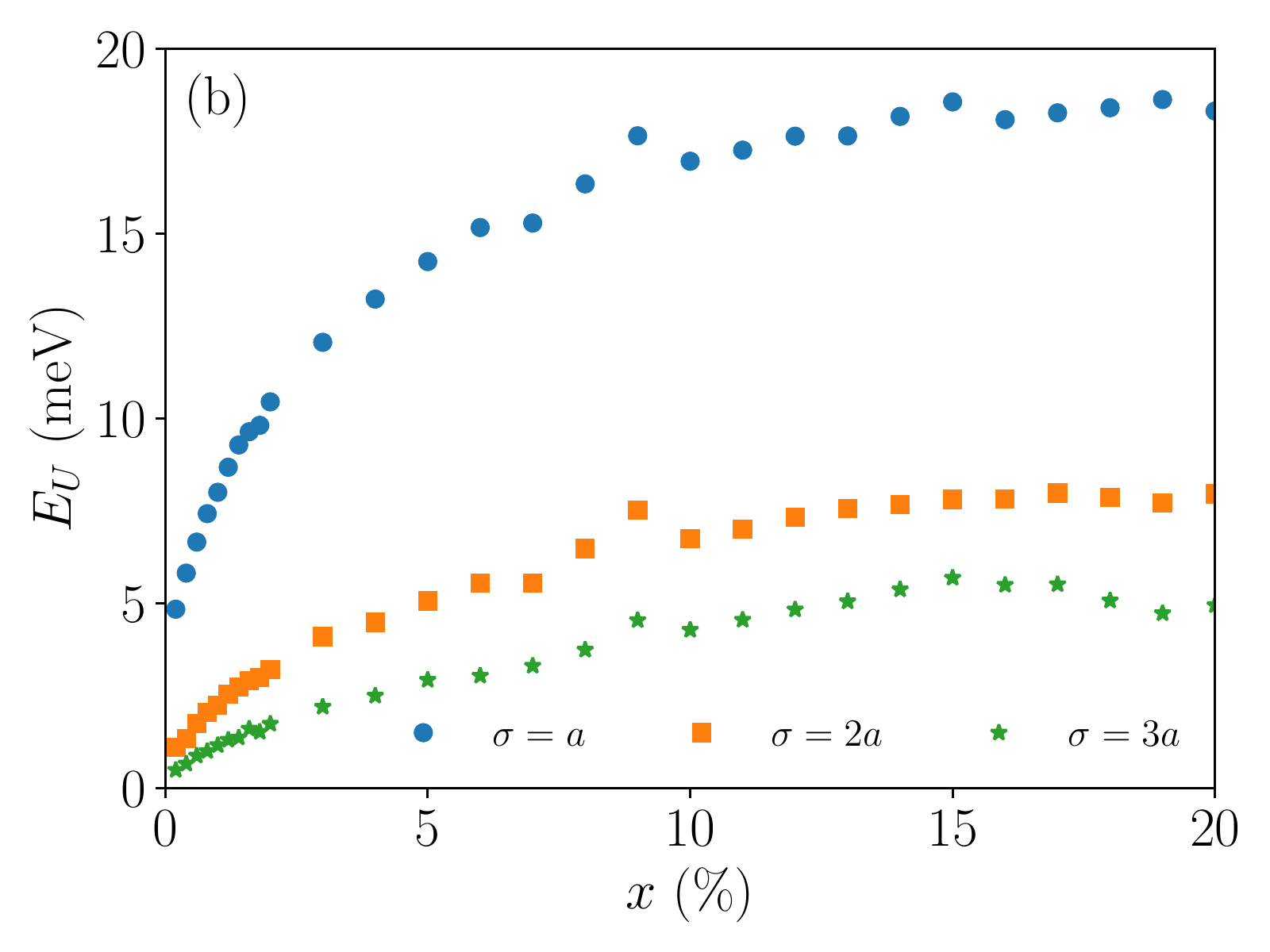}
\caption{ (a) Absorption coefficient as a function of photon energy $\hv$ for different values of the average indium concentration $x$ and of the smearing length $\sigma$.  The shaded area correspond to $\pm 2 \sqrt{\Var{\alpha} / N}$.  (b) Urbach energy as a function of $x$ for different values of $\sigma$.  The data are obtained based on the Wigner-Weyl localization landscape approach.}
\label{fig:Urbach3D}
\end{figure*}

We now turn to three-dimensional systems for which the computation of the eigenstates becomes unpractical for reasonable system sizes.   Figure~\ref{fig:Urbach3D}(a) displays the absorption coefficient spectra obtained with the Wigner-Weyl-landscape model for a few values of the average indium concentration $x$ and for $\sigma = a$, $2a$, and $3a$.  Consistently with our observations for one- and two-dimensional systems, the Urbach tail is less pronounced for larger values of the smearing length $\sigma$. Furthermore, the Urbach energy which controls the decay of the Urbach tail is also smaller than the values obtained for two-dimensional systems for the same value of $\sigma$ [compare Table~\ref{tab3} with Fig~\ref{fig:Urbach3D}(b)]. This is due to the lower variability of the potentials $E_c$ and $E_v$ with the space dimension $d$ [see Eq.(\ref{eq:varscaling})].  Consequently the potentials are less confining. This is particularly true for electrons in the conduction band.  A calculation of a few wave functions for one realization of the alloy (not shown here) shows that the wave functions in the conduction band are delocalized over the entire box and quickly resemble plane waves with increasing energy while the wave functions in the valence band remain localized in local potential wells near the band edge. These observations are in agreement with comparable computations reported in the literature in the \emph{absence} of interface fluctuations in quantum well,  or of the electron-hole Coulomb interaction~\cite{Tanner:2018,David:2019}.\\
 
Figure~\ref{fig:Urbach3D}(b) shows the dependency of the Urbach energy with the average indium concentration $x$ between 0 and 20\%.   We observe that the Urbach energy increases with the indium concentration and decreases with increasing smearing length, as observed for two-dimensional alloys.  It is instructive to compare the values of Urbach energies we have obtained in 2D and 3D, with values obtained experimentally and numerically for quantum wells in Ref.~\cite{Piccardo:2017}.  In Ref.~\cite{Piccardo:2017}, Piccardo \emph{et al.} found values of Urbach energies in the range between 15 and 25~meV for indium concentration varying between 10\% and 30\% by using the technique of bias photocurrent spectroscopy~\cite{Helmers:2013}.  They also found using a model based on the EMA, that a value of $\sigma \approx 2a$ was appropriate to fit the experimental data.  In view of Table~\ref{tab3} and Fig.~\ref{fig:Urbach3D}(b), for $\sigma = 2a$ we have values of Urbach energies which are about 40~meV in 2D and 7~meV in (bulk) 3D for these indium concentrations. Considering that a quantum well is a quasi-two-dimensional system and also the effect of piezo-electric field (which is absent in our calculation),  the fact that the values obtained in Ref.~\cite{Piccardo:2017} fall between the values we have obtained for 2D and 3D systems is quite comforting.  A more detailed comparison between models and experiments is left for a future work.

\begin{figure*}[t]
\centering

\includegraphics[width = 0.32\textwidth, trim = 2.4cm .6cm 0cm 1.6cm,clip]{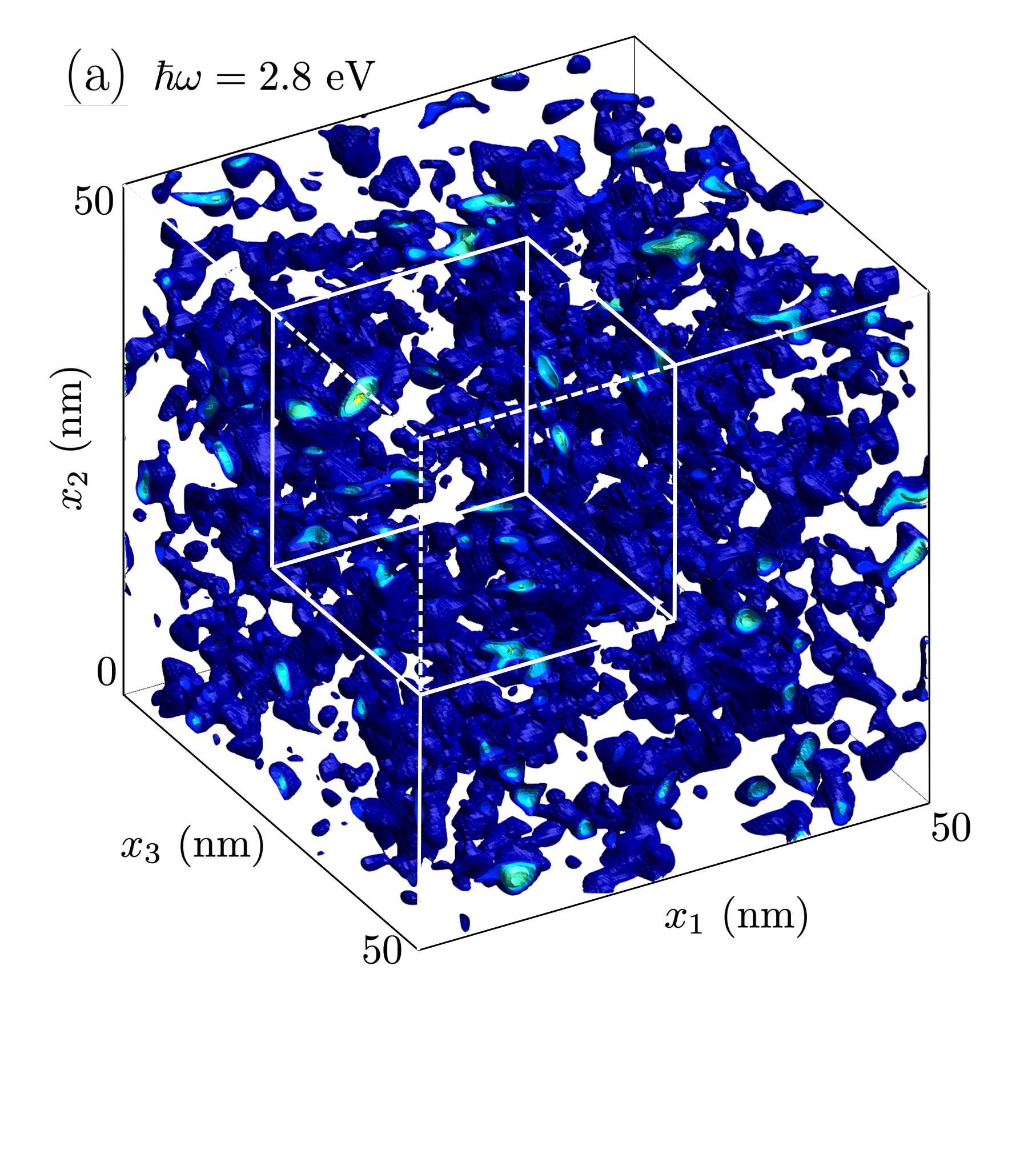}
\includegraphics[width = 0.32\textwidth, trim = 2.4cm .6cm 0cm 1.6cm,clip]{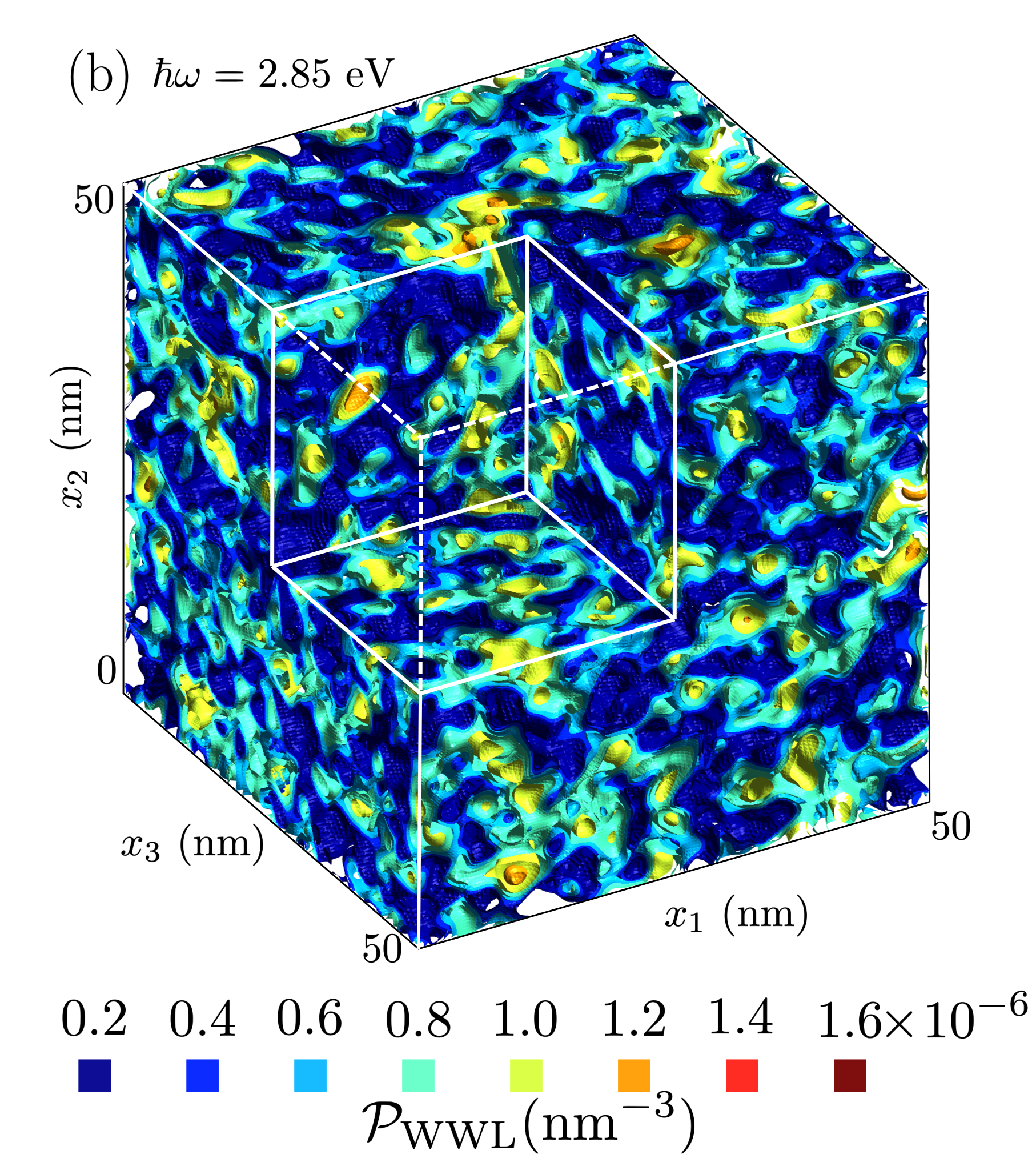}
\includegraphics[width = 0.32\textwidth, trim = 2.4cm .6cm 0cm 1.6cm,clip]{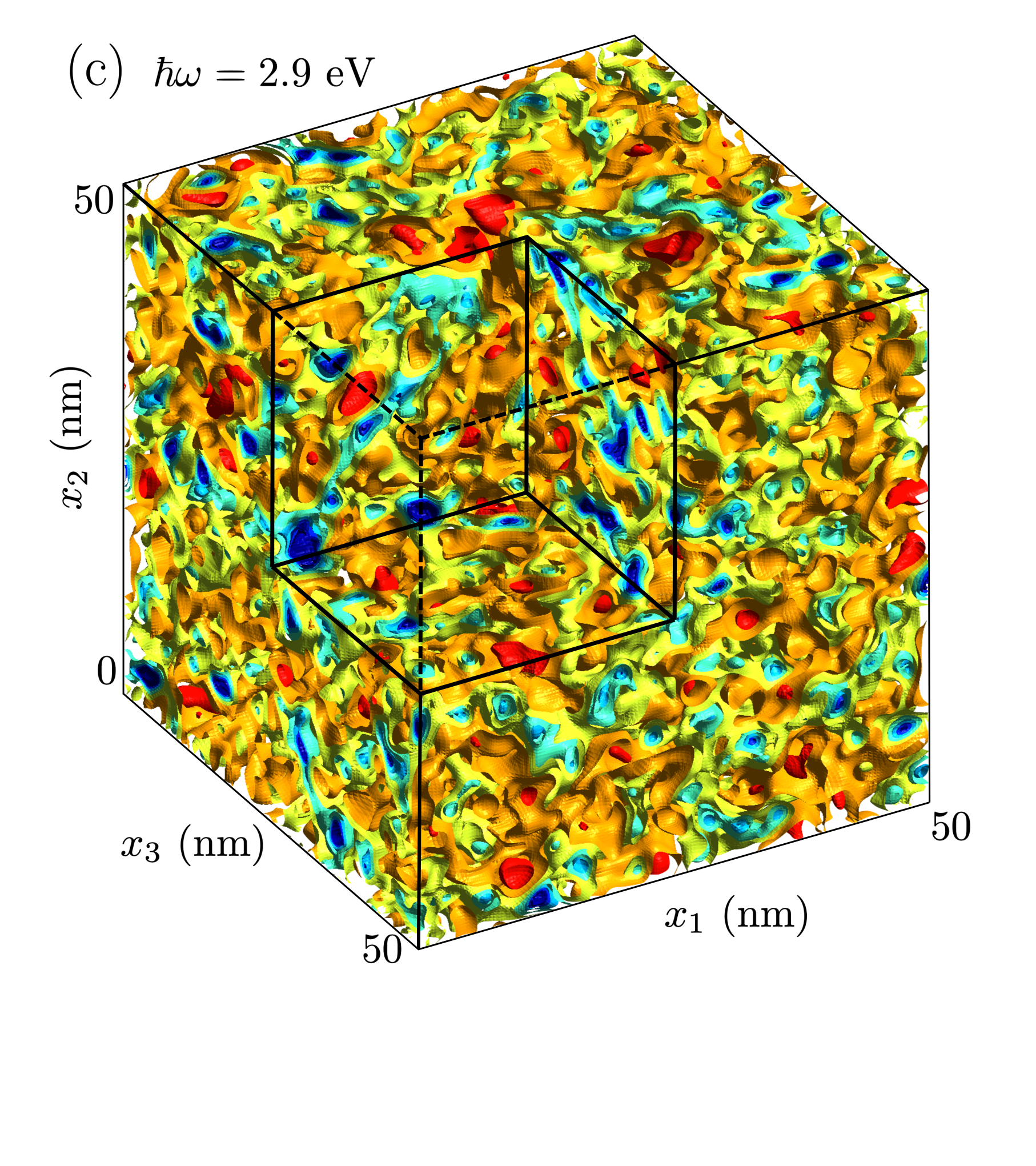}
\caption{ Absorbed power density $\mathcal{P}_\mathrm{WWL}$ for a realization of the alloy for three values of $\hbar \omega$: (a) $\hbar \omega = 2.8$~eV, (b) $\hbar \omega = 2.85$~eV, (c) $\hbar \omega = 2.9$~eV.  The results were obtained with Eq.~(\ref{eq:PowerWLL}) for a size of the computational domain $L \times L \times L = 50$~nm~$ \times 50$~nm~$ \times 50$~nm, with element size $\Delta x =3$~\r{A},  average indium concentration $x = 15\%$ and smearing length $\sigma  = 2 a$.  The eighth top front corner cube is removed to help visualize the inside of the volume. The color scale is common for the three values of $\hbar \omega$ to ease the comparison.}
\label{fig:3D}
\end{figure*}

\subsection{Absorbed power density}

The absorbed power density computed based on Eq.~(\ref{eq:PowerWLL}) is shown in Fig.~\ref{fig:3D} for different photon energies for a domain of $50$~nm side length.  At low enough photon energy,  at the bottom of the Urbach tail [Fig.~\ref{fig:3D}(a)],  we observe that only localized regions contribute to the absorbed power.  As the photon energy increases, an increasingly larger volume contributes to the absorption and with more intensity.  This gives the intuitive picture of an underlying energy landscape,  the effective band gap profile,  which is filled up as the photon energy increases. Furthermore,  with increasing photon energy,  more modes contribute at a given point, a feature which is encoded in the $(\hv - E_g^{(\mathrm{eff})})^{1/2}$ law, in Eq.~(\ref{eq:PowerWLL}), and which is reminiscent of the density of states.

\section{Conclusion}


In summary, we have derived a computationally efficient model for light absorption in disordered semiconductor alloys. The model is based on an original approach in phase space and takes advantage of the localization landscape theory. We have demonstrated that the model gives an accurate prediction for the absorption coefficient over the whole spectrum when compared with the model based on the solution of the Schr\"{o}dinger equations.  The computational speed-up has been estimated for one and two-dimensional systems to be about two orders of magnitude.  Such a speed-up is considerable, especially for three-dimensional systems of relatively large size for which the use of standard methods would be prohibitive.\\


 The presented framework offers unique directions to be investigated. Allowing for fast computation of the absorption coefficient in 3D, the model could be compared to a light absorption experiment for bulk semiconductor alloys like thick layers of InGaN for various indium concentrations,  or more exotic alloys such as perovskites. Moreover, relatively large devices could be simulated such as multiple disordered quantum wells.  \\
 
 In addition, it is well acknowledged that the electron-hole Coulomb interaction can play a significant role in absorption spectra, or more generally in the electronic structure~\cite{David:2019}.  It would be of great theoretical and numerical interest to analyze how we could generalize the presented theoretical approach in phase-space accounting for the electron-hole interaction. \\

  In the present paper, we have restricted ourselves to uncorrelated alloys but the theoretical framework can handle correlated atomic disorder as well.  Exploring the effect of spatial atomic species correlation on the absorption spectrum, for example, due to clustering~\cite{DiVito:2020}, or even more subtle correlations would be both of fundamental and practical interest. 
  Due to the limited precision and efficiency of APT and TEM,  atomic correlations which may hardly be visible with the aforementioned techniques could be complemented by a careful analysis of the absorption tail,  which we believe should be sensitive to atomic correlations.  If one could rely on a precise modeling of the absorption spectrum, in particular, incorporating the Coulomb interaction, deviations from the computed ideal case could be used to assess statistical properties of the alloy based on its physical impact on electronic properties. \\
  
  Furthermore,  increasing interest has emerged in recent years on non-local coupling between extended excitonic states and the electromagnetic field, beyond the dipole approximation,  which may yield significant effects both on the spectra and on the excitonic states life time.  Such a regime becomes relevant for high refractive index and material exhibiting a large scale disorder on a scale of a few nanometers~\cite{Stobbe:2012}.  It would be interesting to see whether our approach can be adapted to go beyond the dipole approximation. \\
   
Beyond the study of light absorption,  the Wigner-Weyl framework is quite general and should apply to a broader class of problems.  The apparently obvious next question to be addressed is that of luminescence phenomena.  Our framework should be easily adaptable to radiative recombination, at least in the assumption of relaxed electrons.  Indeed,  one may assume Fermi-Dirac statistics for the relaxed electrons and holes near the band edges and carry out the derivation presented in the paper by weighting the density of states in phase space with Fermi-Dirac distributions.  Additionally, this would also allow for the study of the effect of temperature on both absorption and luminescence spectra.  Non-radiative recombinations, such as Auger processes, for example,  are also likely to be modeled within the same framework at the expense of coupling three densities of states in phase space instead of two.  Maybe more surprising,  the problem of phonon-assisted transport may also be suitably modeled by the Wigner-Weyl approach in view of the mathematical similarity between the electron-phonon interaction and the electron-photon interaction, although care should be taken with the different wavelength regimes. 

\section{Acknowledgments}

All authors are grateful to Professor Douglas Arnold and Professor James Speck for invaluable discussions.  The authors are also grateful to Myl\`{e}ne Sauty and Abel Thayil for regular and fruitful discussions.
 J.-P.~B., P.~P., and M.~F. are supported by the Simons foundation Grant No. ~601944.  S.~M. is supported by the NSF RAISE-TAQS Grant No. ~DMS-1839077 and the Simons foundation Grant No.~563916. C.~W. is supported by the Simons foundation Grant No.~601954 and NSF Raise-TAQS Grant No.~DMS-1839077.

\appendix

\section{Validity of the effective mass approximation} \label{app:validity}

We would like to make a few remarks on the disordered band approximation, in particular, to motivate its relevance for modeling InGaN. The method is inspired by the so-called \emph{approximation of the envelope wave function}, also known as the \emph{effective mass approximation} (EMA),  for the modeling of quantum wells.  In this context, effective Schr\"{o}dinger equations are written for the different carriers experiencing piece-wise constant or linear potentials on scales of few nanometers, i.e., many lattice parameters. These potentials are constructed via the band gaps of the involved semiconductor layers and potentially electric fields~\cite{Weisbuch:book,Singh_2003}. Although the method is largely validated, and is in fact the state of the art for describing and designing quantum well devices, the validity of the EMA to model \emph{disordered semiconductors} at the sub-nanometer scale is not immediate.  Indeed, why should an alloy even preserve a crystalline band structure? An enlightening answer was given by Popescu and Zunger, who showed numerically that an \emph{effective band structure}, in the sense of a broadening and a deformation of the dispersion curves with increasing alloying concentration can still be defined \emph{but} only for some types of alloys~\cite{Popescu:2010,Popescu:2012}. Depending on the atomic species involved in the alloy, one can observe either a broadening of the band structure for a large range of alloying concentration, or the apparition of impurity states inside the gap at low alloying concentration, to a full population of the gap and destruction of the band structure at higher concentrations. InGaN belongs to the first category which motivates the use of the effective mass approximation~\cite{Popescu:2010,Popescu:2012}.  Furthermore, this approach has been used for modeling disordered quantum well devices with remarkable agreement with experiments,  provided the value of the smearing parameter is adequately chosen~\cite{Filoche:2017:PRB,Piccardo:2017}.

The next question is the choice of EMA parameters and evaluating how well they will lead to a local disordered potential representative of the alloy. This can be done by (i) comparison with other models of disordered semiconductor alloys, supposedly more accurate,  or by (ii) comparison of the resulting computations with some observables of the system.  In a number of cases, EMA has been compared with density functional theory,  such as for Si quantum dots, for which an excellent agreement is obtained between the two methods~\cite{Delley:1995}.  For nitride alloys, comparisons between atomistic models and EMA have also been made, with only small differences~\cite{Watson-Parris:2011,Chaudhuri:2021}.  It is, however,  still difficult to assess these differences to true deficiencies of either computations or to the choice of parameters (e.g.,  the choice of the so-called bowing parameter describing the non-linear variation of the alloy bandgap with alloy composition for the EMA, see, e.g., Caro \emph{et al.}~\cite{Caro:2013}).  Turning to (ii), comparing with experiment, the situation is also somewhat undecisive due to uncertainties on samples quality and geometries~\cite{weisbuch:review:2021}.  For instance,  analysis of the Urbach tails in InGaN quantum wells (QWs) reported by Piccardo~\cite{Piccardo:2017} and David~\cite{David:2019} are significantly different. Both rely on the EMA for analysis.  David \emph{et al.} include the Coulomb interaction to obtain agreement with a single QW absorption data.  In contrast,  Piccardo \emph{et al.} rely on layer thickness fluctuations to fit the larger Urbach tail of multiple QWs samples,  and the impact of Coulomb interaction might be hidden by these fluctuations.

In any case, the use of the EMA is sufficient at this point to generate a representative disordered potential to evaluate the computational approach developed in the present paper.  As observed in Secs.~\ref{sec:benchmark} and \ref{sec:3d},  the final results depend significantly on the value of the smearing length.

\section{Momentum matrix element factorization}\label{app:momentum}

By definition, the matrix element $M_{\mu \nu} = \Bra{\psic} \Vie{A}{0}{} \cdot \Vie{\hat{p}}{}{} \Ket{\psiv}$ can be expanded as
\begin{align}
M_{\mu \nu}  = \, &-i\hbar \Vie{A}{0}{} \cdot \Bigg[ \int_\Omega u_c^{*}(\vb{r})  \nabla u_v (\vb{r})  {\chic}\!^* (\vb{r}) \chiv (\vb{r}) \, \mathrm{d}^3 r \nonumber\\
&+ \int_\Omega u_c^{*}(\vb{r})   u_v (\vb{r})  {\chic}\!^* (\vb{r}) \nabla \chiv (\vb{r}) \, \mathrm{d}^3 r  \Bigg] \: .
\end{align}
Assuming the envelope functions to be slowly varying over the unit cell, the integration over $\Omega$ can be approximated by a sum over unit cells $\Omega_i$ centered on $\Vie{r}{i}{}$ contained in $\Omega$ where the envelope functions are considered constant over each cell, i.e.,
\begin{align}
M_{\mu \nu}  = \, &-i\hbar \Vie{A}{0}{} \cdot \sum_{i \in \mathcal{L}} \Bigg[ {\chic}\!^* (\Vie{r}{i}{}) \chiv (\Vie{r}{i}{})  \int_{\Omega_i} u_c^{*}(\vb{r})  \nabla u_v (\vb{r})  \, \mathrm{d}^3 r \nonumber\\
&+  {\chic}\!^* (\Vie{r}{i}{}) \nabla \chiv (\Vie{r}{i}{}) \int_{\Omega_i} u_c^{*}(\vb{r})   u_v (\vb{r})  \, \mathrm{d}^3 r  \Bigg] \: .
\end{align}
The integral in the second term in the above equation vanishes as it is the scalar product of two unit cell Bloch functions of different bands. The integral in the first term is the momentum matrix element between $u_c$ and $u_v$ and it does not depend on the specific unit cell $\Omega_i$ since $u_c$ and $u_v$ are lattice periodic. Thus, we have
\begin{equation}
 \frac{-i \hbar}{|\Omega_i|} \int_{\Omega_i} u_c^{*}(\vb{r})  \nabla u_v (\vb{r})  \, \mathrm{d}^3 r = \Bra{u_c} \Vie{\hat{p}}{}{} \Ket{u_v} \: ,
\end{equation}
and
\begin{align}
M_{\mu \nu} &= \Bra{u_c} \Vie{A}{0}{} \cdot \Vie{\hat{p}}{}{} \Ket{u_v} \sum_{i \in \mathcal{L}} {\chic}\!^* (\Vie{r}{i}{}) \chiv (\Vie{r}{i}{}) |\Omega_i|  \nonumber\\
&\approx  \Bra{u_c} \Vie{A}{0}{} \cdot \Vie{\hat{p}}{}{} \Ket{u_v} \int_{\Omega} {\chic}\!^* (\vb{r}) \chiv (\vb{r}) \: \mathrm{d}^d r \: ,
\end{align}
which justifies Eq.~(\ref{eq:Mfactorized}).

\section{Homogeneous limit}\label{app:flat:band}

We consider here the limit where either the average indium concentration $x \to 0$ or where the smearing length $\sigma \to \infty$, the disordered conduction and valence bands become constant,   $E_c(\vb{r}) = E_c$ and $E_v(\vb{r})=E_v$,  and so are the effective masses $m_c(\vb{r}) = m_e$ and $m_v(\vb{r}) = - m_h$. In such cases, the envelope functions reduce to plane waves $\chic (\vb{r}) = \exp (i \Vie{k}{\mu}{} \cdot \vb{r}) / \sqrt{|\Omega|}$ and $\chiv (\vb{r}) = \exp (i \Vie{k}{\nu}{} \cdot \vb{r}) / \sqrt{|\Omega|}$. Their scalar product becomes $\BraKet{\chic}{\chiv} = \delta_{\Vie{k}{\mu}{}, \Vie{k}{\nu}{}} $. The wave vectors are given by $\Vie{k}{\mu}{} = \sum_{i=1}^d 2 \pi \mu_i \Vie{e}{i}{} / L$, where $\mu = (\mu_1,\cdots,\mu_d) \in \mathbb{Z}^d$ is a multi-index, and similarly for $\Vie{k}{\nu}{}$. The eigenenergies are given by $E_\mu^{(c)} = E_c + \hbar^2 |\Vie{k}{\mu}{}|^2/ 2 m_e$, and $E_\nu^{(v)} = E_v - \hbar^2 |\Vie{k}{\nu}{}|^2/ 2 m_h$. The spectral coupling density then reads
\begin{align}
\mathcal{C}(\hv) &= \sum_{\mu} \delta \left( \frac{ \hbar^2 |\Vie{k}{\mu}{}|^2 }{2 m_r} + E_g - \hv \right) \nonumber\\
&\approx \frac{|\Omega|}{(2 \pi)^d} \int \delta \left( \frac{ \hbar^2 |\vb{k}|^2 }{2 m_r} + E_g - \hv \right) \: \mathrm{d}^d k \nonumber\\
&=  \frac{d v_d |\Omega|}{2 (2 \pi)^d} \left[ \frac{2 m_r}{\hbar^2} \right]^{d/2} \, \left( \hv - E_g \right)_+^{d/2-1} \: .
\end{align}
Here we have introduced the reduced effective mass  $m_r^{-1} = m_e^{-1} + m_h^{-1}$, used the density of states in $k$ space $|\Omega| / (2 \pi)^d $ and a change of variable $E = \hbar^2 |\vb{k}|^2 / 2 m_r$. The factor $v_d = \pi^d / \Gamma(d/2+1)$ is the volume of the unit ball in dimension $d$ and the $+$ subscript denotes the positive part function $x \mapsto x_+ = \max(x,0)$. The absorption coefficient in 3D is then given by Eq.~(\ref{eq:alphaC}) and becomes
\begin{equation}
\alpha^{(0)}(\omega) = \frac{3 v_3 e^2 E_p \left[ \frac{2 m_r}{\hbar^2} \right]^{3/2} }{ 2 (2 \pi)^{2} m_0 \varepsilon_0 \omega c_0 n(\omega)} \,  \, \left( \hv - E_g \right)_+^{1/2} \: .
\label{eq:alpha:homo}
\end{equation}

\section{Marginal distributions of $\mathcal{D}^{(c)}$}\label{app:IDOS}

We prove here the identity in Eq.~(\ref{eq:LDOSP:IDOS}). 
First, by definition of $\mathcal{D}^{(c)}(\vb{r},\vb{k},\varepsilon)$, we readily have
\begin{equation}
\int_{-\infty}^E \mathcal{D}^{(c)}(\vb{r},\vb{k},\varepsilon) \: \mathrm{d}\varepsilon = \sum_\mu W_{\chic}(\vb{r},\vb{k}) \, \Theta(E - E_\mu^{(c)}) \: ,
\end{equation}
where $\Theta$ is the Heaviside step function. Integrating the above quantity over phase space yields
\begin{align}
&\iint \int_{-\infty}^E \mathcal{D}^{(c)}(\vb{r},\vb{k},\varepsilon) \: \mathrm{d}\varepsilon \, \frac{\mathrm{d}^d r \, \mathrm{d}^d k}{(2\pi)^d} = \nonumber\\
& \sum_\mu \iint W_{\chic}(\vb{r},\vb{k}) \: \frac{\mathrm{d}^d r \, \mathrm{d}^d k}{(2\pi)^d} \, \Theta (E - E_\mu^{(c)}) \: .
\label{eq:LDOSP:IDOS_proof}
\end{align}
Finally, the states $\chic$ being $L^2$ normalized and in virtue of the property of the Wigner transform for the marginal density~\cite{Mallat:ch4}
\begin{equation}
\int W_{\chic}(\vb{r},\vb{k}) \frac{\mathrm{d}^d k}{(2\pi)^d} = |\chic (\vb{r})|^2 \: ,
\end{equation}
we have
\begin{equation}
\iint W_{\chic}(\vb{r},\vb{k}) \frac{\mathrm{d}^d r \, \mathrm{d}^d k}{(2\pi)^d} = 1 \: .
\end{equation}
Inserting the above equation in Eq.~(\ref{eq:LDOSP:IDOS_proof}) completes the proof since
\begin{align}
\iint \int_{-\infty}^E \mathcal{D}^{(c)}(\vb{r},\vb{k},\varepsilon) \: \mathrm{d}\varepsilon \, \frac{\mathrm{d}^d r \, \mathrm{d}^d k}{(2\pi)^d} &= \sum_\mu \Theta (E - E_\mu^{(c)}) \nonumber\\
&= \mathrm{IDOS}^{(c)}(E) \: ,
\end{align}
the last line being the definition of the IDOS.  As a side note, it is also straightforward to show that we have the following relationships between the  quasi-density of states in phase space and the density of states (DOS), the local density of states (LDOS) and density of states in momentum space (MDOS) also known as the spectral function for plane waves:
\begin{align}
\iint \mathcal{D}^{(c)}(\vb{r},\vb{k},E) \: \frac{\mathrm{d}^d r \, \mathrm{d}^d k}{(2\pi)^d} &= \sum_\mu \delta (E - E_\mu^{(c)}) \nonumber\\
&= \mathrm{DOS}^{(c)}(E) \: ,
\end{align}
\begin{align}
\int \mathcal{D}^{(c)}(\vb{r},\vb{k},E) \: \frac{\mathrm{d}^d k}{(2\pi)^d} &= \sum_\mu |\chic (\vb{r})|^2 \delta (E - E_\mu^{(c)}) \nonumber\\
&= \mathrm{LDOS}^{(c)}(\vb{r}, E) \: ,
\end{align}
and
\begin{align}
\int \mathcal{D}^{(c)}(\vb{r},\vb{k},E) \: \mathrm{d}^d r &= \sum_\mu |\hat{\chi}_{\mu}^{(c)} (\vb{k})|^2 \delta (E - E_\mu^{(c)}) \nonumber\\
&= \mathrm{MDOS}^{(c)}(\vb{k}, E) \: .
\end{align}

In other words,  all the usual densities of states can be recovered as marginal distributions of the quasi-density of states in phase space since the latter inherits the properties on the marginal distributions of the Wigner transform by construction.

\section{The two-particle picture}\label{app:deriv2}

We present in this appendix an alternative derivation of Eq.~(\ref{eq:C:weyl:final}), which gives a complementary physical picture to the problem. First, we recast the inner product in Eq.~(\ref{eq:coupling_density}) as follows:
\begin{align}
\BraKet{\chic}{\chiv} &= \int {\chic}^* (\vb{r}) \chiv(\vb{r}) \: \mathrm{d}^d r \nonumber\\
&= \iint {\chic}^*(\vb{r}) \chiv(\Vie{r}{}{\prime}) \, \delta(\vb{r} - \Vie{r}{}{\prime}) \: \mathrm{d}^d r^\prime \mathrm{d}^d r \nonumber\\
&= \BraKet{ \chi_{\mu \nu}^{(c,v)}}{ \delta_\mathrm{diag}  } \: ,
\end{align} 
where the last bracket denotes a $2d$-dimensional inner product (in fact, a duality bracket in a space of distributions). Here we have defined the state $\Ket{\chi_{\mu \nu}^{(c,v)}} = \Ket{\chi_{\mu}^{(c)}} \otimes \Ket{\chi_{\nu}^{(v)}}^* $ whose wave function is given by $\chi_{\mu \nu}^{(c,v)}(\vb{r},\Vie{r}{}{\prime}) =  {\chic}(\vb{r}) {\chiv}^*(\Vie{r}{}{\prime})$, and $ \delta_\mathrm{diag}(\vb{r},\Vie{r}{}{\prime}) = \delta(\vb{r} - \Vie{r}{}{\prime})$ is the diagonal Dirac distribution. Equation~(\ref{eq:coupling_density}) can thus be recast as
 \begin{equation}
\mathcal{C} (\hv) = \sum_{\mu, \nu} \Big| \BraKet{ \chi_{\mu \nu}^{(c,v)} }{ \delta_\mathrm{diag} }  \Big|^2 \, \delta \Big( E_{\mu \nu}^{(c,v)} - \hbar \omega \Big) = A_{\delta_\mathrm{diag}} (\hv) \:  ,
\label{eq:coupling_density:2d}
 \end{equation}
with $E_{\mu \nu}^{(c,v)} = E_{\mu}^{(c)} - E_{\nu}^{(v)}$. The doubling of variables suggests interpreting Eq.~(\ref{eq:coupling_density:2d}) as the so-called \emph{spectral function}, $A_{\delta_\mathrm{diag}}$,  associated to the distribution $\delta_\mathrm{diag}$, which corresponds to two particles found at the same position $\vb{r} = \Vie{r}{}{\prime}$, for the $2d$-dimensional Hamiltonian $\hat{H} = \hat{H}_c \otimes \hat{I} - \hat{I} \otimes \hat{H}_v$ of independent particles. Indeed, since $\chic$ and $\chiv$ are eigenfunctions of $\hat{H}_c$ and $\hat{H}_v$ with eigenenergies $E_\mu^{(c)}$ and $E_\nu^{(v)}$, the product $\chi_{\mu \nu}^{(c,v)}(\vb{r},\Vie{r}{}{\prime}) =  {\chic}(\vb{r}) {\chiv}^*(\Vie{r}{}{\prime})$  is an eigenfunction of the Hamiltonian $\hat{H}$ with eigenenergy $E_{\mu \nu}^{(c,v)}$.

The spectral function Eq.~(\ref{eq:coupling_density:2d}) can thus be interpreted as the energy distribution associated to the two-particle state $\delta_\mathrm{diag}$ evaluated at energy $E = \hv$. In particular, any moment of the energy in this state, $\left\langle E^n \right\rangle$, is given by
\begin{equation}
\Bra{ \delta_\mathrm{diag}} \hat{H}^n \Ket{ \delta_\mathrm{diag}} = \int E^n \, A_{\delta_\mathrm{diag}} (E) \: \mathrm{d}E \: .
\end{equation}
The left-hand side of the above equation can also be expressed in phase space by using the Wigner-Weyl formalism. We denote the Wigner transform of a function $\psi$ in $2d$ dimension as
\begin{widetext}
\begin{equation}
W_{\psi} (\vb{r}, \Vie{r}{}{\prime}, \vb{k}, \Vie{k}{}{\prime}) = \iint \psi^* \Big(\vb{r} - \frac{\Vie{x}{}{}}{2}, \Vie{r}{}{\prime} - \frac{\Vie{x}{}{\prime}}{2} \Big) \psi \Big(\vb{r} + \frac{\Vie{x}{}{}}{2}, \Vie{r}{}{\prime} + \frac{\Vie{x}{}{\prime}}{2} \Big) \, \exp \Big( - i \vb{k} \cdot \Vie{x}{}{} - i \Vie{k}{}{\prime} \cdot \Vie{x}{}{\prime} \Big) \: \mathrm{d}^d x \, \mathrm{d}^d x^\prime \: .
\end{equation}
\end{widetext}
Note that here, we have explicitly expressed the Wigner transform in the $4d$-dimensional phase space associated to our problem, hence the variables $\vb{r}$, $\Vie{r}{}{\prime}$, $\vb{k}$, and $\Vie{k}{}{\prime}$.
The Wigner transform of the diagonal state $\delta_\mathrm{diag}$, which will be useful below, can be easily computed and reads
\begin{equation}
W_{\delta_\mathrm{diag}} (\vb{r},\Vie{r}{}{\prime},\vb{k},\Vie{k}{}{\prime}) = (2\pi)^d \delta(\vb{r} - \Vie{r}{}{\prime}) \delta(\vb{k} + \Vie{k}{}{\prime}) \: .
\label{eq:wigner:D}
\end{equation}
The expectation value of an operator $\hat{M}$ for a state $\psi$, i.e., $\Bra{ \psi} \hat{M} \Ket{\psi}$ can be expressed in terms of the Wigner transform of $\psi$ and the Weyl transform $M$ associated to the operator $\hat{M}$ as \cite{Case}
\begin{align}
\Bra{\psi} \hat{M} \Ket{\psi } = & \frac{1}{(2\pi)^{2d}} \, \int_{\mathbb{R}^{4d}} W_\psi (\vb{r},\Vie{r}{}{\prime},\vb{k},\Vie{k}{}{\prime}) \nonumber\\
& \times M(\vb{r},\Vie{r}{}{\prime},\vb{k},\Vie{k}{}{\prime}) \: \mathrm{d}^d r \, \mathrm{d}^d r^\prime \, \mathrm{d}^d k \, \mathrm{d}^d k^\prime \: , 
\label{eq:EV}
\end{align}
where the Weyl transform of the operator is given by
\begin{widetext}
\begin{equation}
M(\vb{r},\Vie{r}{}{\prime},\vb{k},\Vie{k}{}{\prime}) =  \iint  \left\langle \vb{r} + \frac{\Vie{x}{}{}}{2}, \Vie{r}{}{\prime} + \frac{\Vie{x}{}{\prime}}{2} \right| \hat{M}  \left| \vb{r} - \frac{\Vie{x}{}{}}{2}, \Vie{r}{}{\prime} - \frac{\Vie{x}{}{\prime}}{2} \right\rangle \, \exp \Big( - i \vb{k} \cdot \Vie{x}{}{} - i \Vie{k}{}{\prime} \cdot \Vie{x}{}{\prime} \Big) \: \mathrm{d}^d x \, \mathrm{d}^d x^\prime \: .
\end{equation}
\end{widetext}
Of particular interest in our study is the Hamiltonian $\hat{M} = \hat{H}$.  Provided the effective masses vary weakly, the Weyl transform of the Hamiltonian $\hat{H}$ reads
\begin{equation}
H (\vb{r},\Vie{r}{}{\prime},\vb{k},\Vie{k}{}{\prime}) = \frac{\hbar^2 k^2}{2 m_c(\vb{r}) } - \frac{\hbar^2 {k^\prime}^2}{2 m_v(\Vie{r}{}{\prime}) } + E_c(\vb{r}) - E_v(\Vie{r}{}{\prime}) \: .
\end{equation}
Hence, in view of Eq.~(\ref{eq:EV}) and Eq.~(\ref{eq:wigner:D}), the expectation value of the energy in the state $\delta_\mathrm{diag}$ is
\begin{equation}
\Bra{\delta_\mathrm{diag}} \hat{H} \Ket{\delta_\mathrm{diag}} = \iint \left[ \frac{\hbar^2 k^2}{2 m_r (\vb{r}) } + E_g (\vb{r}) \right] \: \frac{ \mathrm{d}^d r \, \mathrm{d}^d k}{(2 \pi)^d} \: .
\label{eq:dHd}
\end{equation}
The above equation may seem at first sight problematic since, mathematically speaking, the integral on the right-hand side clearly diverges. However, we must notice that the left-hand side also diverges as can be seen from the definition of the diagonal Dirac distribution (a state which is perfectly localized has a non-normalizable energy spectrum).  To make our calculations rigorous, one would need to regularize the diagonal Dirac distribution or, equivalently, introduce a high energy cut off and study an appropriate limit.  Here, we are rather interested in manipulating Eq.~(\ref{eq:dHd}) formally.  We may rewrite the right-hand side as
\begin{align}
\Bra{\delta_\mathrm{diag}} \hat{H} \Ket{\delta_\mathrm{diag}} &= \int E \int_{E < \frac{\hbar^2 k^2}{2 m_r (\vb{r}) } + E_g (\vb{r}) < E + \mathrm{d}E }  \: \frac{\mathrm{d}^d r \, \mathrm{d}^d k}{(2 \pi)^d} \nonumber\\
&= \int E \, f(E) \: \mathrm{d}E \: ,
\label{eq:PDF}
\end{align}
where we have formally written
\begin{equation}
f(E) \: \mathrm{d}E = \int_{E < \frac{\hbar^2 k^2}{2 m_r (\vb{r}) } + E_g (\vb{r}) < E + \mathrm{d}E }  \: \frac{\mathrm{d}^d r \, \mathrm{d}^d k}{(2 \pi)^d} \: .
\label{eq:PDF:def}
\end{equation}
Equation~(\ref{eq:PDF}) means that the average energy can be written as the integral of the energy variable against the function $f$ which plays the role of an energy probability density and is given by the Lebesgue measure in phase space of an elementary shell about $\frac{\hbar^2 k^2}{2 m_r (\vb{r}) } + E_g (\vb{r}) = E$ as shown in Eq.~(\ref{eq:PDF:def}). This result is reminiscent of what we have obtained with the quasi-density of states in phase space Eq.~(\ref{eq:C:weyl}). Note that $f$ is \emph{not} the probability density of energy in the state $\delta_\mathrm{diag}$ but only an approximation inducted from the expectation value of the energy. This is to be linked to the plateau function approximation in the point of view of the quasi-density of states in phase space. Finally, another approximation can be obtained by replacing $E_g$ by $E_g^{(\mathrm{eff})}$.


%
\bibliography{biblio}

\end{document}